\documentclass[twocolumn]{aastex61}

\def\vector#1{\mbox{\boldmath $#1$}}

\received{2017 May 18}
\revised{2017 October 30}
\accepted{2017 October 31}

\shorttitle{Mode conversion in NS thermal emission}
\shortauthors{Yatabe \& Yamada}

\begin{document}

\title{Systematic analysis of the effects of mode conversion on thermal radiation from neutron stars}

\correspondingauthor{Akihiro Yatabe}
\email{yatabe@heap.phys.waseda.ac.jp}

\author{Akihiro Yatabe}
\affiliation{Advanced Research Institute for Science and Engineering, Waseda University, 3-4-1, Okubo, Shinjuku, Tokyo 169-8555, Japan}

\author{Shoichi Yamada}
\affiliation{Advanced Research Institute for Science and Engineering, Waseda University, 3-4-1, Okubo, Shinjuku, Tokyo 169-8555, Japan}

\begin{abstract}
In this paper, we systematically calculate the polarization in soft X-rays emitted from magnetized neutron stars, which are expected to be observed by the next-generation X-ray satellites. Magnetars are one of the targets for these observations. This is because thermal radiation is normally observed in the soft X-ray band, and it is thought to be linearly polarized because of different opacities for two polarization modes of photons in the magnetized atmosphere of neutron stars and the dielectric properties of the vacuum in strong magnetic fields. In their previous study, Taverna et al. illustrated how strong magnetic fields influence the behavior of the polarization observables for radiation propagating in vacuo without addressing a precise, physical emission model. In this paper, we pay attention to the conversion of photon polarization modes that can occur in the presence of an atmospheric layer above the neutron star surface, computing the polarization angle and fraction and systematically changing the magnetic field strength, radii of the emission region, temperature, mass, and radii of the neutron stars. We confirmed that if plasma is present, the effects of mode conversion cannot be neglected when the magnetic field is relatively weak, $B \sim 10^{13} \mathrm{G}$. Our results indicate that strongly magnetized ($B \gtrsim 10^{14} \mathrm{G}$) neutron stars are suitable to detect polarizations, but not-so-strongly magnetized ($B \sim 10^{13} \mathrm{G}$) neutron stars will be the ones to confirm the mode conversion.
 
\end{abstract}


\section{Introduction} \label{sec:intro}

X-ray polarimetry will be realized in the near future. In fact, the {\it Imaging X-ray Polarimetry Explorer (IXPE)} was recently selected as the next Small Explorer astrophysics mission of NASA recently and is planned to be launched in 2020 \citep{2013SPIE.8859E..08W}. There are other satellite-borne X-ray polarimetry projects, such as the {\it X-ray Imaging Polarimetry Explorer (XIPE)} \citep{2016SPIE.9905E..15S} and the {\it enhanced X-ray Timing and Polarimetry (eXTP)} \citep{2016SPIE.9905E..1QZ}, which, if approved will advance X-ray astronomy substantially.

Neutron stars are among the targets in some proposed observations in the soft X-ray band, $\sim \mathrm{a \ few \ keV}$. Thermal radiation has been detected from isolated neutron stars such as X-ray dim isolated neutron stars (XDINSs) and magnetars. The polarization of this thermal radiation, if observed, will provide us with an important clue to the physical properties of neutron stars near the surface, as well as the possible configurations of their magnetic fields.

Another aim of the polarimetry is the validation of strong-field quantum electrodynamics (QED), a quantum theory for electrons and photons in the supra-critical electromagnetic fields with strengths $\gtrsim 4.4 \times 10^{13} \mathrm{G}$ in the case of magnetic fields. The strong-field QED has been studied theoretically for a long time \citep{eulerheisenberg,schwinger51,2000pqve.book.....D}: it predicts, for instance, that the vacuum becomes birefringent and a single photon may split into two photons in the presence of strong electromagnetic fields, both of which are absent in the ordinary vacuum and are of purely quantum origin. Although high-intensity laser is supposed to be a promising probe into QED in the strong-field regime~\citep{2006OptCo.267..318H,2006PhRvL..96k0406Z,2007PhRvL..99l9901Z,KingHeinzl2016}, the currently attainable field strength is still much smaller than the critical one \citep{hercules}, and the strong-field QED effects are yet to be observed in laser experiments. In contrast, some neutron stars are believed to possess very strong magnetic fields, which are comparable to or stronger than the critical field \citep{2008A&ARv..15..225M} and may hence be the only realistic possibility to study the strong-field QED for the moment. Recently, a hint of the vacuum polarization effect is obtained in the optical observation of polarizations in the thermal emissions from an XDINS \citep{2017MNRAS.465..492M}.

Photons emitted thermally from the surface of a magnetized neutron star propagate through its magnetosphere. They may be polarized in the atmosphere, and their polarization state will be further modified in the magnetosphere. It is well known that there are generally two elliptical polarization modes for photons propagating in magnetized plasmas \citep{1992hrfm.book.....M}. One is called the ordinary mode ($O$-mode), in which the major axis of the ellipse for the electric field of the photon is parallel to the \vector{k}-\vector{B} plane, with \vector{k} and \vector{B} being the wave vector and the external magnetic field, respectively. The other mode is referred to as the extraordinary mode ($E$-mode), in which the ellipse is perpendicular to the \vector{k}-\vector{B} plane. These situations are not changed if one takes into account the vacuum polarization. Note, however, that the helicities of these modes are changed as the plasma density varies. In fact, when the plasma is dominant, the $O$-mode is left-handed, whereas it becomes right-handed if the vacuum polarization is more important \citep{1979PhRvD..19.3565M,2003ApJ...588..962L}. Incidentally, the two modes are linearly polarized in the limit of the vanishing plasma density. 

For ionized hydrogen atmospheres, which may cover the neutron star surface in a gas state, the opacity is different between the two modes \citep{1974ApJ...190..141L}. In fact, it is lower for the $E$-mode than for the $O$-mode, because the scattering with electrons is suppressed for the former owing to gyration motions of electrons around magnetic field lines. The $E$-mode photons are hence emitted from deeper and hotter regions in the atmosphere than the $O$-mode photons and are dominant when they get out of the atmosphere. Then, the polarization vector of the surface emission is expected to be perpendicular to the $\vector{k}$-$\vector{B}$ plane. Such polarizations may be significantly reduced when integrated over the neutron star surface, however, since the magnetic field is not uniform on the surface and, as a result, the polarizations originated from different parts will cancel each other \citep{2000ApJ...529.1011P}.

Note, in contrast, that the polarization changes adiabatically thereafter during the passage through the magnetosphere of the neutron star \citep{2002PhRvD..66b3002H}. Although such evolutions of the polarization along the photon trajectories were computed and the light curves were obtained by~\cite{2003MNRAS.342..134H}, configurations of the neutron star considered in their paper were limited. \cite{2015MNRAS.454.3254T} conducted more systematic study on the evolution of the polarization in the magnetosphere but with simplifications: they considered QED effects only for photons propagating in vacuo, assuming that all photons are emitted in one of the linearly polarized states. If propagation in a sufficiently dense medium is also considered, conversions of the polarization modes, which are one of the important effects caused by QED, become important. \cite{2003PhRvL..91g1101L} and~\cite{2006MNRAS.373.1495V} took into account both the mode conversion and the radiative transfer in the atmosphere to find the polarization properties. Unfortunately, they considered emissions from a small hot spot alone, which may not be applicable to some neutron stars.

Although it is not considered in this study, the resonant cyclotron scattering occurs in the magnetosphere if the density of charged particles is not low there, and its effect on the polarization was discussed \citep{2008MNRAS.386.1527N,2011ApJ...730..131F,2014MNRAS.438.1686T}. While we pay attention only to the persistent emission from neutron stars in this article, transient phenomena such as the bursts and flares of magnetars were investigated actively these days \citep{2015ApJ...815...45Y,2016MNRAS.461..877V,2017arXiv170501130T}.

Once such polarization features are observed, possibly by the planned satellite-borne detectors, then we may be able to obtain new insights not only into the configuration of the magnetic fields of a neutron star and the thermodynamic state at the neutron star surface but also into the strong-field QED. In fact, \cite{2015MNRAS.454.3254T} calculated the fraction and position angle of polarization for various configurations of a rotating magnetized neutron star, accounting for the vacuum polarization in the magnetosphere as well as geometrical effects. \cite{2016MNRAS.459.3585G} applied the same method with realistic surface emission models to XDINSs and compared the results with observations \citep{2017MNRAS.465..492M}. They detected a possible imprint of the vacuum polarization in strong magnetic fields.

They considered two possibilities for the thermodynamic state of the neutron star surface, i.e., the normal gaseous state and the condensed state. It has been argued that the latter may occur via a phase transition at $T \lesssim 0.1 \mathrm{keV}$ for neutron stars endowed with relatively strong magnetic fields, $B_{p} \gtrsim 10^{13} \mathrm{G}$ such as XDINSs \citep{turolla2004bare,2012A&A...546A.121P}. The polarization properties of the thermal radiation from the bare surface in the condensed state are different from those from the gas atmosphere, and \cite{2016MNRAS.459.3585G} and \cite{2017MNRAS.465..492M} claimed that they will be distinguished in polarimetric observations of the soft X-rays.

Although the dielectric effect of the vacuum polarization and resonant features in the radiative opacities at the vacuum resonance were considered in these papers, the mode conversion at the vacuum resonance was neglected. It may be irrelevant for photons with energies less than 1keV, which are dominant in the thermal emissions from XDINSs, but it cannot be neglected for photons with higher energies of $\sim$ a few keV, which may be radiated as a thermal component in magnetars.

The aim of this paper is to study the polarizations of thermal radiation from isolated rotating magnetized neutron stars more systematically, taking the mode conversion at the vacuum resonance properly into account properly in the formulation of~\cite{2015MNRAS.454.3254T}; we explore a large number of configurations systematically. Inhomogeneities on the neutron star surface, i.e., the possible existence of a hot spot, are also investigated.

The paper is organized as follows. We first describe our method in Section~\ref{methods_and_models}. In Section~\ref{results_and_discussions} we first make some comparisons with the previous study~\citep{2015MNRAS.454.3254T} to validate our method and then show the main results, with a particular emphasis on the vacuum resonance and the hot-spot effects. Some discussions are also given in this section. We summarize this paper in Section~\ref{summary}.

\section{Methods and Models} \label{methods_and_models}
\subsection{Theoretical Overview}

We first summarize some theoretical basics on the behaviors of photons in strongly magnetized plasmas and vacuum and the polarization properties of thermal radiation in X-ray bands from magnetized rotating neutron stars. In the magnetosphere, it suffices to consider the vacuum polarization alone, whereas the contributions from magnetized plasmas also need to be taken into account in the neutron star atmosphere, in which photospheres are located in the case of our current interest.

X-ray photons have two elliptically polarized normal modes in the magnetized plasma, i.e., $O$-mode and $E$-mode. This is also true of the magnetized vacuum. As mentioned already, the $O$-mode has the electric field that traces the ellipse, the major axis of which is parallel to the \vector{k}-\vector{B} plane, whereas for the $E$-mode, it is perpendicular to the plane. What is interesting is that the $O$-mode ($E$-mode) photons in the plasma-dominant regime have the same helicity as the $E$-mode ($O$-mode) photons in the vacuum-dominant regime. As a result of this property, when a photon propagates from the inner atmosphere of neutron star, where the plasma effect is dominant, through the outer part to the magnetosphere, where the vacuum polarization is dominant, the so-called mode conversion may occur from the $O$-mode photon to the $E$-mode and vice versa \citep{1979PhRvD..19.3565M}.

This is also referred to as the vacuum resonance, since the conversion takes place at the resonance point, at which the plasma and vacuum polarizations become comparable to each other. This resonant mode conversion proceeds adiabatically if the following condition is satisfied:
\begin{eqnarray}
E \gg E_{\mathrm{ad}} = 1.49 ( f \tan \theta_B | 1- u_i |)^{2/3} \left( \frac{5 \mathrm{cm}}{H_{\rho}} \right) ^{1/3} \mathrm{keV}, \hspace{4mm}  \label{adiabatic_energy}
\end{eqnarray}
where $f$ is a factor of the order of unity and will be explained below separately; $E$ is the photon energy; $\theta _B$ is the angle between \vector{k} and \vector{B}; $u_i = ( E_{ci} / E )^2$, with $E_{ci} = \hbar eB/m_p c$ being the cyclotron energy of the proton; and $H_{\rho}$ is the density scale height, i.e., $H_{\rho} \simeq 2kT/(m_p g \cos \theta ) = 19.1 \: T_{1} / (g_{14} \cos \theta) \mathrm{cm}$, for the ionized hydrogen atmosphere with a temperature $kT = 1 \: \mathrm{keV} \: T_1$, a surface gravity $g = 10^{14} \: \mathrm{cm \ s^{-2}}  \: g_{14}$, and the angle $\theta$ between \vector{k} and the surface normal \citep{2002ApJ...566..373L,2003MNRAS.338..233H,2003PhRvL..91g1101L,2003ApJ...588..962L}.

\begin{figure}
  \includegraphics[width=\columnwidth]{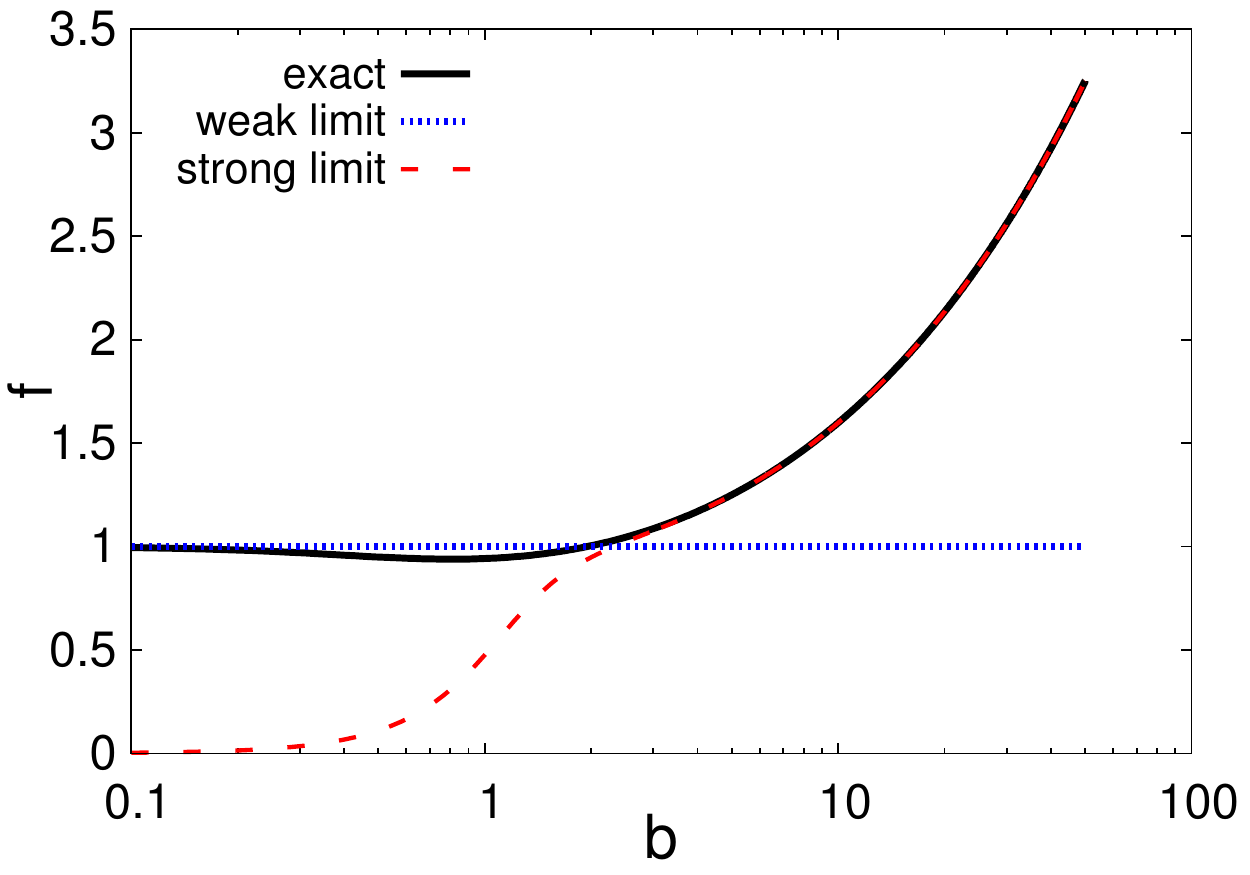}
  \caption{Comparison of the approximate expressions for $f$. The solid black line is the exact result, whereas the blue dotted line and the red dashed line show the results in the weak (Equations (\ref{q_weak}) and (\ref{m_weak})) and strong limits (Equations (\ref{q_strong}) and (\ref{m_strong})) limits, respectively.}
\label{ffactor}
\end{figure}

The factor $f$ in Equation (\ref{adiabatic_energy}) is expressed as $f = [ 3 \delta / (q+m)]^{1/2}$, where $\delta = ( \alpha /45 \pi ) b^2$, with $\alpha = 1/137$ being the fine structure constant and $b = B /B_Q$ being the field strength normalized with the critical field strength, given as $B_Q = m_e ^2 c^3 / e \hbar = 4.4 \times 10^{13} \mathrm{G}$. Parameters $q$ and $m$ are defined in the following formulae \citep{1997PhRvD..55.2449H,1997JPhA...30.6485H}:
\begin{eqnarray}
q &=& \int ^{\infty} _{0} d \eta \frac{e^2 e^{- \eta}}{48 b \pi ^2 \eta ^2 \sinh ^2 (b \eta)} \nonumber \\
 &\times& \left\{ 6 b \eta + ( -3 + 2 b^2 \eta ^2) \sinh (2b \eta) \right\},  \label{q_factor} \\
 m &=& \int ^{\infty} _{0} d \eta \frac{e^2 e^{- \eta}}{32 b \pi ^2 \eta ^2 \sinh ^2 ( b \eta)} \nonumber \\
 &\times& \left\{ -4 b \eta + (1 + 8 b^2 \eta ^2 ) \frac{\cosh (b \eta)}{\sinh ( b \eta)} - \frac{\cosh ( 3 b \eta)}{\sinh (b \eta )} \right\}, \label{m_factor}
\end{eqnarray}
the derivation of which is given in Appendix \ref{derivation_of_parameters}, but they can be well approximated as
\begin{eqnarray}
& &  q \simeq 7 \delta , \label{q_weak} \\
& & m \simeq -4 \delta , \label{m_weak}
 \end{eqnarray}
for $B \ll B_Q$ and as
\begin{eqnarray}
& & q \simeq - \frac{\alpha}{2 \pi} \left[ - \frac{2}{3} b + 1.272 - \frac{1}{b} ( 0.307 + \ln b ) - \frac{0.7}{b^2} \right], \label{q_strong} \\
& & m \simeq - \frac{\alpha}{2 \pi} \left[ \frac{2}{3} + \frac{1}{b} ( 0.145 - \ln b ) - \frac{1}{b^2} \right] , \label{m_strong}
\end{eqnarray}
for $B \gtrsim B_Q$~\citep{2002ApJ...566..373L}. We compare these approximate expressions for $f$ with the exact one in Fig.~\ref{ffactor}. In our calculations, we employ Equations (\ref{q_weak}) and (\ref{m_weak}) for $b < 0.1$, whereas we adopt Equations (\ref{q_strong}) and (\ref{m_strong}) for $b \geq 50$. In between, we use the exact expressions (Equations (\ref{q_factor}) and (\ref{m_factor})).

For $E \sim E_{\mathrm{ad}}$, the adiabatic approximation is no longer valid. The mode conversion occurs only partially, and its probability may be given approximately \citep{2003PhRvL..91g1101L} as
\begin{eqnarray}
 P_{\mathrm{con}} = 1 - \exp \left[ - \left( \frac{\pi}{2} \right) \left( \frac{E}{E_{\mathrm{ad}}} \right) ^3 \right]. \label{conversion_rate}
\end{eqnarray}

The surface radiation of neutron stars is thought to be strongly polarized. This is because the opacity for the $E$-mode photon is smaller in magnetized plasma compared with that of the $O$-mode photon, $\kappa _E \sim ( E/E_{ce} )^2 \kappa _O$, where $E_{ce} = \hbar e B / m_e c$ is the electron cyclotron energy \citep{1974ApJ...190..141L}. The photosphere of the $E$-mode is hence located inside the photosphere of the $O$-mode; i.e., the $E$-mode photons are emitted from deeper and hotter regions in the atmosphere than the $O$-mode photons. As a result, emergent photons are dominated by the $E$-mode photons. Since we focus on how the mode conversion affects the photon polarization, and solving the radiative transfer of photons in the atmosphere is outside the scope of this paper, we assume for the sake of simplicity that photons are all in the $E$-mode at the top of the atmosphere in the absence of the mode conversion.

The mode conversion modifies the polarization produced in the surface radiation. It is the relative positions of the vacuum resonance point with respect to the photospheres that are relevant here. When the magnetic field is not so strong and satisfies the condition
\begin{eqnarray}
 B < B_l \simeq 4.9 \times 10^{13}  \ \mathrm{G} \ f T_{1} ^{-1/8} E_{1} ^{-1/4} G^{-1/4}, \label{B_l}
\end{eqnarray}
where $E = 1 \mathrm{keV} \: E_{1}$ and $G = 1 - e^{-E / kT}$, with $E$ being the photon energy, the vacuum resonance point lies outside the photospheres for both the $E$- and $O$-modes. If the magnetic field is stronger, in contrast, and the following condition holds, 
\begin{eqnarray}
 B_l < B < B_h \simeq 2.8 \times 10^{16}  \ \mathrm{G} \ f^2 T_1 ^{-1/4} E_1 ^{-3/2} G^{-1/2}, \hspace{3mm} \label{B_h}
\end{eqnarray}
the vacuum resonance point is still located outside the $E$-mode photosphere but now lies inside the $O$-mode photosphere \citep{2003ApJ...588..962L}. It follows, then, that when $B< B_l$, both the $E$- and $O$-modes photons experience mode conversion, and the $O$-mode, into which the originally dominant $E$-mode is converted, becomes predominant as long as the photon energy satisfies the adiabaticity condition: $E \gtrsim E_{\mathrm{ad}}$ \citep{2003PhRvL..91g1101L}. If $B_l < B < B_h$ is met, in contrast, the $E$-mode photons emitted from their photosphere transform into the $O$-mode photons at the vacuum resonance point. Since this point is inside the $O$-mode photosphere, the $O$-mode photons thus converted cannot escape immediately and diffuse out until the O-mode photosphere is reached. The $E$-mode photons generated at the vacuum resonance point, in contrast, can escape as soon as they are produced, since matter is transparent for them there. This implies that the vacuum resonance point behaves as the effective $E$-mode photosphere, whereas the $O$-mode photosphere is essentially intact; as a result, the $E$-mode is dominant in this case \citep{2003ApJ...588..962L}. In this paper, we assume that all photons are initially emitted in the $E$-mode from their photosphere if $B<B_l$ and from the resonance point if $B>B_l$. We also explicitly take into account the mode conversion only for the former, although even in the case of $B > B_l$, the mode conversion occurs in the atmosphere between the photospheres of the two modes.

The polarization is further modified in the magnetosphere according to the equation
\begin{eqnarray}
\frac{d}{dZ} \left(
\begin{array}{c}
A_X \\
A_Y \\
\end{array}
\right) = \frac{i k_0 \delta}{2} \left(
\begin{array}{cc}
M & P \\
P & N \\
\end{array}
\right) \left(
\begin{array}{c}
A_X \\
A_Y \\
\end{array}
\right) \label{polarization_evolution}
\end{eqnarray}
for photons propagating in the $Z$-direction, where $\vector{A} = ( A_X , A_Y )$ are the $X$- and $Y$-components of the electric-field amplitude of the photon with an angular frequency $\omega$, $k_0 = \omega / c$, $\delta = ( \alpha /45 \pi ) b^2$. Here $M$, $N$, and $P$ are given as $M = 7 \hat{B}_X \hat{B}_X + 4 \hat{B}_Y \hat{B}_Y$, $N = 4 \hat{B}_X \hat{B}_X + 7 \hat{B}_Y \hat{B}_Y$, and $P = 3 \hat{B}_X \hat{B}_Y$, where $\hat{B}_X = {\bf B} \cdot \hat{{\bf e}}_X / |{\bf B}|$ and $\hat{B}_Y = {\bf B} \cdot \hat{{\bf e}}_Y  / |{\bf B}|$ are the $X$- and $Y$-components of the unit vector aligned with the magnetic field, respectively. The above equation is the same as Equations (21) and (22) in \cite{2011ApJ...730..131F}, except that those authors assumed that the magnetic field lies in the $X$-$Z$ plane, which is not assumed in our paper for numerical convenience (see also \cite{2014MNRAS.438.1686T,2015MNRAS.454.3254T}). Note that these expressions of $M$, $N$, and $P$ are valid in the weak-field limit ($B \ll B_Q$), which is well satisfied in the magnetosphere in the present case. There are two length scales of relevance in these equations: one is the scaled wavelength of the photon, $l_A = 2 / k_0 \delta$, and the other is the scale height of the magnetic field in the direction of the wave vector, $l_B = |\vector{B}| |\vector{k}| / | \vector{k} {\bf \cdot} \nabla \vector{B}| \sim r$, where $r$ is the radial distance. If the wavelength of the photon is short and/or the magnetic field is strong, satisfying $l_A \ll l _B$, then the polarization varies adiabatically as the direction of the external magnetic field changes slowly. If the opposite is true, $l_A \gg l_B$, in contrast, the polarization cannot follow the rapid change of the magnetic field
and is unchanged. This means that the polarization is essentially fixed at the point corresponding to the so-called polarization-limiting radius, at which $l_A = l_B$ is satisfied.

This point is somewhat far from the surface if the magnetic field is strong, $B \sim 10^{14} \mathrm{G}$ and is given, for example, as
\begin{eqnarray}
 \frac{r_{pl}}{R_{\mathrm{NS}}} \simeq 74 \left( \frac{B_p}{10^{14} \mathrm{G} } \right) ^{2/5} \left( \frac{E}{1 \mathrm{keV}} \right) ^{1/5} \left( \frac{R_{\mathrm{NS}}}{10 \mathrm{km}} \right) ^{1/5} , \hspace{4mm} \label{polarization_limiting_radius}
\end{eqnarray}
on the symmetry axis of a dipolar magnetic field, where $B_p$ is the field strength at the magnetic pole and $R_{\mathrm{NS}}$ is the radius of the neutron star. If one considers an imaginary surface that is formed by the polarization-limiting radii and referred to hereafter as the polarization-limiting surface, the photons reaching a distant observer should pass through a small patch on the surface. Since the magnetic field is fairly uniform on the patch, the superposition of radiation coming from different portions on the neutron star surface does not cancel the polarizations \citep{2002PhRvD..66b3002H}.

Although the evolution of polarization in the magnetosphere is obtained by solving Equation (\ref{polarization_evolution}) in principle, we use the adiabatic approximation; i.e., the polarization state follows the change in the eigenvectors of the matrix in Equation (\ref{polarization_evolution}): ${\bf b}_E = (-\hat{B}_Y , \hat{B}_X)$ and ${\bf b}_O = (\hat{B}_X , \hat{B}_Y)$, which correspond to the $E$- and $O$-mode, respectively. It is true that the adiabaticity is violated near the limiting radius, but we ignore it for simplicity and apply the approximation down to the limiting radius, at which we evaluate the final polarization state \citep{2015MNRAS.454.3254T}.

\subsection{Method} \label{method}

\begin{figure}
  \begin{center}
  \includegraphics[width=6cm]{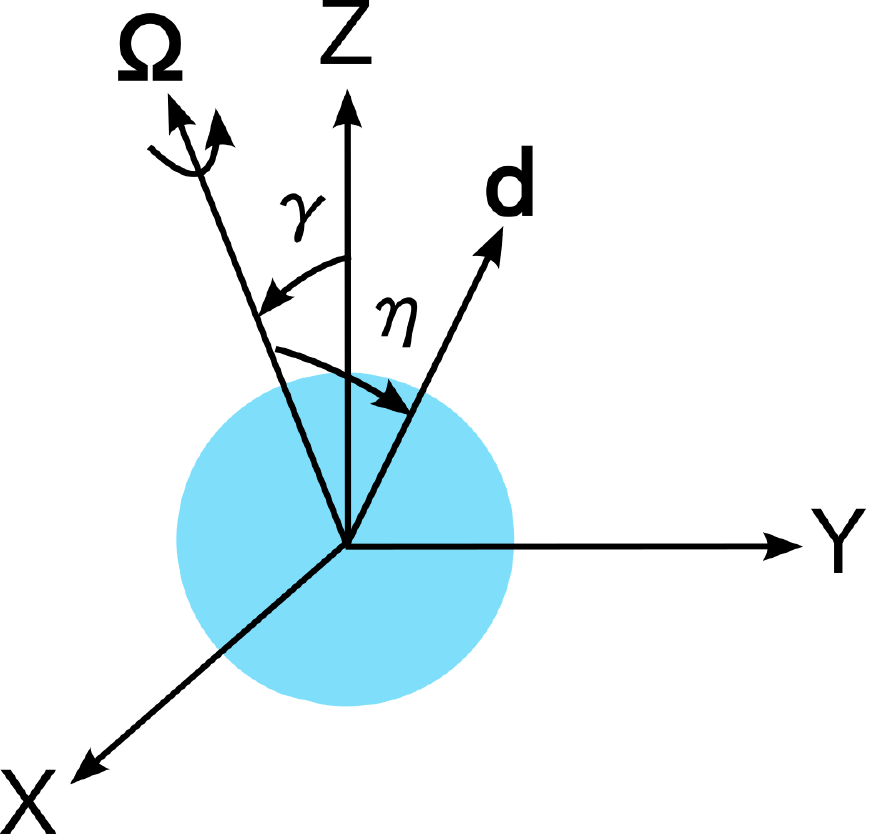}
  \caption{Configuration of a magnetar. The rotation axis and magnetic dipole are denoted by ${\bf \Omega}$ and ${\bf d}$, respectively. The angles that ${\bf \Omega}$ makes with the $Z$-axis and ${\bf d}$ are denoted by $\gamma$ and $\eta$, respectively. The observer is assumed to be sitting at infinity on the $Z$-axis.}
   \label{config}
  \end{center}
\end{figure}

We now explain the procedure to obtain the polarization angle and fraction of X-rays emitted from magnetars based on the picture just mentioned. We first specify the configuration of the magnetic field. In this paper, we consider only dipole magnetic fields, although the formulation is applicable to other configurations as well. We introduce coordinates as shown in Figure \ref{config}. In this frame, an observer is assumed to be sitting at an infinite distance on the positive $Z$-axis. We assume without loss of generality that the spin axis of the magnetar (${\bf \Omega}$) lies in the $X$-$Z$ plane and that the angle between the $Z$-axis and the spin axis ${\bf  \Omega}$ is $\gamma$. The magnetic dipole \vector{d} is assumed to be tilted from the rotation axis by an angle $\eta$. Its rotation around ${\bf \Omega}$ is specified by another angle $\psi$, which is measured from the $X$-$Z$ plane. The magnetic dipole moment $\vector{d}$ in this reference frame is expressed as
\begin{eqnarray}
 \vector{d} = R_Y ( \gamma ) R_Z ( \psi ) R_Y ( \eta ) \vector{d}_Z, \label{magnetic_moment}
\end{eqnarray}
where
\begin{eqnarray}
 R_Y ( \theta ) = \left( 
\begin{array}{ccc}
\cos \theta & 0 & \sin \theta \\
0 & 1 & 0 \\
- \sin \theta & 0 & \cos \theta \\
\end{array}
\right) , \ R_Z ( \theta ) = \left( 
\begin{array}{ccc}
\cos \theta & - \sin \theta & 0 \\
\sin \theta & \cos \theta & 0 \\
0 & 0 & 1 \\
\end{array}
\right) \nonumber 
\end{eqnarray}
are rotational matrices around the $Y$- and $Z$-axes, respectively, and
$\vector{d}_Z = B_p R_{\mathrm{NS}}^3 ( 0, 0, 1/2 )$.

The initial polarization is determined by the magnetic field at the photosphere. As explained earlier, if the condition $B > B_l$ is satisfied, we assume that the radiation is completely in the $E$-mode, though the mode conversion occurs inside the $O$-mode photosphere. If, in contrast, the surface magnetic field satisfies $B < B_l$, then the originally dominant $E$-mode is converted to a mixture of the $O$- and $E$-modes according to Equation (\ref{conversion_rate}). As a result, the radiation generally contains in general both polarized and unpolarized parts, and we consider the former alone in the following. The fraction of the polarized part is $| 1 - 2 P_{\mathrm{con}}|$.

\begin{figure}
  \begin{center}
  \includegraphics[width=\columnwidth]{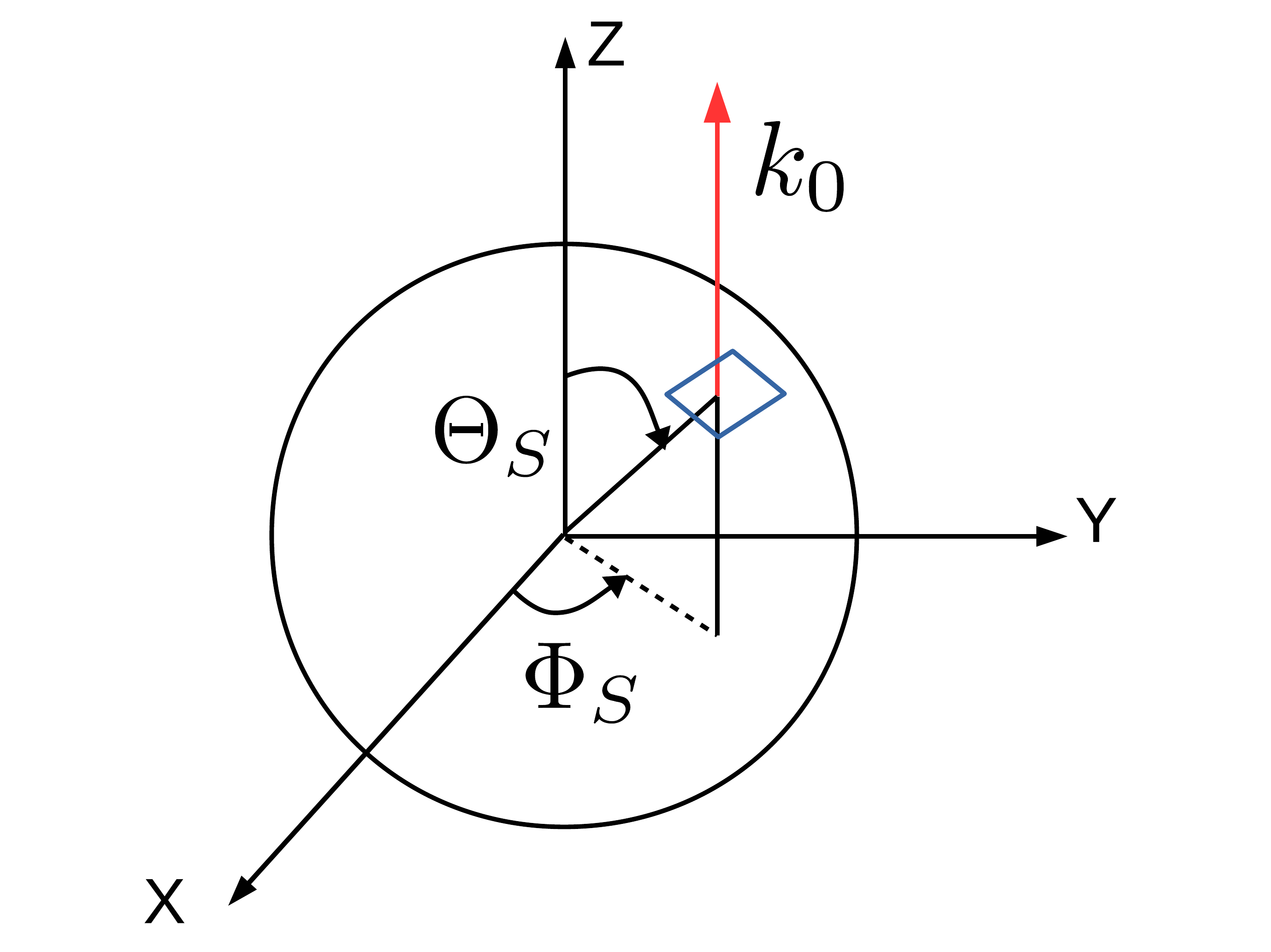}
  \caption{Emission (red arrow) from a patch (blue square) on the (effective) photosphere. The location of the patch is specified by the zenith and azimuth angles $\Theta _S$ and $\Phi _S$. The observer is assumed to be located at infinity on the positive $Z$-axis.}
   \label{polarization-limiting}
  \end{center}
\end{figure}

As mentioned above, we employ the adiabatic approximation in solving Equation (\ref{polarization_evolution}). Then, the solution is expressed as follows:
\begin{eqnarray}
 {\bf A} ({\bf r}) = A_E {\bf b}_E ( {\bf r}) + A_O {\bf b}_O ( {\bf r}),
\end{eqnarray}
in which ${\bf b}_E ( {\bf r})$ and ${\bf b}_O ({\bf r})$ are the eigenvectors of the coefficient matrix in Equation (\ref{polarization_evolution}) at point ${\bf r}$. Since the matrix depends on the magnetic field at each point on the photon trajectory, the eigenvectors also change along the path. Since we assume in this paper that the polarization state is finally fixed at the polarization-limiting radius, it is given by the coefficients $A_E$ and $A_O$ determined at the (effective) photosphere and the eigenvectors at the limiting radius. We neglect gravitational effects such as redshifts and ray bendings other than those on the scale height $H _{\rho}$ of the atmosphere. Observed polarizations are the sum of individual polarizations obtained in the fashion described just now for emissions from different portions of the (effective) photosphere, which are specified by the zenith and azimuth angles, $\Theta_S$ and $\Phi_S$, as shown in Figure \ref{polarization-limiting} \citep{2015MNRAS.454.3254T}.

To derive the polarization angle and fraction, we utilize the Stokes parameters, $Q$, $U$, and $V$, which describe the linear and circular polarizations. They are expressed as
\begin{eqnarray}
& & Q = A^p _X A_X ^{p*} - A^p _Y A_Y ^{p*} , \\
& & U = A^p _X A_Y ^{p*} + A^p _Y A_X ^{p*} , \\
& & V = i ( A^p _X A_Y ^{p*} - A^p _Y A_X ^{p*} ),
\end{eqnarray}
where $\vector{A}^p = ( A^p _X , A^p _Y) $ is the amplitude of the polarized component. The other Stokes parameter, $I$, is nothing but the intensity of the emission. The polarization angle and fraction are finally derived from the Stokes parameters as
\begin{eqnarray}
& &  \chi_p = \frac{1}{2} \mathrm{arctan} \left( \frac{U}{Q} \right),  \\
& & \Pi _L = \frac{\sqrt{Q^2 + U^2}}{I} \label{polarization_fraction_Stokes} .
\end{eqnarray}

Note that the Stokes parameters are additive quantities and are hence used in calculating the polarization properties of spatially and/or temporally integrated radiation. It should be also mentioned that we assume in this paper that the $E$- and $O$-modes are completely uncorrelated with each other. In reality, however, circular polarizations will be produced by the partial mode conversion and they are expected to be correlated \citep{2003PhRvL..91g1101L}. They will also be produced if the magnetic field near the polarization-limiting radius changes rapidly and the polarization cannot catch up. Such situations may occur if the polarization-limiting surface is close to the neutron star \citep{2002PhRvD..66b3002H}. Although the superposition of radiation emitted from different points on the neutron star surface will reduce the circular polarization in general, quantitative investigations are certainly interesting and will be conducted in the future.

\section{Results and Discussion} \label{results_and_discussions}

\subsection{Comparison with Previous Study}

We now apply the formalism developed so far to concrete models. We begin with a comparison with the work by~\cite{2015MNRAS.454.3254T}, in which they studied the polarization of the emissions from the surface of a neutron star with a mass and radius of $M = 1.4 M_{\odot}$ and $R_{\mathrm{NS}} = 10 \mathrm{km}$, respectively. They assumed that the surface temperature is given as $ T ( \theta _{\mathrm{NS}} ) = \mathrm{max} ( T_p | \cos \theta _{\mathrm{NS}} |^{1/2} , T_e )$, where $T_p = 150 \mathrm{eV}$, $T_e = 100 \mathrm{eV}$, and $\theta _{\mathrm{NS}}$ is the zenith angle measured from the north pole of the core-centered dipole field \citep{1983ApJ...271..283G,1995ApJ...442..273P}; the surface emission was assumed to be in the $E$-mode initially. Ignoring the mode conversion entirely, they calculated the phase-resolved polarization fraction and angle, as well as the phase-averaged polarization fraction and semi-amplitude (defined to be half the range of variations in the polarization angle during a single rotation) for different field strengths. They also considered the ray bending and modifications of the dipole magnetic field by the strong gravity of the neutron star.

\begin{figure}[htbp]
    \begin{tabular}{cc}
      \begin{minipage}[t]{0.45\hsize}
        \centering
        (a) No mode conversion
        \includegraphics[width=4cm]{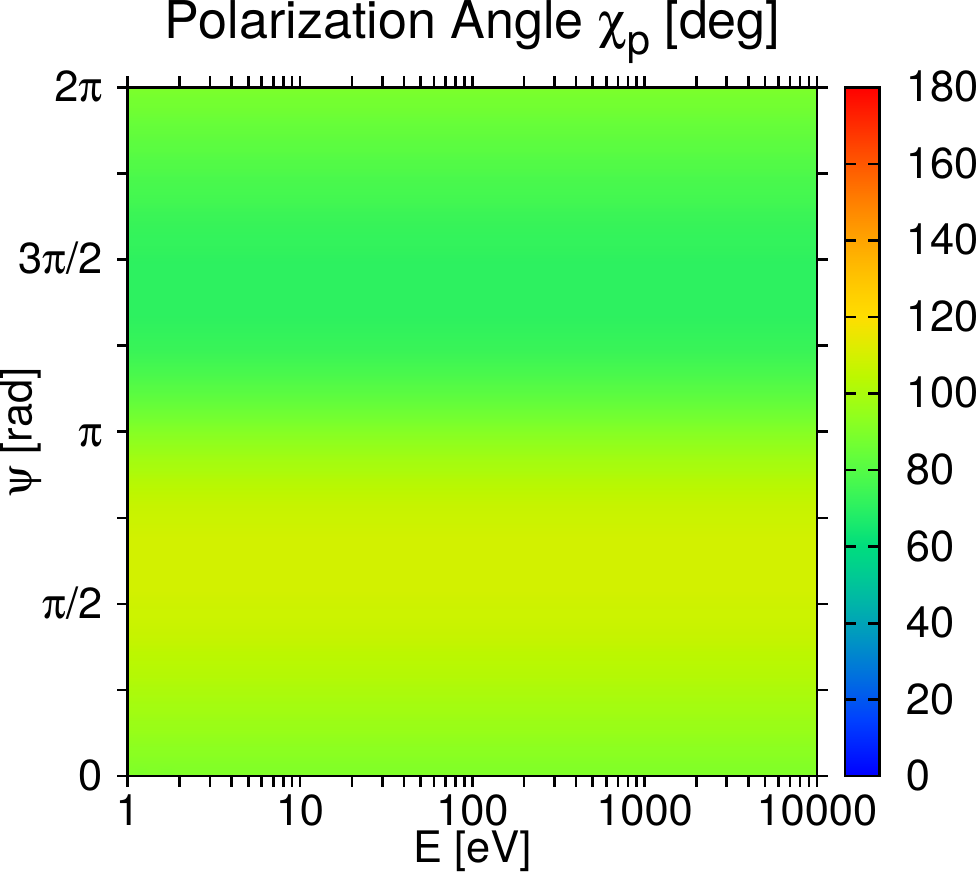} \\
        \includegraphics[width=4cm]{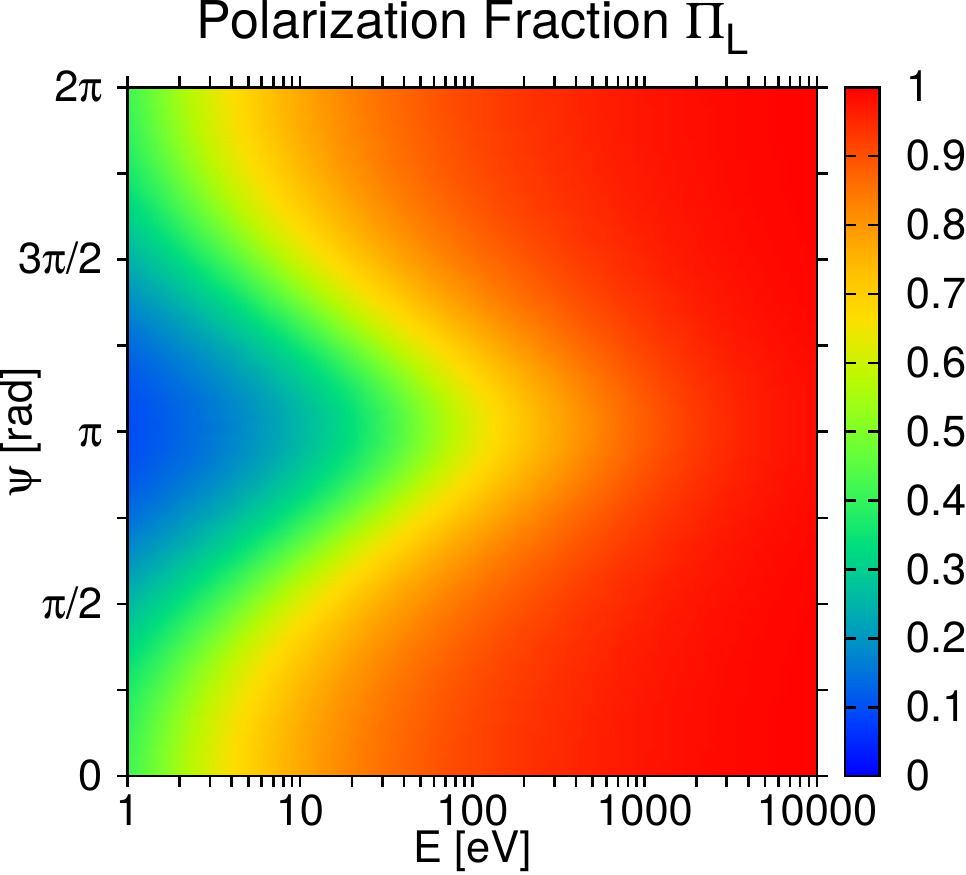} 
      \end{minipage} &
      \begin{minipage}[t]{0.45\hsize}
        \centering
        (b) With mode conversion
        \includegraphics[width=4cm]{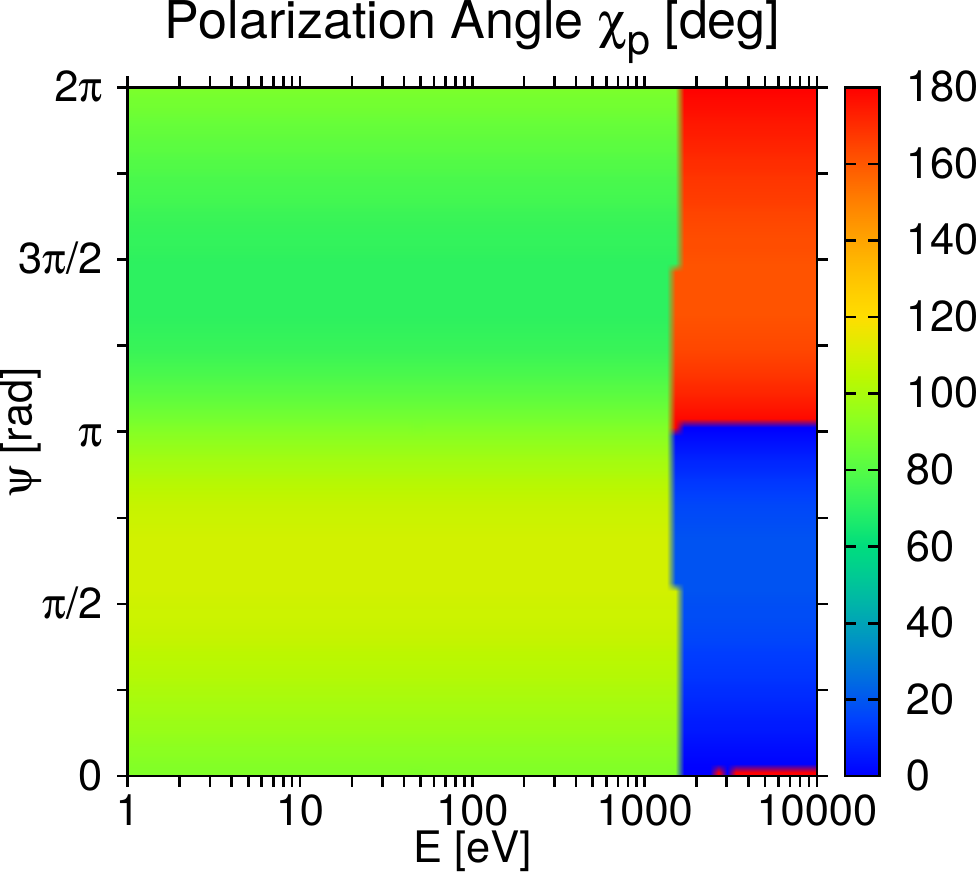} \\
        \includegraphics[width=4cm]{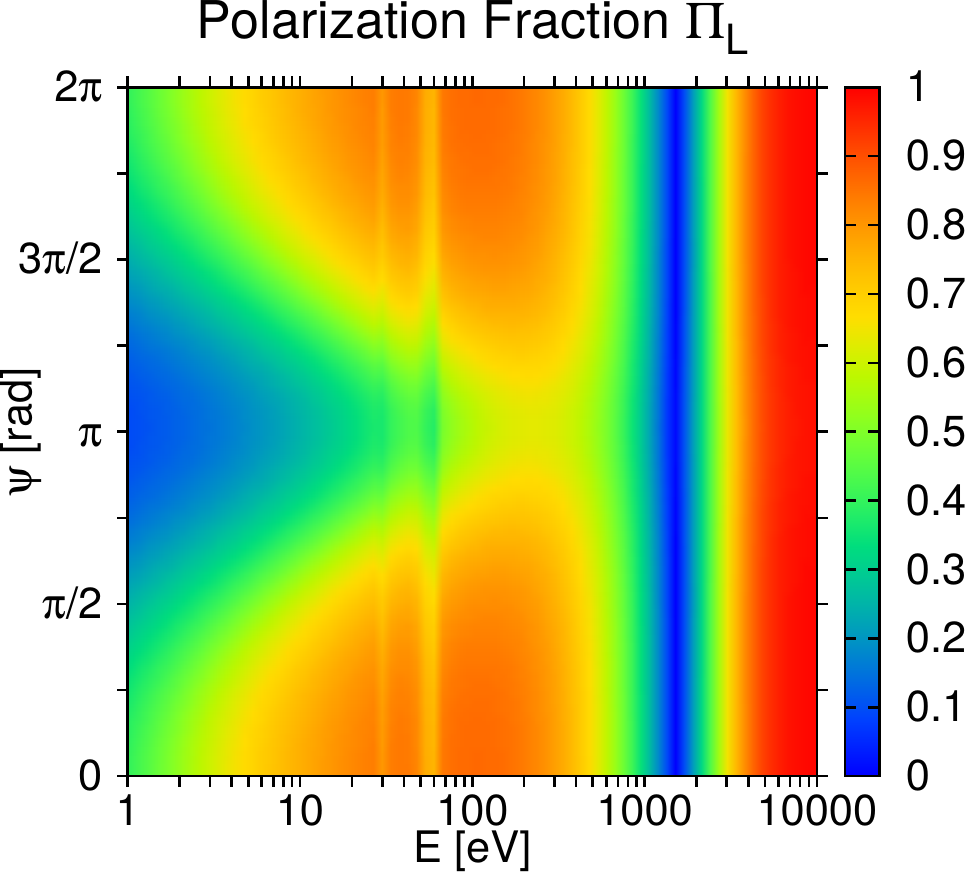}
      \end{minipage}
    \end{tabular}
\caption{Phase-resolved polarization angles (upper panels) and fractions (lower panels) for the same condition as in \cite{2015MNRAS.454.3254T}, except for the relativistic ray bending and the modifications of the magnetic field by the strong gravity near the neutron star. (a) The mode conversion is ignored, as in \cite{2015MNRAS.454.3254T}. (b) The mode conversion is considered.}
  \label{t+tvr}
  \end{figure}

Let us start with the phase-resolved quantities. We apply our method to the same model with $B_p = 10^{13} \mathrm{G}$, $\gamma = 15 ^{\circ}$, and $\eta = 5 ^{\circ}$. Note that $\chi$, $\xi$, and $\gamma$ in~\cite{2015MNRAS.454.3254T} correspond to $\gamma$, $-\eta$, and $-\psi$ in our notation. For comparison, we first neglect the mode conversion. The results are shown in the left panels of Figure \ref{t+tvr}. The upper and lower panels present the polarization angle $\chi _p$ and fraction $\Pi _L$, respectively, as color contours in the $E-\psi$ plane, which are to be compared with Figure 5 in \cite{2015MNRAS.454.3254T}. We find a good agreement between them.

\begin{figure}[htbp]
\begin{center}
 \includegraphics[width=\columnwidth]{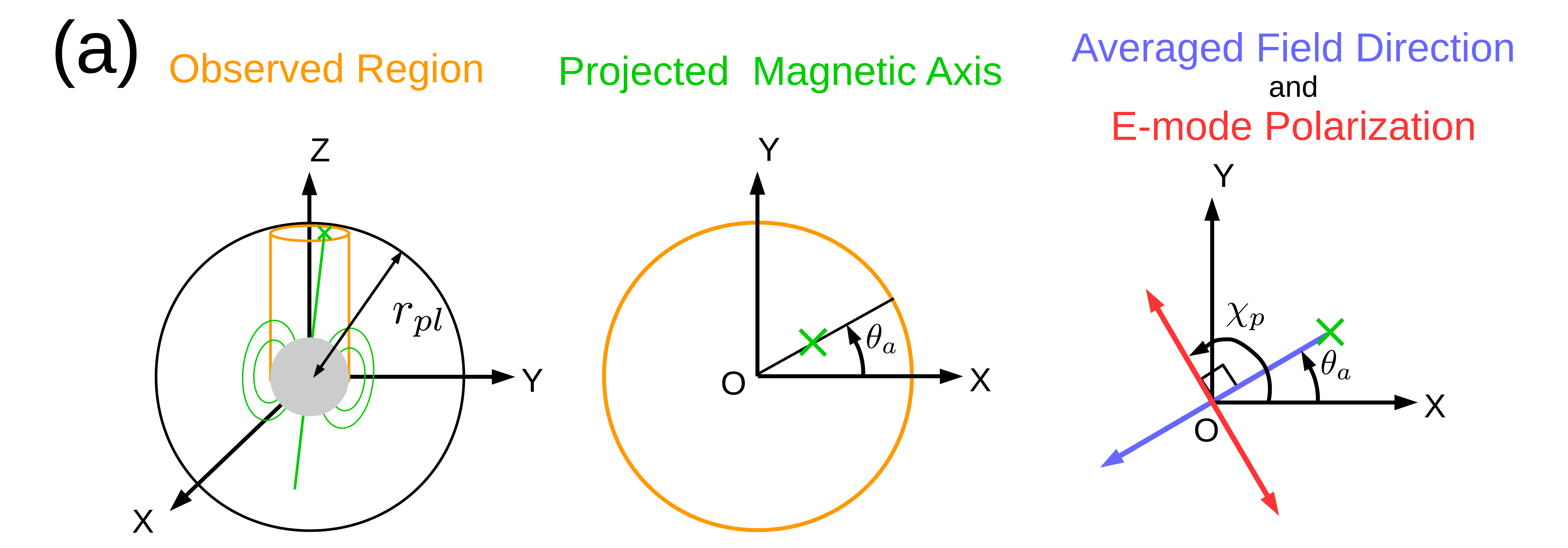} 
\end{center}
\begin{center}
 \includegraphics[width=\columnwidth]{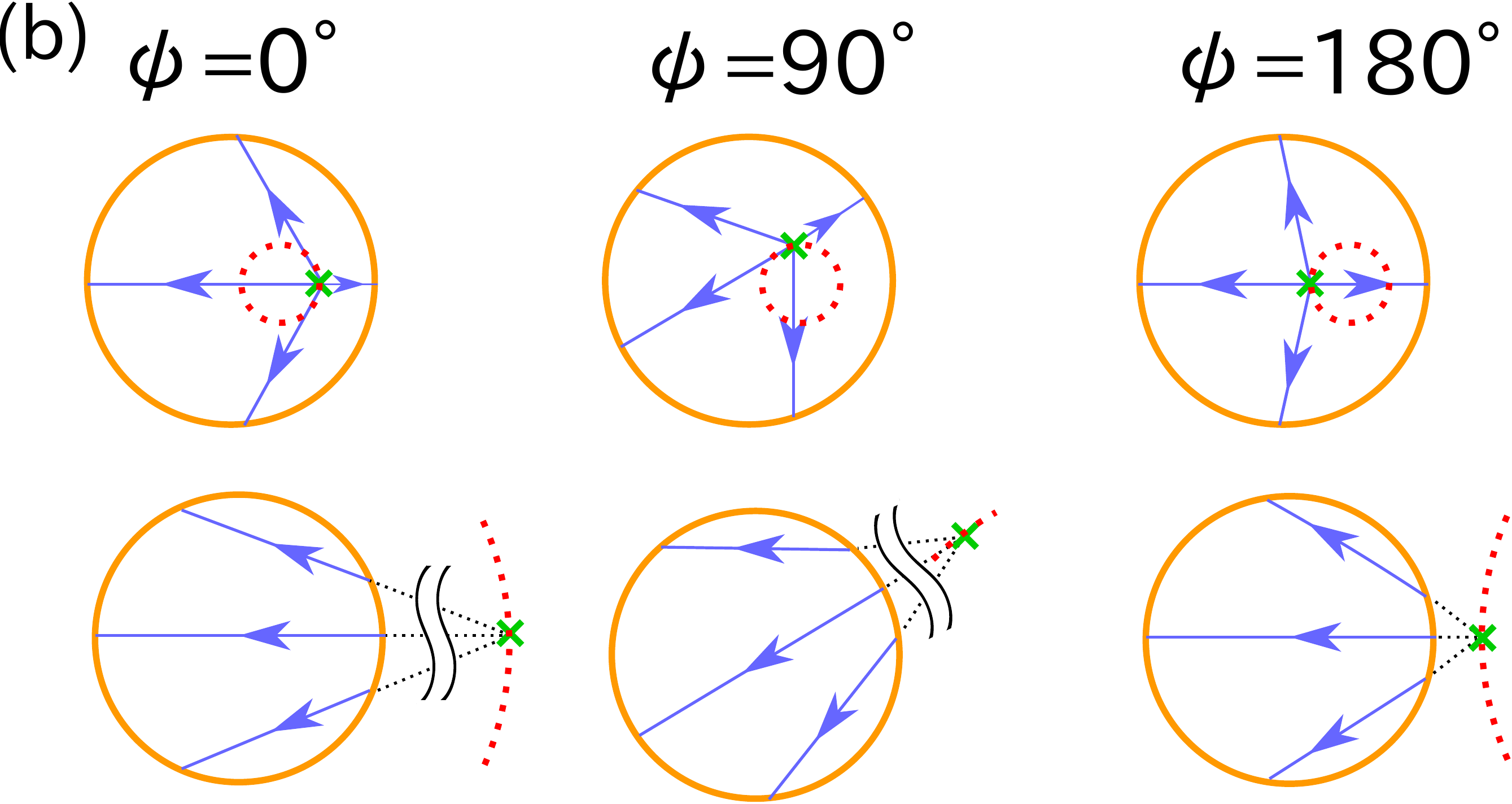} 
\end{center}
 \caption{Schematic pictures to explain how the polarization is determined. (a) The left panel is a snapshot of the configuration of the neutron star considered in Figure \ref{t+tvr}. The central gray sphere is the neutron star, and the green arrow and curves are the star magnetic axis and some field lines, respectively. The orange cylinder is a bundle of light rays emitted from the neutron star parallel to the $Z$-axis; the outer sphere of radius $r_{pl}$ is the polarization-limiting surface. The middle panel is a top view of the patch on the polarization-limiting surface cut out by the cylinder; the green cross is the point where the star magnetic axis meets the polarization-limiting surface, and $\theta_a$ is its angle from the $X$-axis. The right panel is the same as the middle one but for the average magnetic field (blue arrow) and polarization direction (red arrow); the polarization angle is denoted by $\chi_p$. See the text for definitions. (b) The upper three panels show the projected magnetic field lines on the patch at different rotational phases; the green cross is again the magnetic north pole on the polarization-limiting surface. The red dotted circle indicates the locus of the magnetic north pole. The lower three panels show the case with larger $r_{pl}$, in which the magnetic north pole is located outside the observed patch.}
 \label{PAabstract}
\end{figure}

The behavior of the polarization angle is understood from Figure \ref{PAabstract} as follows. In the upper left panel, we draw a schematic picture of a snapshot of the neutron star we are considering now. The central gray sphere is the neutron star, with the green arrow and curves being the magnetic dipole and some field lines, respectively. The outer sphere with the radius $r_{pl}$ is the polarization-limiting surface. Note that the surface is not a sphere in general. Photons reaching the distant observer on the positive $Z$-axis should propagate in the cylinder drawn in orange. 

It is the configuration of the projected magnetic field on the patch of the polarization-limiting surface cut out by this fictitious cylinder that finally determines the polarization angle. In the top middle panel, we schematically depict this patch as the orange circle and mark with the green cross the point at which the star magnetic axis meets the polarization-limiting surface. The angle of this point from the $X$-axis is denoted by $\theta_a$. Note that, depending on the configuration of the neutron star and the radius of the polarization-limiting surface, the green cross may sit outside the orange circle, the radius of which is equal to that of the neutron star (see also the bottom panels). 

The top right panel is the same as the top middle panel, except that the magnetic field averaged over the patch and the corresponding polarization direction are exhibited in blue and red, respectively, instead of the circle to indicate the patch. We find that the average magnetic field, which is defined to be the integral of the (projected) magnetic fields over the observed patch of the polarization-limiting surface divided by its area, is actually directed from the green cross to the origin of the patch from the symmetry of the projected magnetic fields around the green cross. In fact, the angle between the projection of the magnetic axis and the $X$-axis $\theta _a$ is given as $\theta _a = \mathrm{arctan} ( d_2 / d_1)$ from the magnetic dipole momentum expressed as $\vector{d} = (d_1, d_2, d_3)$; then, the orientation from the green cross to the origin on the patch is given by the angle $\theta _a + 180^{\circ}$ from the $X$-axis, which is found to be almost identical to the direction of the averaged magnetic field obtained numerically from the surface integral. In the case of $\gamma = 15^{\circ}, \ \eta = 5^{\circ}$, for example, we find $\theta _a + 180^{\circ} = 198.^{\circ}02$, whereas the numerically obtained value is $198.^{\circ}68$ for $\psi = \pi/2$; they are identical at $\psi = 0$. In the same figure, we assume that the photons are all in the $E$-mode and hence the polarization direction, which is specified by the electric field of the photon, is perpendicular to the (averaged) magnetic field. Then, the polarization angle is given as $\chi_p = \theta_a + 90^{\circ} \ ( \mathrm{mod} \ 180^{\circ})$.

In Figure \ref{PAabstract} (b), we display some (projected) field lines on the observed patch at different rotational phases. As mentioned above, the location of the green cross, i.e., the (extension of the) magnetic north pole to the polarization-limiting surface, may be inside (upper panels) or outside (lower panels) the observed patch. It moves around on the surface, as indicated in red, owing to the rotation of the neutron star. The radius of the trajectory depends on the angles $\gamma$ and $\eta$ (see Figure \ref{config}), which we assume here to be $\gamma=15 ^{\circ}$ and $\eta=5 ^{\circ}$. In this case, it is not very large, and the polarization angle does not change much, as confirmed in the upper left panel of Figure \ref{t+tvr}.

Using the same figure, we can also understand the behavior of the polarization fractions shown in the lower left panel of Figure \ref{t+tvr}. In fact, it is clear from the upper panels of Figure \ref{PAabstract} (b) that the polarization is somewhat canceled when averaged over the observed patch if the green cross, or the magnetic north pole, is located inside the patch. This happens if the magnetic field is weak and/or the photon energy is low, and, as a consequence, the polarization-limiting surface is rather close to the neutron star (see Equation (\ref{polarization_limiting_radius})). Such cancellations do not occur if the polarization-limiting surface is distant from the neutron star and the magnetic north pole sits outside the observed patch on the surface (see the bottom panels of Figure \ref{PAabstract} (b)).

In the lower left panel of Figure \ref{t+tvr}, the polarization fraction is $\sim 1$ at high photon energies, since the polarization-limiting surface is far away from the neutron star and the magnetic north pole is always outside the observed patch during the entire rotation period. As the energy is decreased, this is no longer the case, and the pole enters the patch at some rotational phases near $\psi = \pi$. As a result, the cancellation occurs, and the polarization fraction is reduced there. At very low energies, the north pole stays inside the patch at all times, and the polarization fraction is always low accordingly. 

This is the essential picture in the absence of the mode conversion. We now consider how it is modified by the mode conversion, using the same model.

In the right column of Fig.~\ref{t+tvr}, the results with the mode conversion are displayed. The density scale height is set to $H_{\rho} \simeq 10.3 T_{1} / \cos \theta \ \mathrm{cm}$ in this calculation. The upper and lower panels are for $\chi _p$ and $\Pi_L$, respectively. One can see that they are different from the previous ones for high-energy photons with $E \gtrsim 1 \mathrm{keV}$. The most remarkable is the abrupt change in the polarization angle $\chi_p$ by $\simeq 90^{\circ}$ at $E \simeq 2 \mathrm{keV}$, which indicates that the dominant polarization mode changes from the $E$-mode at low energies to the $O$-mode when the photon energy exceeds 2keV. At this energy, $E \sim E_{\mathrm{ad}}$ is satisfied. The adiabatic mode conversion occurs at the vacuum-resonance points above this energy, whereas the conversion is suppressed below it \citep{2003PhRvL..91g1101L}.

The effects of the mode conversion on the polarization fraction are shown in the bottom right panel. It is remarkable that there is a blue strip at $E \simeq 2 \mathrm{keV}$, where the polarization fraction is very small. As mentioned above, this energy corresponds to the adiabatic energy given by Equation (\ref{adiabatic_energy}). The mode conversion occurs nonadiabatically below this energy, and both the $E$- and $O$-mode photons are emitted according to Equation (\ref{conversion_rate}). In the blue strip of the panel, in particular, both modes are almost equally mixed, and the polarization fraction becomes very small as observed. At much smaller energies, the mode conversion is essentially frozen, and the polarization fraction returns to the original value at emission.

One can recognize, however, that other vertical strips exist at $E \sim 30 \mathrm{eV}$ and $\sim 60 \mathrm{eV}$, where the polarization fraction is somewhat reduced again. These energies are special, corresponding to the cyclotron energies of protons for the magnetic fields of $5 \times 10^{12} \mathrm{G}$ at the (magnetic) equator and $10^{13} \mathrm{G}$ at the (magnetic) pole, respectively. Note that when the photon energy equals the proton cyclotron energy and $E_{\mathrm{ad}} = 0$, the completely adiabatic conversion occurs again for this particular energy of photons. As a result, the $O$-mode photons increase at this energy, reducing the polarization fraction. Note also that the magnetic pole and equator are the two main contributors to the surface emissions in the current configuration (see the explanations given later).

\begin{figure}
    \begin{tabular}{cc}
      \begin{minipage}[t]{0.45\hsize}
	 \centering
        (a) no mode conversion
        \includegraphics[width=4cm]{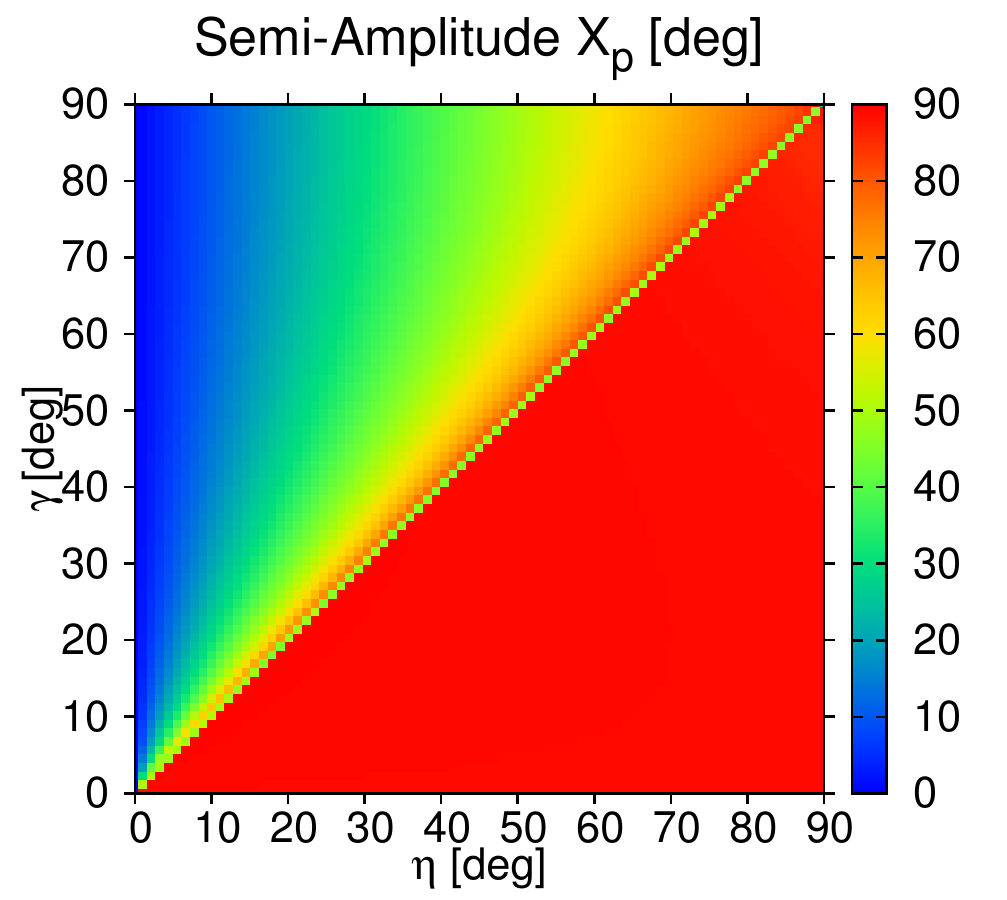}
        \includegraphics[width=4cm]{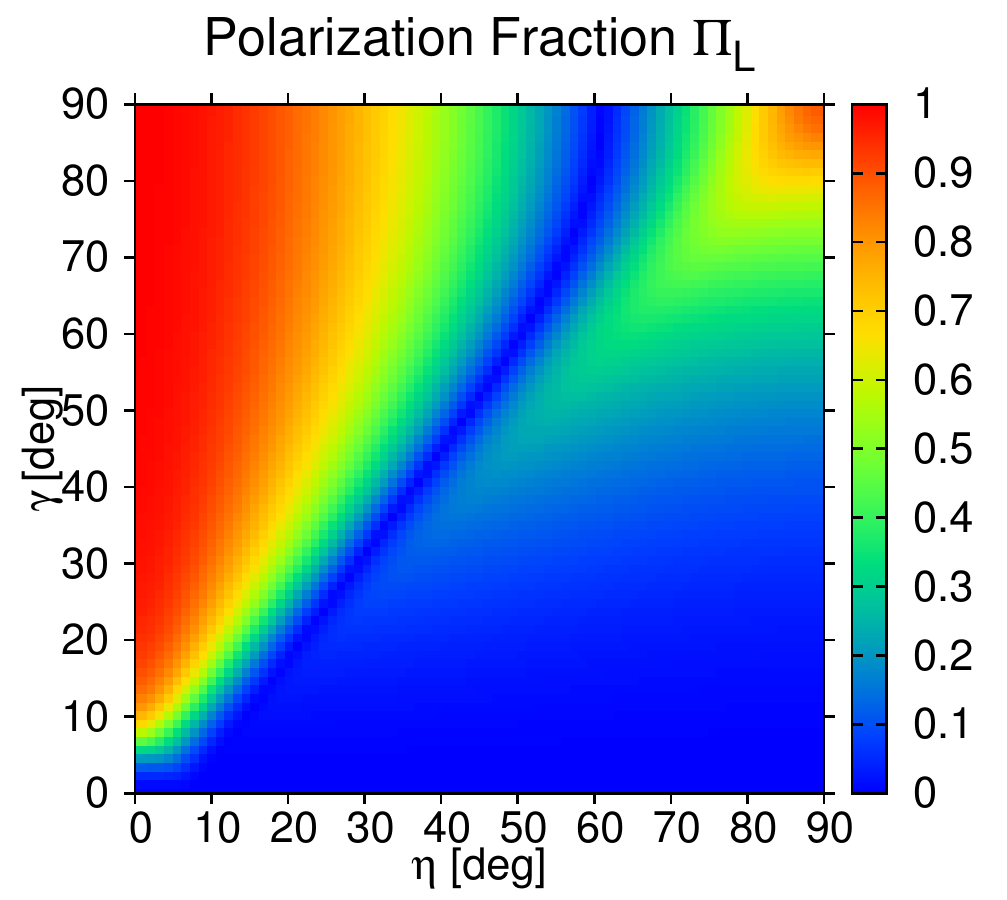}
      \end{minipage} &
      \begin{minipage}[t]{0.45\hsize}
        \centering
        (b) with mode conversion
        \includegraphics[width=4cm]{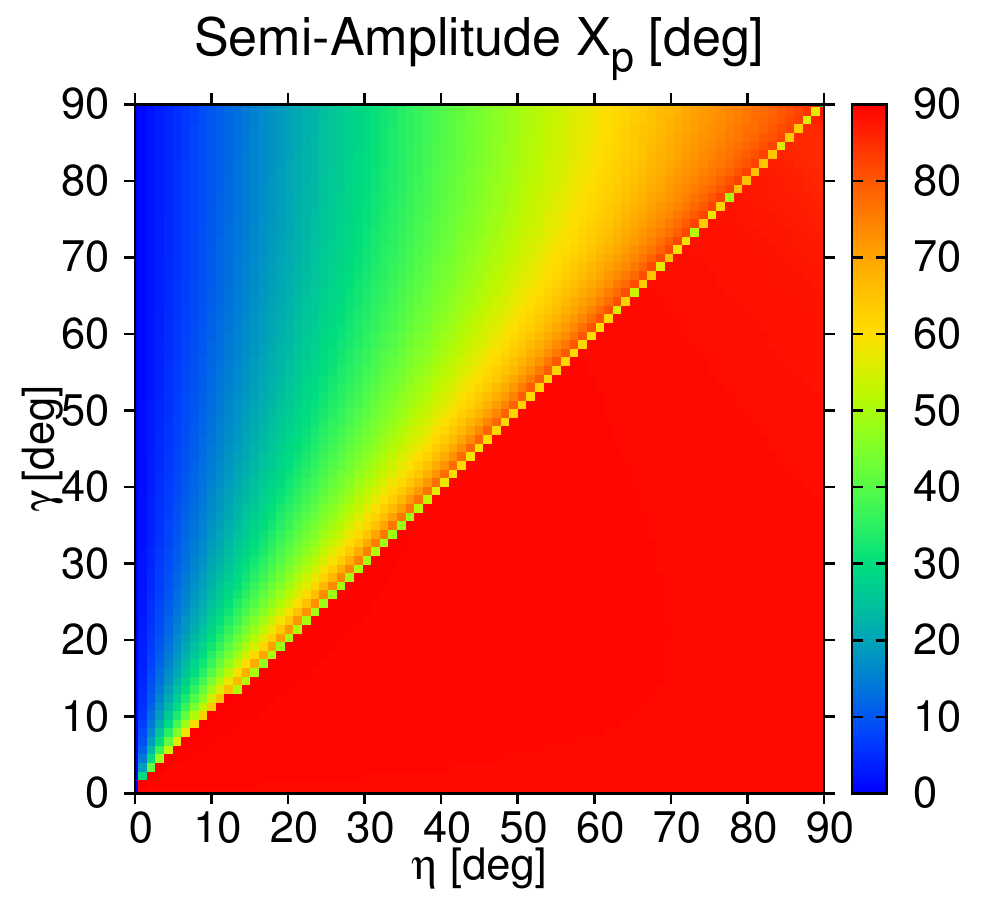}
        \includegraphics[width=4cm]{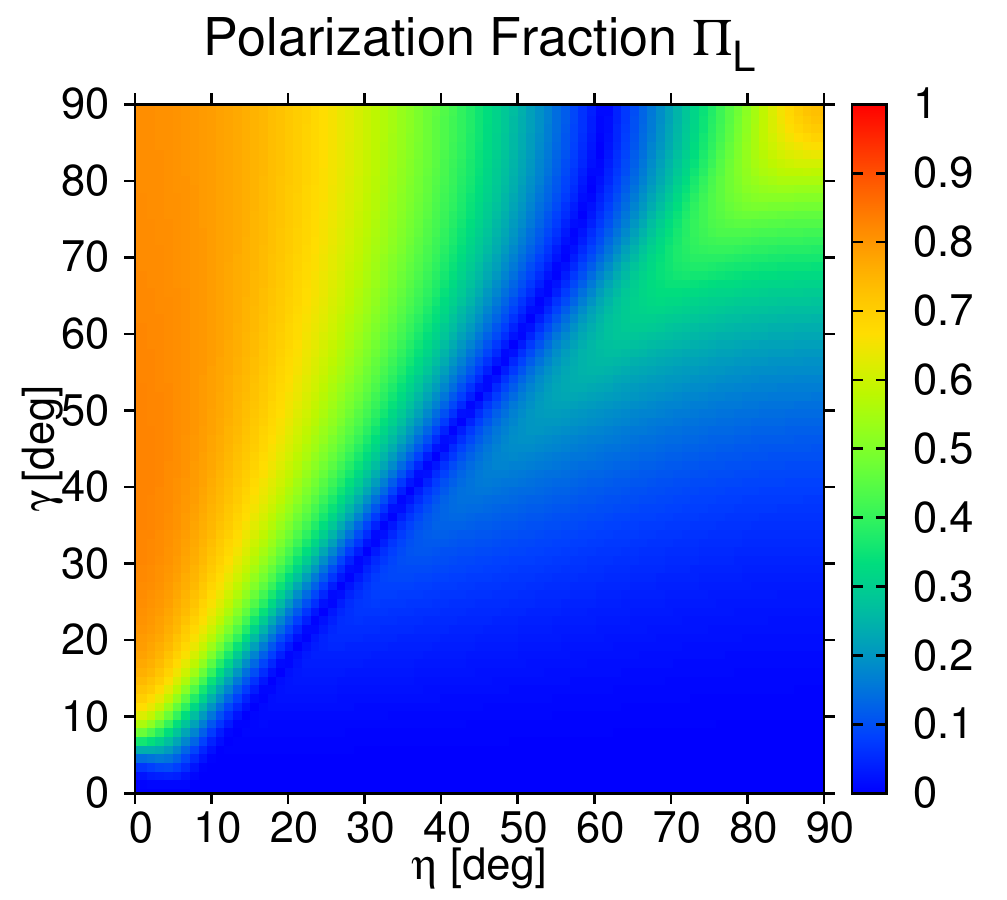}
      \end{minipage}
    \end{tabular}
\caption{Semi-amplitudes of the polarization angle $\chi _p$ (upper panels) and phase-averaged polarization fractions (lower panels) for different configurations of a rotating neutron star. The left and right columns correspond to the cases without and with the mode conversion, respectively. The photon energy is set to $E = 300 \mathrm{eV}$.} 
  \label{t+tvr_a}
\end{figure}

We next discuss the phase-averaged quantities. The semi-amplitudes and polarization fractions are shown in the upper and lower panels of Figure \ref{t+tvr_a}, respectively. Note that, rigorously speaking, the semi-amplitude is not a phase-averaged quantity, but we consider it here just for comparison. In this paper, the semi-amplitude is defined as the quantity related to the total variation of the polarization angle during the rotational period divided by four, which is expressed as
\begin{eqnarray}
& & X_p = \frac{1}{4} \sup _P \left[ \sum ^{n_P - 1} _{i = 0} \mathrm{min} \left( (\Delta \chi _{p})_i , 180^{\circ} - (\Delta \chi _{p})_i  \right) \right] , \nonumber \\
& & ( \Delta \chi_{p})_i = | \chi _p ( \psi_{i+1}) - \chi _p ( \psi_{i} ) |, \label{semi_amplitude}
\end{eqnarray}
where the supremum is taken over all possible partitions of the range $[0, 2 \pi]$ for $\psi$. This definition coincides with that given in \cite{2015MNRAS.454.3254T} in most cases but not always (see below). The photon energy is set to $E=300 \mathrm{eV}$ following \cite{2015MNRAS.454.3254T}. In the left column, the mode conversion is neglected on purpose for comparison with the previous work, whereas it is incorporated in the right column. The results without mode conversion are consistent with those in \cite{2015MNRAS.454.3254T}. The discontinuous suppression of the semi-amplitude on the diagonal line observed in our result (see the top left panel of Fig.~\ref{t+tvr_a}) but absent in their result is mainly due to the fact that our definition of the semi-amplitude is not completely the same as theirs.

\begin{figure}[htbp]
\begin{center}
 \includegraphics[width=\columnwidth]{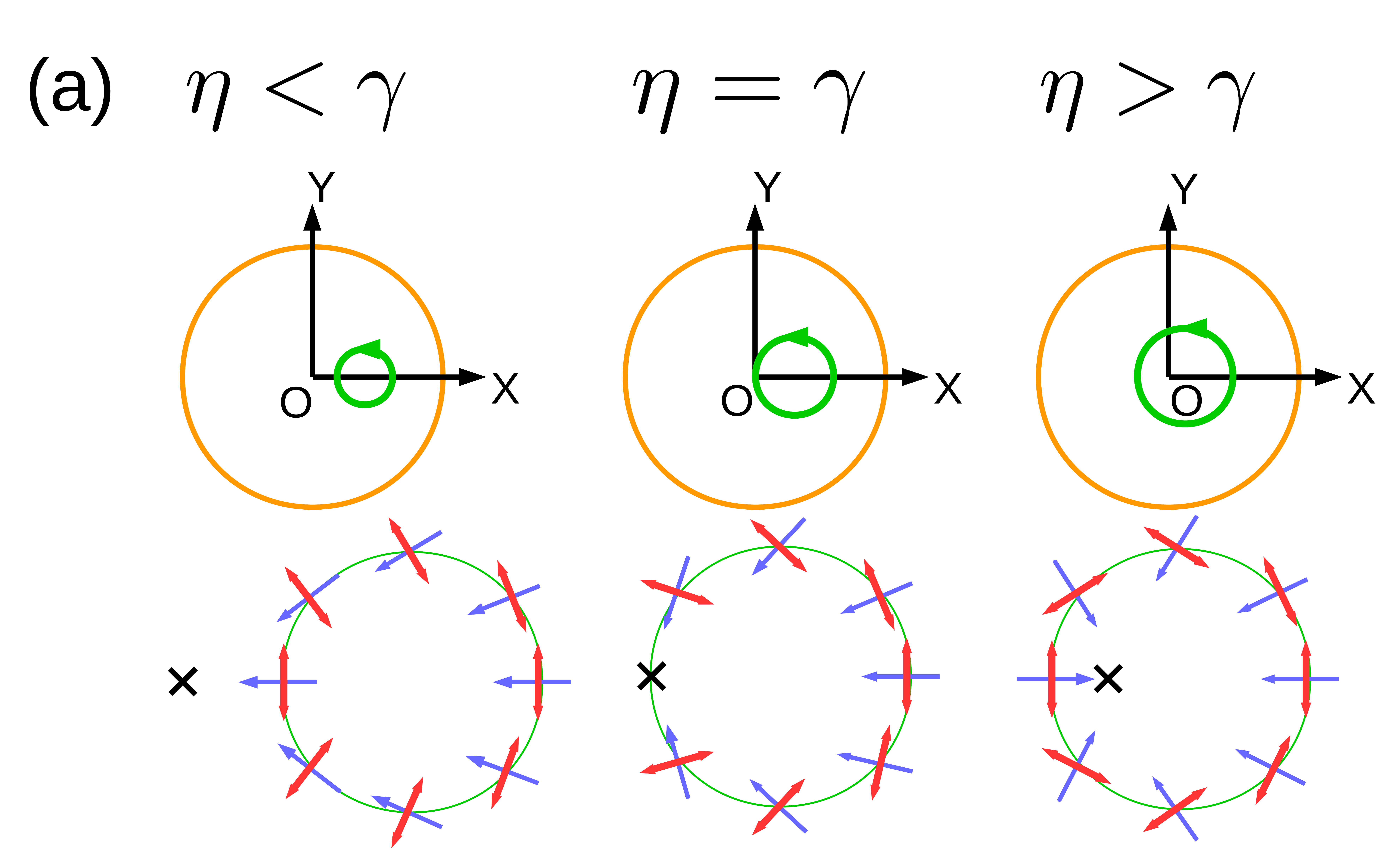} 
\end{center}
 \begin{center}
 \includegraphics[width=\columnwidth]{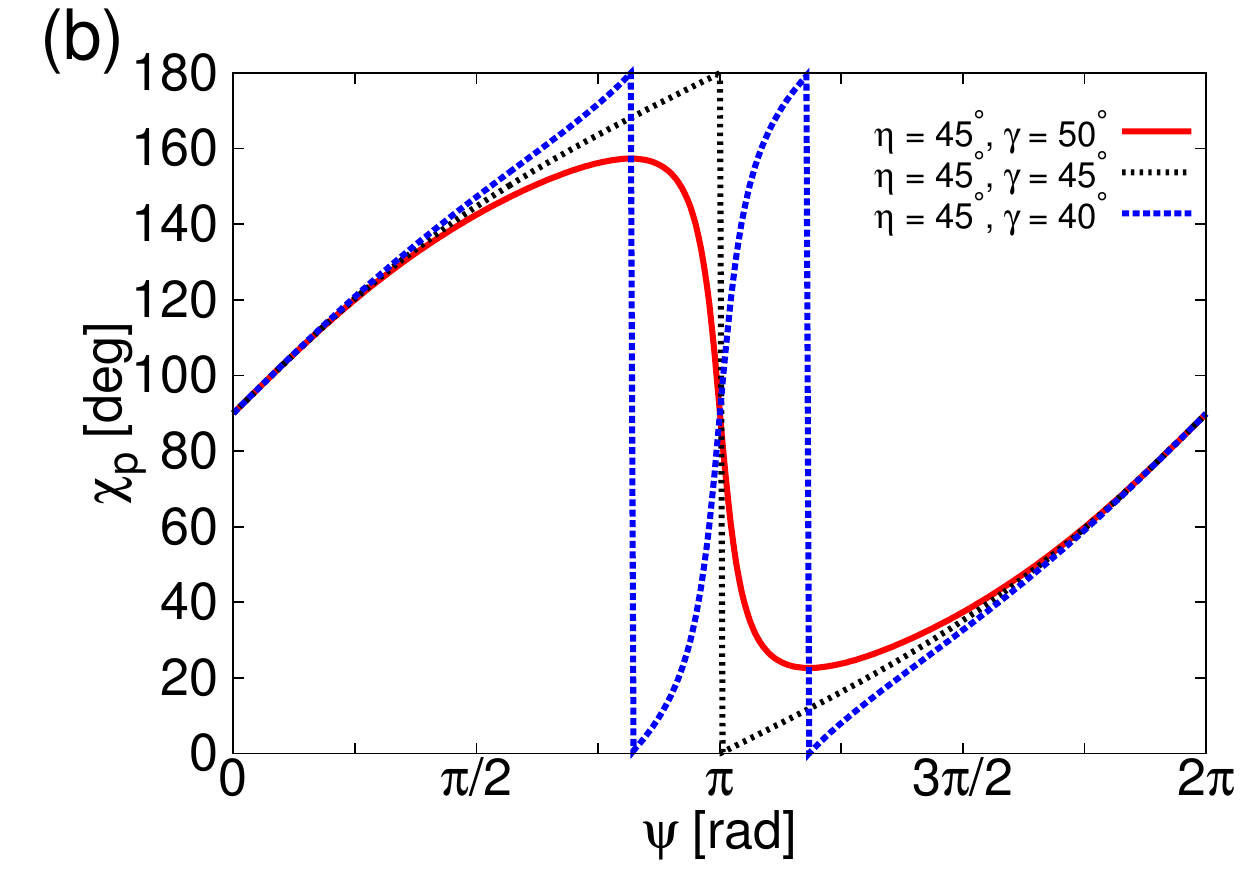}
 \caption{(a) Top views of the observed patches (orange circles) on the polarization-limiting surface in the middle row and the polarization directions (red arrows) at different rotational phases in the bottom row for $\eta < \gamma$ (left), $\eta = \gamma$ (center), $\eta > \gamma$ (right). The trajectories of the (extended) magnetic north pole are drawn in green. Blue arrows and black crosses in the bottom panels indicate the directions of the average magnetic fields at the different rotational phases and the position of the coordinate origin, respectively. (b) Polarization angles as functions of the rotational phase. The angle $\eta$ is fixed to $\eta = 45 ^{\circ}$, whereas the angle $\gamma$ is changed: $\gamma = 40^{\circ}, 45^{\circ}, 50^{\circ}$.}
   \label{par_300eV}
  \end{center}
\end{figure}

\begin{figure*}[htbp]
 \begin{minipage}[t]{0.45\hsize}
  \centering
  \includegraphics[width=6cm]{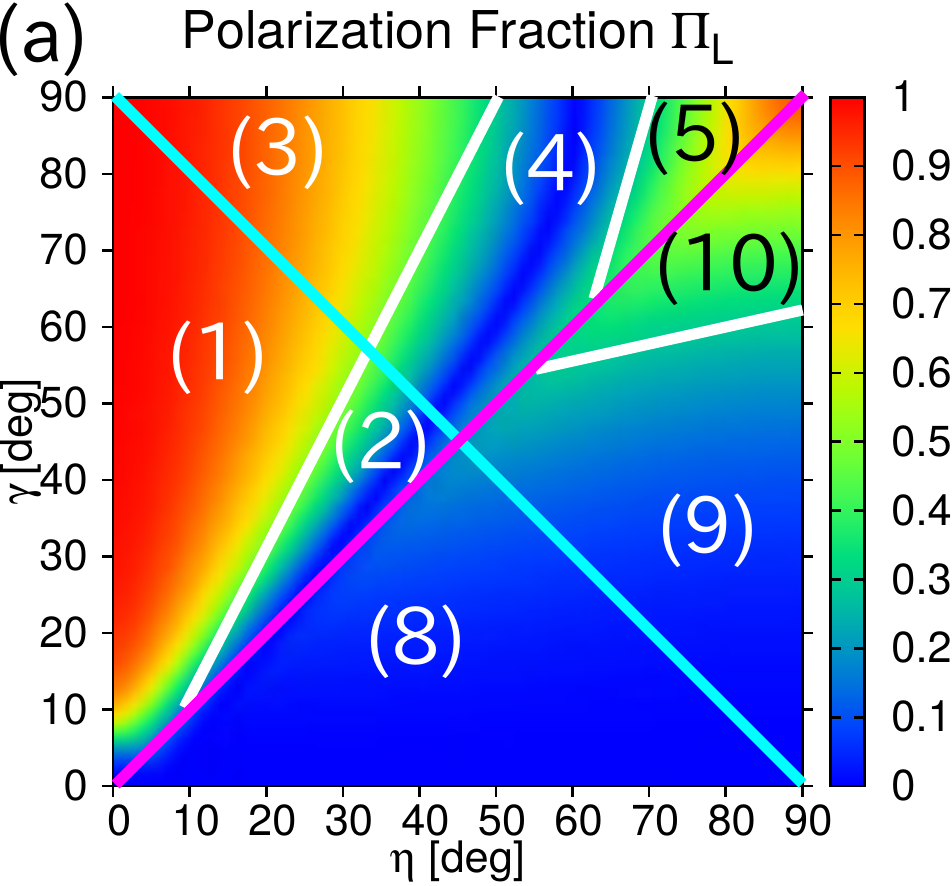}
 \end{minipage}
 \begin{minipage}[t]{0.5\hsize}
  \includegraphics[width=10cm]{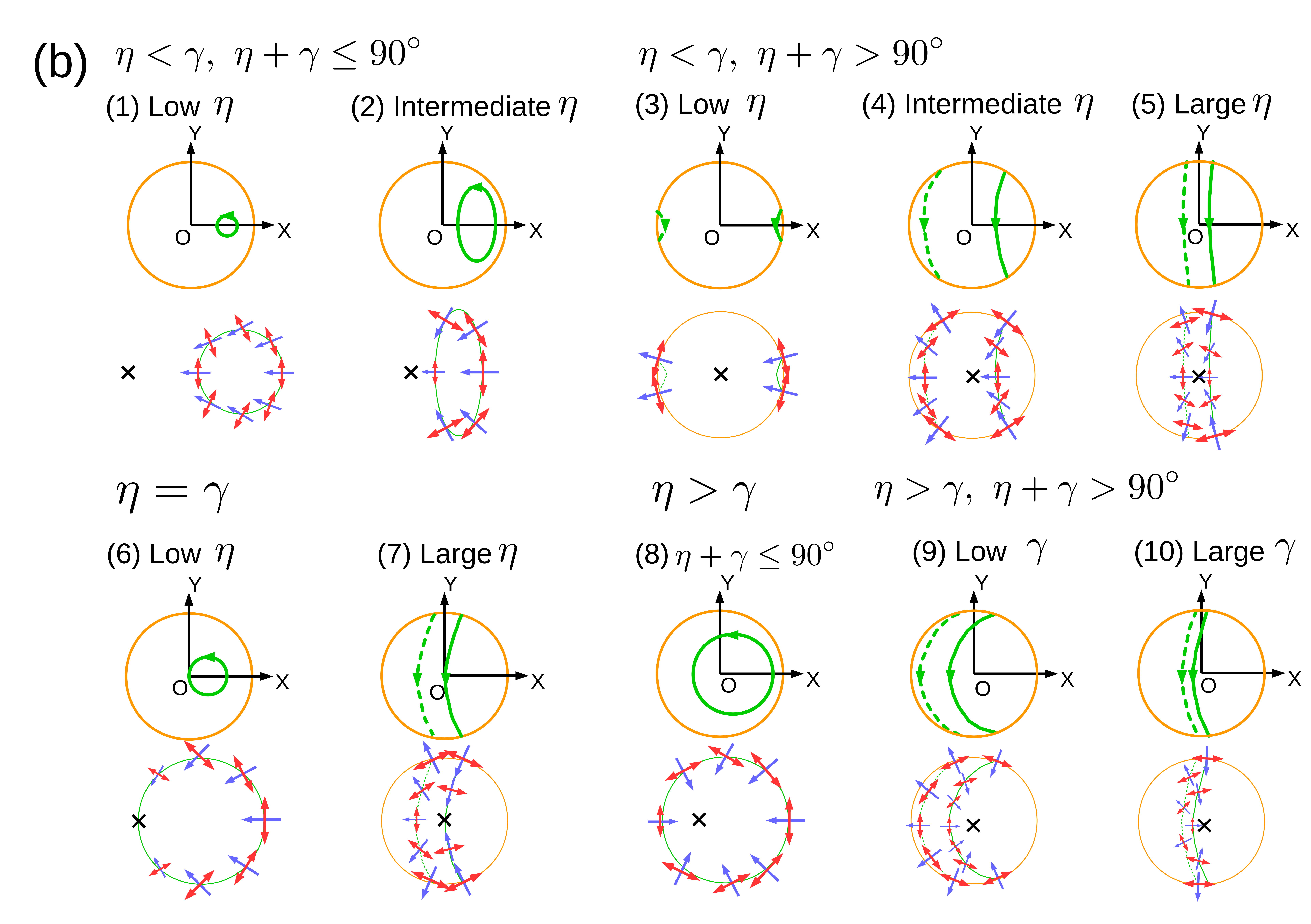}
 \end{minipage}
 \caption{(a) Classification of configurations in the $\eta - \gamma$ plane. Ten cases are distinguished.  Regions (6) and (7) are the lower ($< 45^{\circ}$) and upper ($> 45^{\circ}$) halves of the diagonal line of $\eta = \gamma$, shown in magenta. (b) Trajectories (green) of the (extended) magnetic north pole on the observed patch (orange) of the polarization-limiting surface and the corresponding directions of the average (projected) magnetic field (blue) and the polarization (red) of $E$-mode. The length of these arrows represents either the amplitude of the magnetic field or the polarization fraction. Note that the phase-averaged polarization fraction is not the average of the phase-resolved polarization fraction but is obtained from the Stokes parameters integrated over the entire rotational phase. The dashed lines are the trajectories of the south pole. The black crosses are the center of the observed patch.}
   \label{PF}
\end{figure*}

We explain this in more detail using Figure \ref{par_300eV}, which shows how the polarization angle changes with the rotational phase for different combinations of $\eta$ and $\gamma$.

In the middle three panels of Figure \ref{par_300eV} (a), we schematically draw the top views of the observed patch on the polarization-limiting surface for $\eta < \gamma$, $\eta = \gamma$, and $\eta > \gamma$. The green circles indicate the trajectories of the (extended) magnetic north pole on this surface. Note that they are not exact circles in general. In the bottom panels, we give the corresponding average magnetic fields (blue arrows) and polarization directions (red arrows), which are estimated from the relative locations of the magnetic north pole and the origin on the observed patch as explained earlier. Again, we assume that the photons are all in the $E$-mode. It is apparent from the middle panels and easily understood from the configurations that the coordinate origin is sitting outside, on, and inside the green circle for $\eta < \gamma$, $\eta = \gamma$, and $\eta > \gamma$, respectively. Then, it should be also be clear that the average magnetic fields and polarization angles behave as exhibited in the bottom panels. 

In the case of $\eta < \gamma$ (left column), the average magnetic field is always directed leftward, or in the negative $X$-direction. As a result, the polarization angle is limited in a certain range less than $\pi$. This is confirmed in Figure \ref{par_300eV} (b), in which we show the polarization angles as functions of the rotational phase for three combinations of $\eta$ and $\gamma$. The red solid line for $\eta = 45^{\circ}$ and $\gamma = 50^{\circ}$ corresponds to the current case. The polarization angle changes continuously and is indeed limited between $23^{\circ}$ and $157^{\circ}$. Note that $\chi_p$ changes more rapidly near $\psi=\pi$ as $\eta$ approaches $\gamma$. In fact, it becomes discontinuous at $\eta = \gamma$, as demonstrated by the black dotted line in Figure \ref{par_300eV} (b). In this case, the polarization angle changes by $| \Delta \chi_p | = 180^{\circ}$ at $\psi = \pi$, indicating the reverse of the magnetic field direction there. This is indeed confirmed in the bottom center panel of Figure \ref{par_300eV} (a). As a matter of fact, the average magnetic field vanishes at that point. Although $\eta = \gamma$ is a limit of $\eta < \gamma$, Equation (\ref{semi_amplitude}) gives a discontinuity to the semi-amplitude at $\eta = \gamma$. The semi-amplitude defined in~\cite{2015MNRAS.454.3254T} is continuous, on the contrary. This is the reason for the apparent discrepancy we mentioned earlier.  

When $\eta > \gamma$ is satisfied, in contrast, the direction of the average magnetic field changes continuously again and rotates by $360^{\circ}$ in this case, as demonstrated in the bottom right panel of Figure \ref{par_300eV} (a). As a result, the polarization direction also varies by the same amount continuously. This is confirmed as the blue dotted line in Figure \ref{par_300eV} (b), although the polarization angle $\chi_p$ is given modulo $180^{\circ}$ and looks discontinuous at two values of $\psi$. Note also that even in this case, $\chi_p$ changes rapidly around $\psi =\pi$.

Next, we shift our attention to the phase-averaged polarization fraction in the same model. It is calculated according to Equation (\ref{polarization_fraction_Stokes}) from the Stokes parameters integrated over the entire rotational phase. Note that it is not equal to the average of the phase-resolved polarization fractions. For the understanding of this quantity, it is not sufficient to distinguish the three cases, $\eta < \gamma$, $\eta = \gamma$, and $\eta > \gamma$, as for the semi-amplitude, but it is necessary to divide the cases further according to the values of $\eta$ and $\gamma$. In fact, we distinguish 10 cases, as shown in Figure \ref{PF} (a). Note that regions (6) and (7) are the lower ($\eta < 45^{\circ}$) and upper ($\eta > 45^{\circ}$) halves of the diagonal line of $\eta = \gamma$, shown in magenta. The other diagonal line, $\eta + \gamma = 90^{\circ}$, is shown in cyan.

We consider each regime in turn, referring to Figure \ref{PF}(b). Region (1) is a regime with $\eta \ll \gamma$ and $\eta + \gamma < 90^{\circ}$. As shown in the upper left panel of Figure \ref{PF} (b), the north pole is always inside the observed patch but is not very close to the origin. It does not move very much during a rotation, either. As a result, the average magnetic field is directed in the $-X$ direction, having similar amplitudes. This then leads to the facts that the phase-averaged polarization fraction is very high and that the polarization angle remains $\sim 90^{\circ}$.

As $\eta$ approaches $\gamma$, we enter region (2). The typical situation is displayed in the second panel from the left in the upper row of Figure \ref{PF} (b). In this case, the north pole still remains inside the observed patch but moves over a wider region. As a result, the polarization angle changes more widely with the rotational phase, leading to the cancellation of polarizations. Note that the magnetic field averaged over the observed patch nearly vanishes when the north pole comes close to the origin. 

We move on to the regimes still with $\eta < \gamma$ but with $\eta + \gamma > 90^{\circ}$, i.e., regions (3)-(5). In these cases, not only the north pole but also the south pole comes into sight. If $\eta$ is small, i.e., region (3), the rotation axis is almost perpendicular to the line of sigh,t and the typical situation is depicted in the third panel from the left in the upper row. It is evident that the polarization angle is nearly $90^{\circ}$ at all phases, irrespective of which pole is visible. Since there is no cancellation in the averaging of the magnetic field over the observed patch, the phase-averaged polarization fraction is high. At intermediate $\eta$ values in region (4), the variation of $\chi_p$ becomes large. In the fourth panel from the left in the upper row of the figure, it changes between $\chi_p \sim 45^{\circ}$ and $\chi_p \sim 135^{\circ}$. As a result, the phase-averaged polarization fraction is lowered by the cancellation. At even larger $\eta$ values in region (5), the polarization angle does not change much again, lingering at $\chi_p \sim 0^{\circ}$, and the phase-averaged polarization fraction returns to higher values.

We now consider the case of $\eta = \gamma$. In the case of low $\eta$ values, the leftmost panel in the lower row of Figure \ref{PF}(b) shows the typical situation. The polarization angle changes substantially, and the cancellation leads to low values of the phase-averaged polarization fraction. Although the variation of the polarization angle still exists at large $\eta$ values, the cancellation is much reduced, and the phase-averaged polarization fraction becomes higher in region (7). 

Finally, we look at the regions with $\eta > \gamma$. Region (8) corresponds to the one with $\eta + \gamma \leq 90^{\circ}$. As demonstrated in the middle panel in the lower row of Figure \ref{PF} (b), the north pole goes around the origin in the observed patch, and, as a result, the polarization angle also rotates by $360^{\circ}$. The polarization is mostly canceled when averaged over the rotational phase in this case. Regions (9) and (10), where $\eta + \gamma > 90^{\circ}$, are distinguished by the value of $\gamma$. For low values of $\gamma$, i.e., region (9), neither the north pole nor the south pole comes close to the origin and the polarizations are large at all phases, while the polarization angle changes by large amounts. The severe cancellation still occurs, and the phase-averaged polarization fraction remains low. At high $\gamma$ values in region (10), in contrast, the polarization angle still varies by large amounts, but the polarization itself becomes very small when the poles come close to the origin, where $\chi_p \sim 90^{\circ}$. When averaged over the rotational phase, this leads to higher polarization fractions where the phase-averaged polarization angle is either $\sim 0^{\circ}$ or $180^{\circ}$, which are, in fact, almost the same.

We now consider the effect of the mode conversion on these phase-averaged quantities. The semi-amplitude is little affected for the case shown in Figure \ref{t+tvr_a}. This is simply because the polarization angle is not modified at the energy of 300eV in the figure, which is evident in the upper right panel of Figure \ref{t+tvr}. Then, the above discussion is not changed by the mode conversion. The polarization fraction, in contrast, tends to be reduced. It is particularly clear in regions (1) and (3). This is because the $O$-mode photons that are partially converted from the $E$-mode cancel the polarization. See the bottom right panel of Figure \ref{t+tvr}, in which the phase-resolved polarization fractions are shown for different photon energies. Since the energy of 300eV assumed in Figure \ref{t+tvr_a} is a bit lower than $E \sim E_{\mathrm{ad}} \sim 1 \mathrm{keV}$, the adiabatic mode conversion at the resonance point is partially suppressed, leading to the mixture of $O$- and $E$-mode photons just mentioned.

\subsection{Phase-resolved Quantities for Various Configurations with Different Magnetic Field Strengths}

 \begin{figure*}[htbp]
\centering
\includegraphics[width=17.2cm]{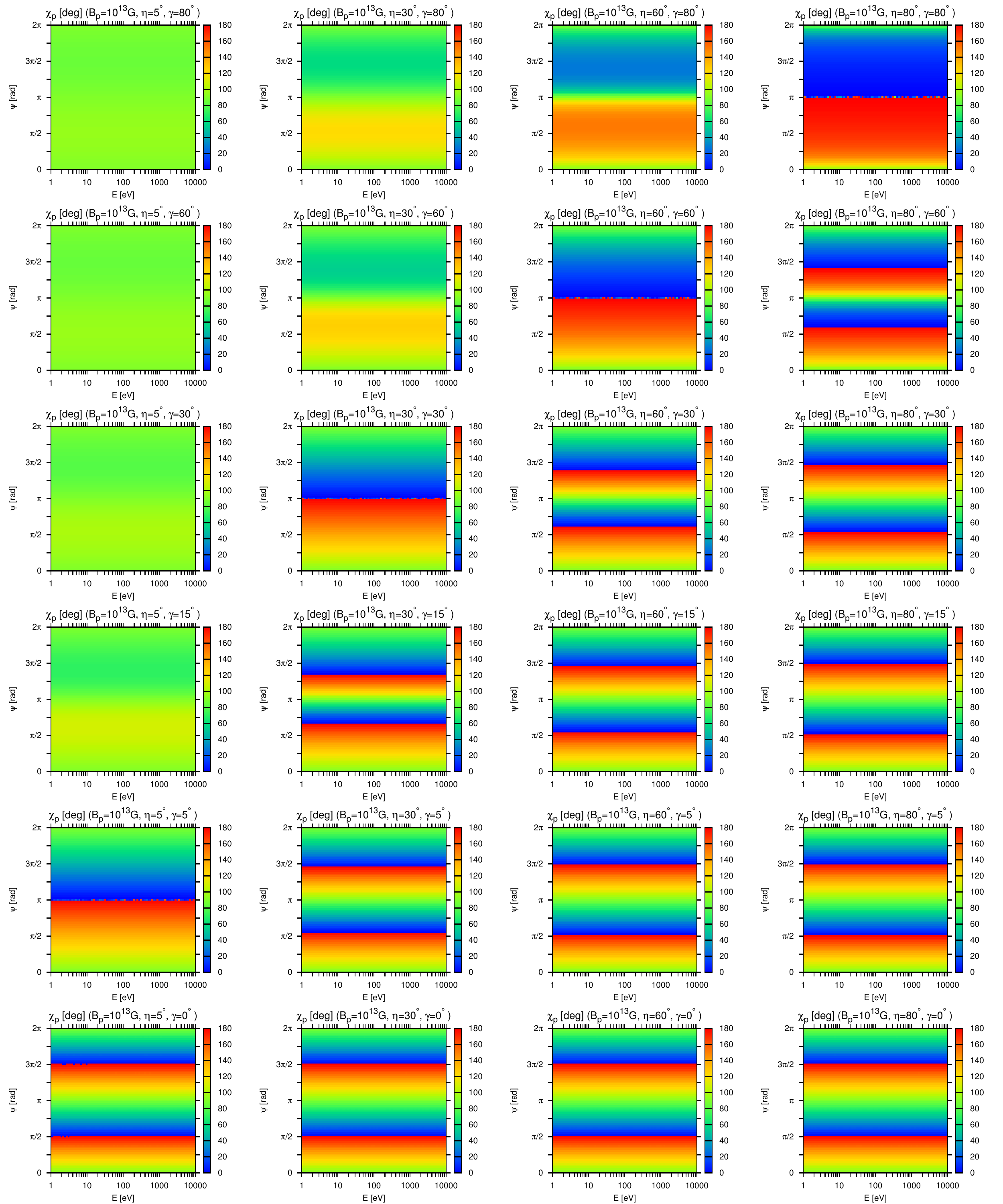}
  \caption{Phase-resolved polarization angles for different combinations of $\eta$ and $\gamma$. The mode conversion is neglected in this figure. The magnetic field strength is fixed to $B_p=10^{13} \mathrm{G}$.}
  \label{nmc_polarization_angle_1d13G}
\end{figure*}

 \begin{figure*}[htbp]
\centering
\includegraphics[width=17.2cm]{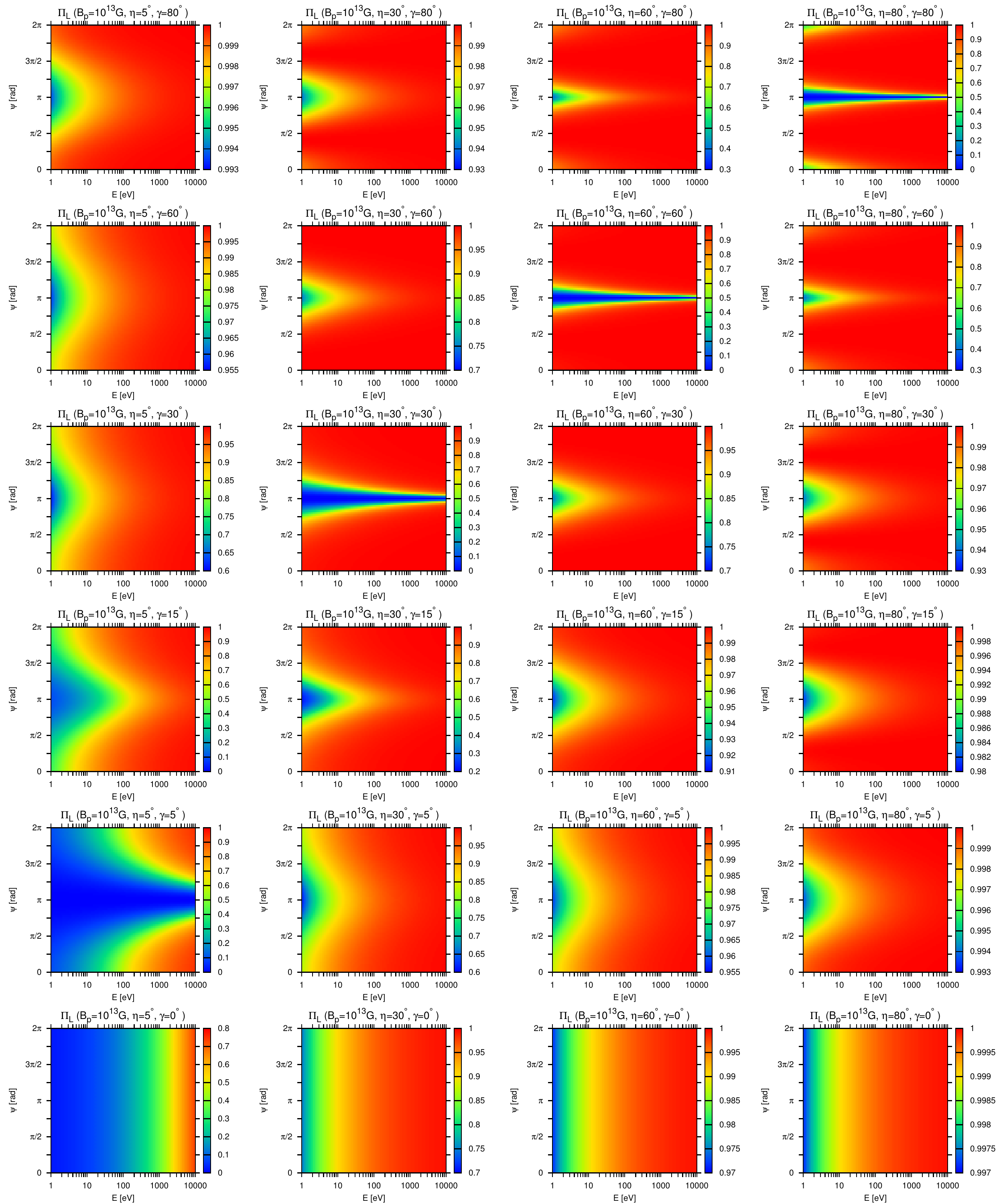}
\caption{Same as Figure \ref{nmc_polarization_angle_1d13G} but for the phase-resolved polarization fraction. Note the color scale in each panel.}
\label{nmc_polarization_fraction_1d13G}
 \end{figure*}

 \begin{figure*}[htbp]
\centering
\includegraphics[width=17.2cm]{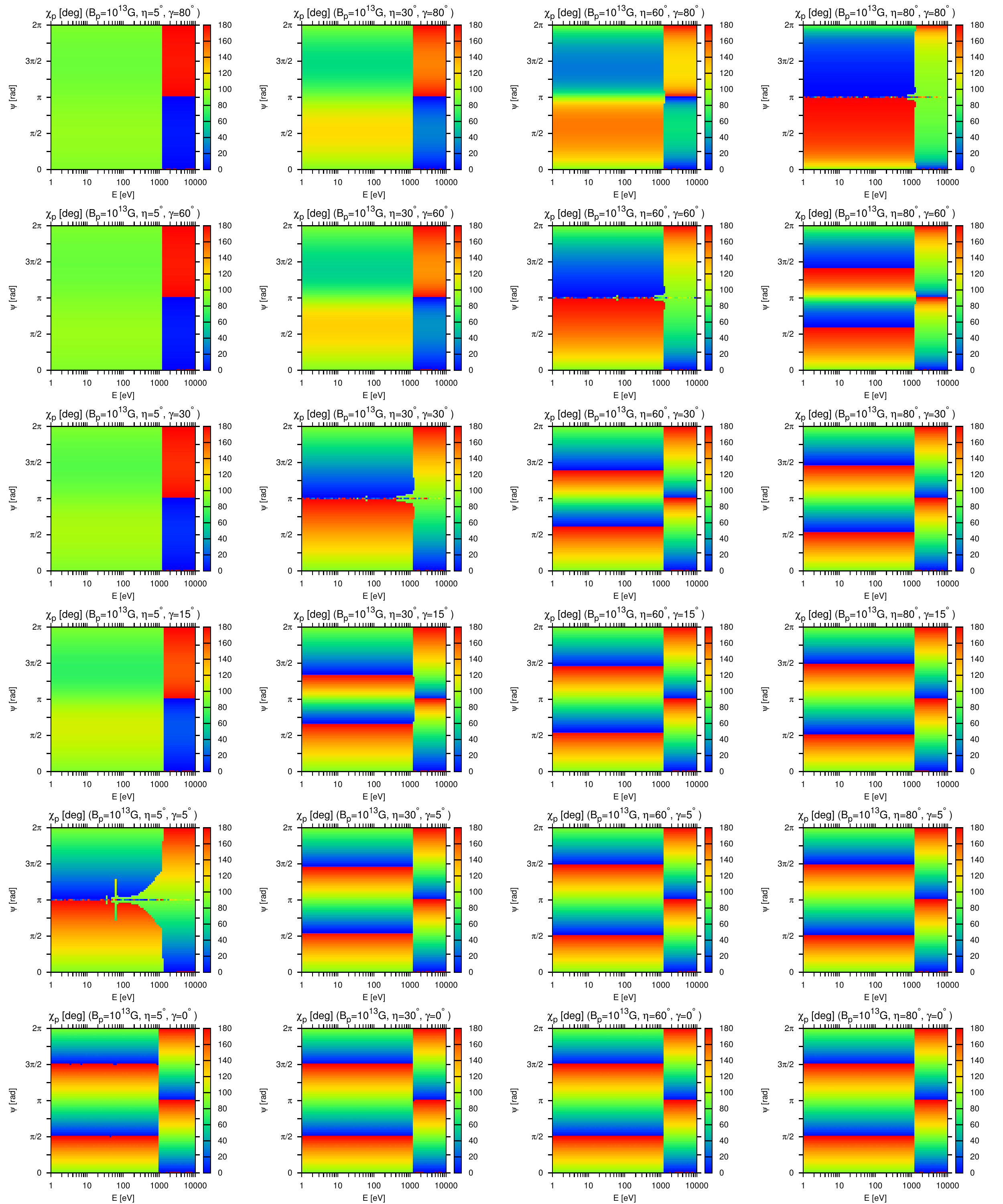}
  \caption{Same as Figure \ref{nmc_polarization_angle_1d13G} but with the mode conversion taken into account.}
  \label{polarization_angle_1d13G}
\end{figure*}

 \begin{figure*}[htbp]
\centering
\includegraphics[width=17.2cm]{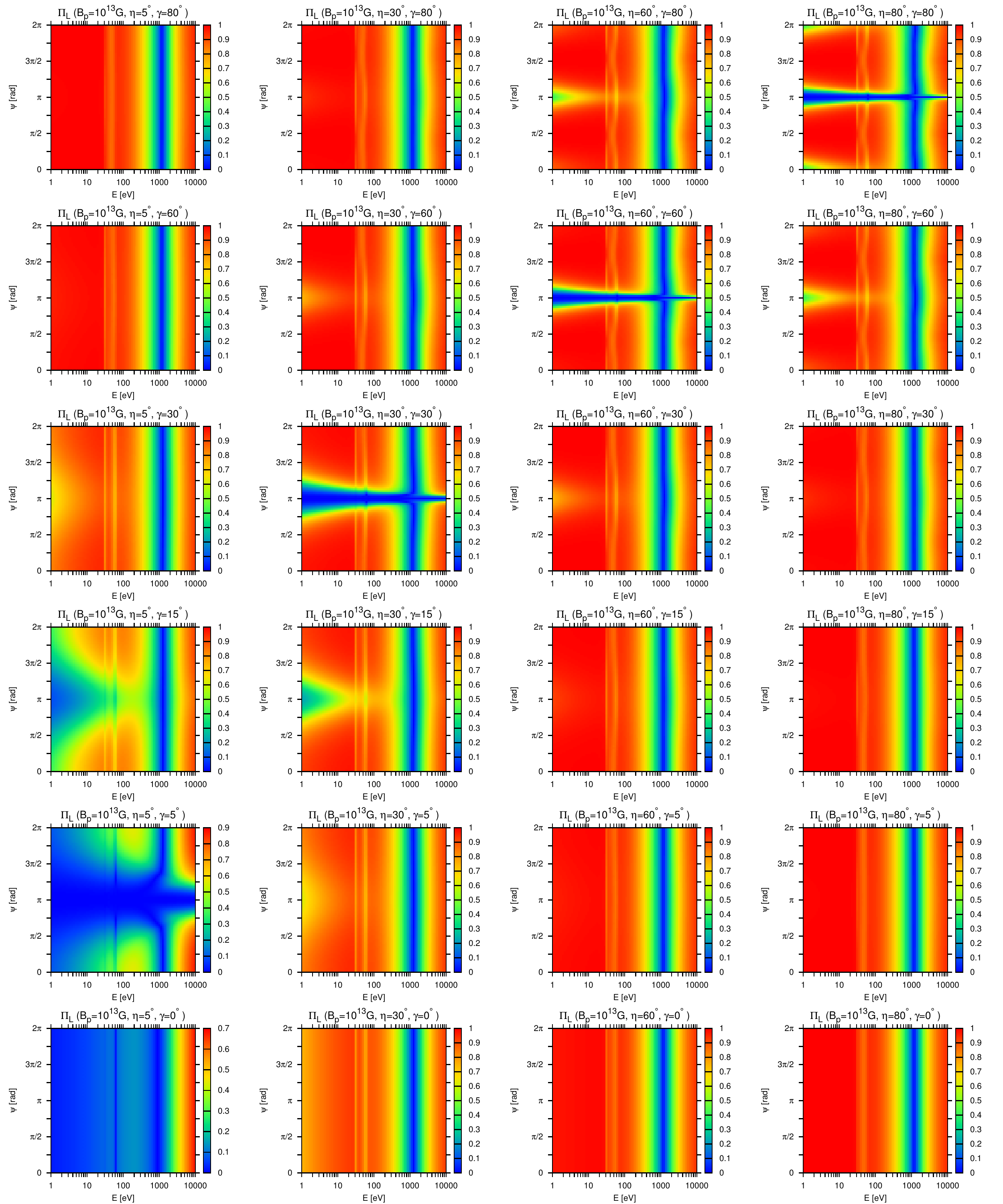}
\caption{Same as Figure \ref{polarization_angle_1d13G} but for the polarization fraction.}
\label{polarization_fraction_1d13G}
 \end{figure*}

 \begin{figure*}[htbp]
\centering
\includegraphics[width=17.2cm]{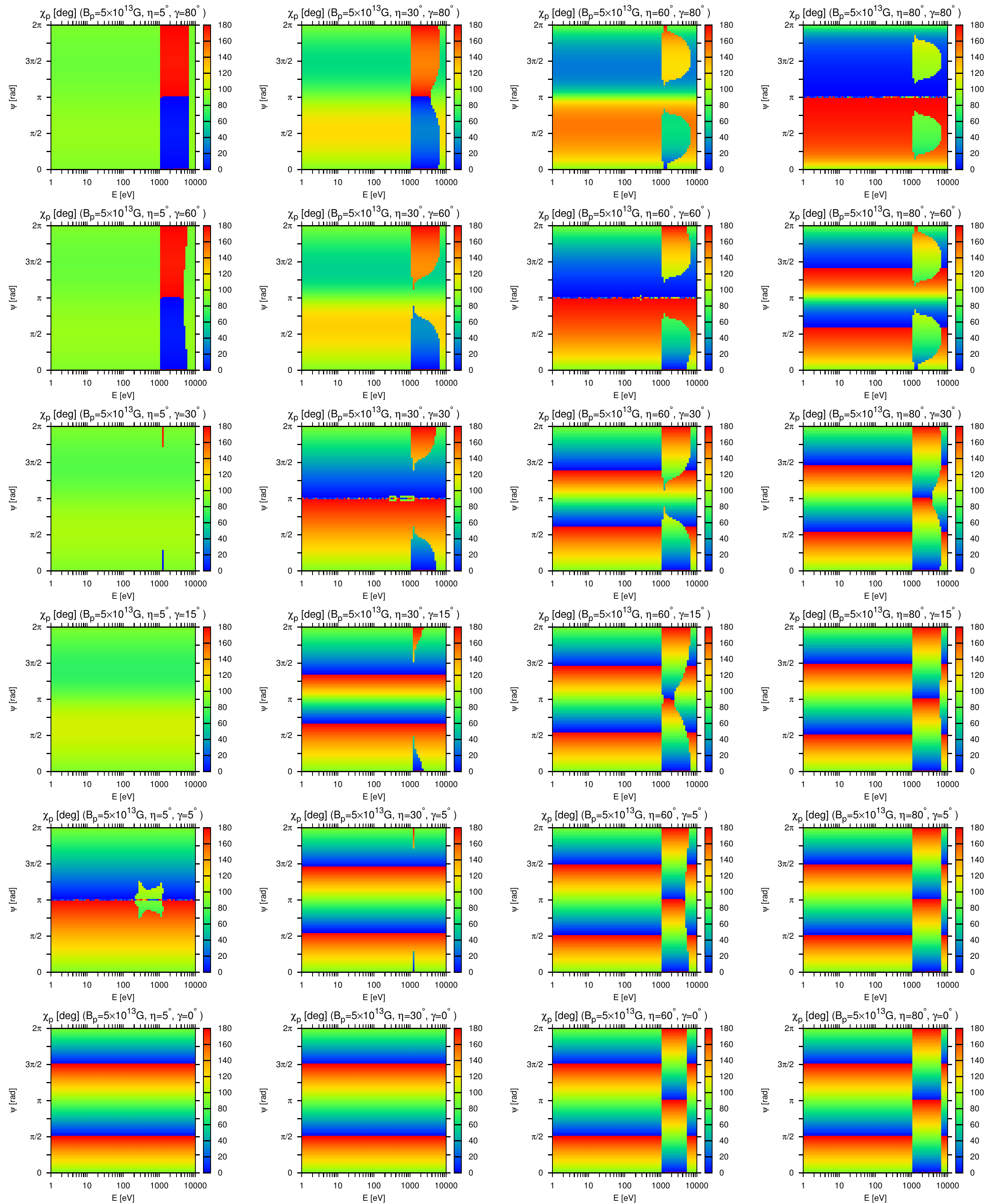}
  \caption{Same as Figure \ref{polarization_angle_1d13G} but for $B_p = 5 \times 10^{13} \mathrm{G}$.}
\label{polarization_angle_5d13G}
 \end{figure*}

 \begin{figure*}[htbp]
\centering
\includegraphics[width=17.2cm]{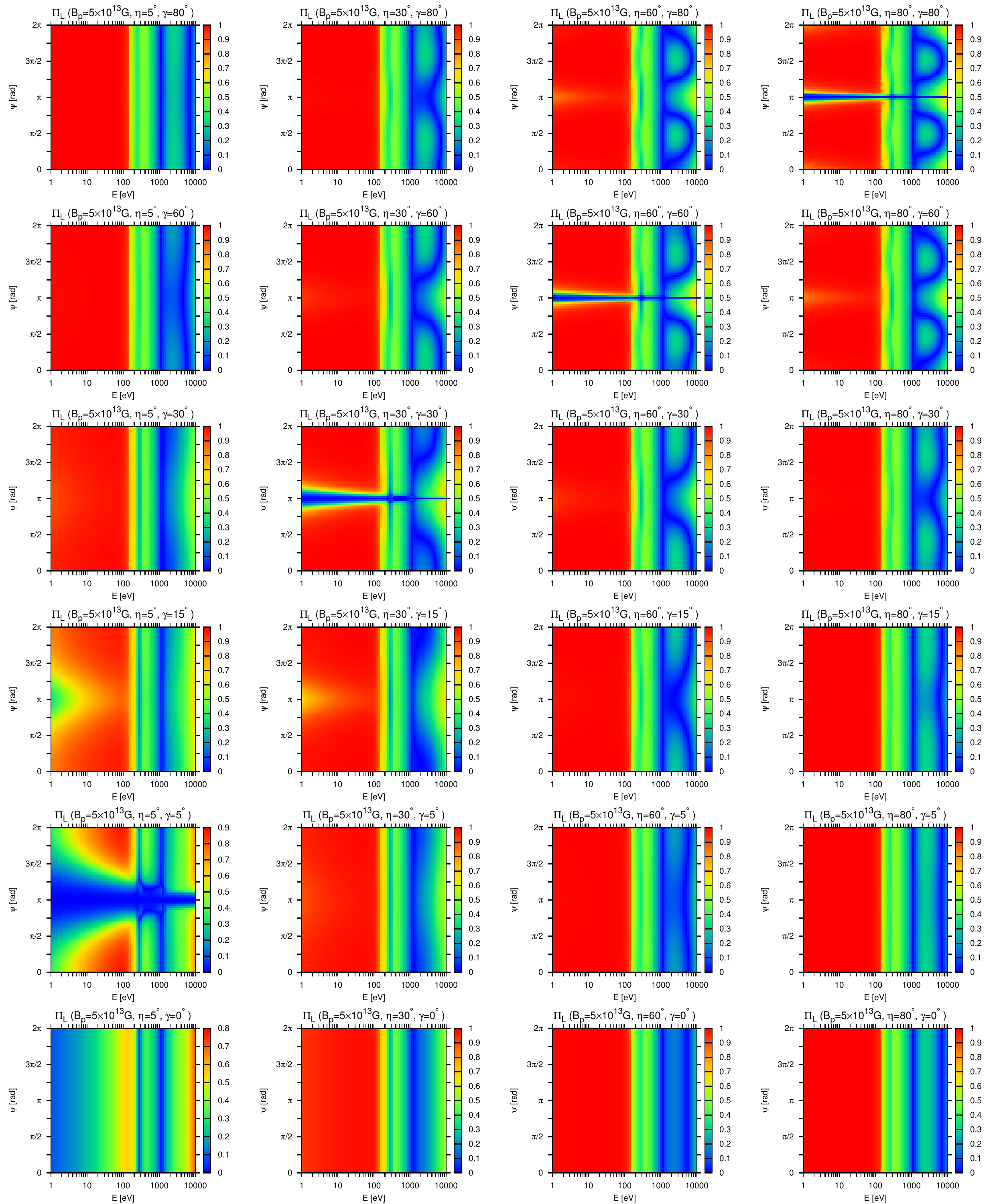}
  \caption{Same as Figure \ref{polarization_fraction_1d13G} but for $B_p = 5 \times 10^{13} \mathrm{G}$.}
  \label{polarization_fraction_5d13G}
 \end{figure*}

 \begin{figure*}[htbp]
\centering
\includegraphics[width=17.2cm]{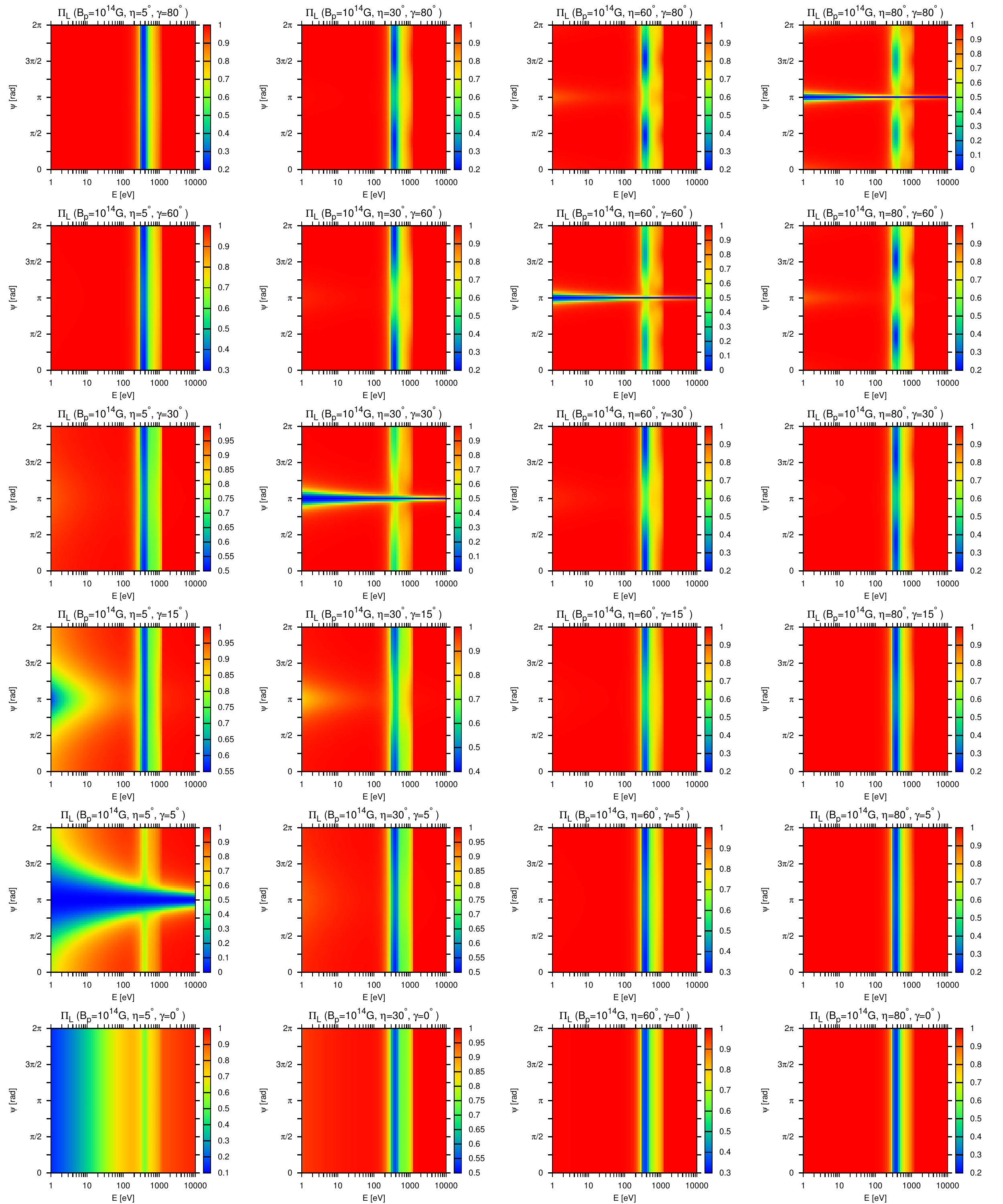}
  \caption{Same as Figure \ref{polarization_fraction_1d13G} but for $B_p = 10^{14} \mathrm{G}$.}
  \label{polarization_fraction_1d14G}
\end{figure*}

 \begin{figure*}[htbp]
\centering
\includegraphics[width=17.2cm]{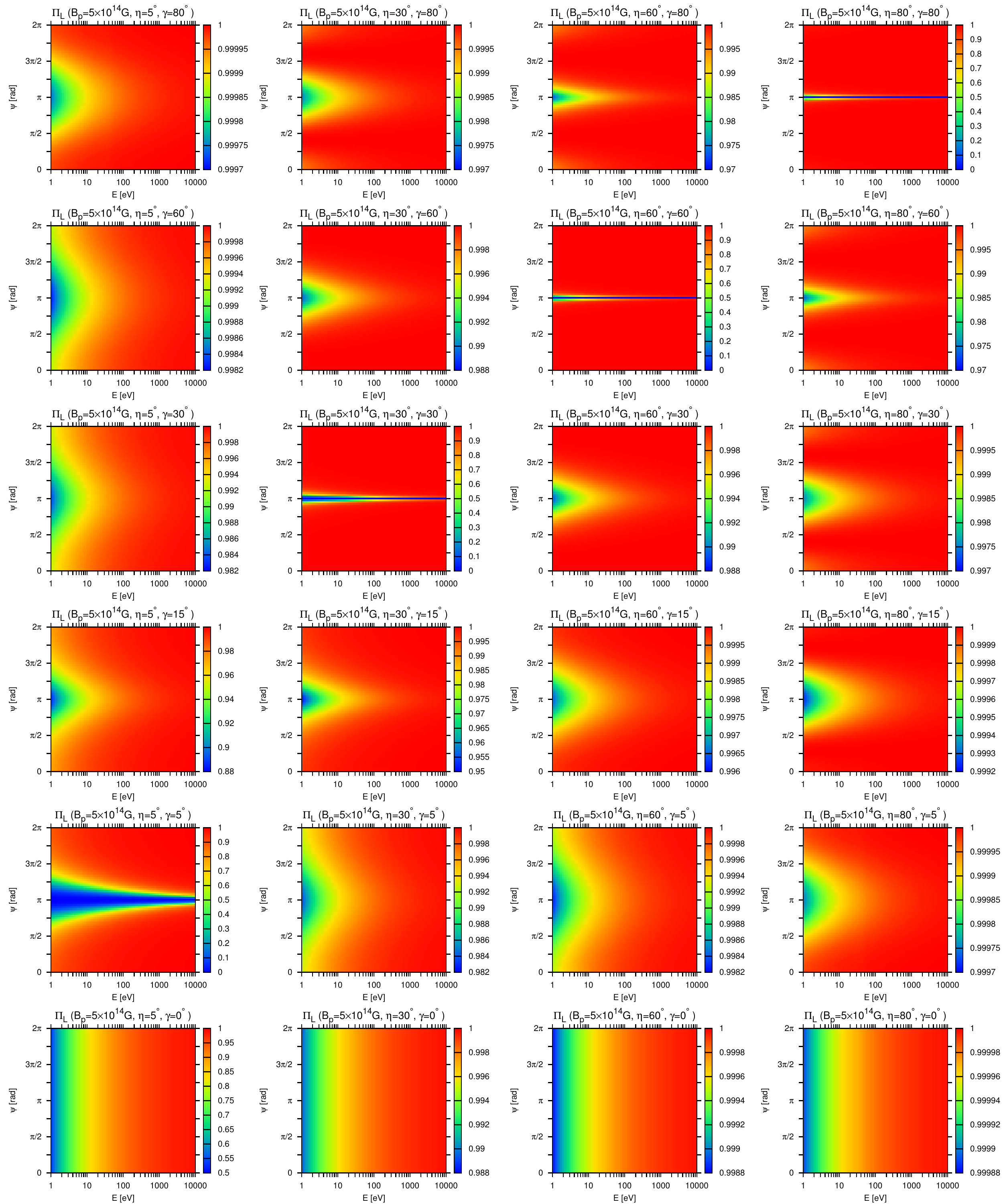}
  \caption{Same as Figure \ref{polarization_fraction_1d13G} but for $B_p=5 \times 10^{14} \mathrm{G}$.}
  \label{polarization_fraction_5d14G}
 \end{figure*}

It should be evident from the results given in the previous section that we need to study more systematically the phase-resolved polarization angle and fraction for various configurations of the rotating magnetized neutron star, different photon energies, and magnetic field strengths, based on the classification given in Figures \ref{par_300eV} and \ref{PF}. This is particularly true of the mode-conversion effects, since they are sensitive to the photon energy. 

We should begin without the mode conversion, however. We can then assume that the photons are all in the $E$-mode. Although we vary it later, we set the strength of the magnetic field to $B_p = 10^{13} \mathrm{G}$ here. The parameters on the neutron star are fixed to $R_{\mathrm{NS}} = 10 \mathrm{km}$, $M_{\mathrm{NS}} = 1.4 M_{\odot}$, and $T = 0.4 \mathrm{keV}$, though they are not relevant as long as the mode conversion and general relativistic effects are ignored.

The phase-resolved polarization angles are displayed as a function of the photon energy and rotational phase for 24 combinations of $\eta$ and $\gamma$ in Figure \ref{nmc_polarization_angle_1d13G}. It is clear at a glance that the results do not actually depend on the photon energy. Small glitches on some boundaries between different colors are just artifacts in drawing pictures. As explained in Figure \ref{par_300eV}, these cases can be understood by dividing them into the three regimes: $\eta < \gamma$, $\eta = \gamma$, and $\eta > \gamma$. In the first case, i.e., the upper left panels in Figure \ref{nmc_polarization_angle_1d13G}, the polarization angle oscillates around $\chi_p = 90^{\circ}$. It becomes $\chi_p = 90^{\circ}$ exactly when the rotational phase is $0$, $\pi$, and $2 \pi$. The color maps in this case are hence rather featureless. In the case of $\eta = \gamma$, in contrast, the polarization angle changes by $180^{\circ}$ during a single rotation. Note that the sharp boundary between blue and red is an artifact from the mod $( \pi )$ nature of the polarization angle, and it actually changes continuously there as well. Finally, for $\eta > \gamma$, the polarization angle varies by more than $180^{\circ}$ in general (see cases (8), (9), and (10) in Figure \ref{PF}), and, as a result, the polarization is mostly canceled, as is evident from Figure \ref{PF} (a). The color maps in this case (lower right panels in Figure \ref{nmc_polarization_angle_1d13G}) are characterized by the two horizontal sharp boundaries between blue and red, which are, again the artifact that occurs when the polarization angle exceeds $180^{\circ}$. It actually changes continuously just as in between the boundaries. The polarization angle becomes $90^{\circ}$ exactly at $\psi = 0$, $\pi$, and $2 \pi$.

The phase-resolved polarization fraction is mainly determined by the position of the (extended) north or south pole in the observed patch on the polarization-limiting surface. Various cases are summarized in Figure \ref{nmc_polarization_fraction_1d13G}. As the photon energy increases, the radius of the polarization-limiting surface gets larger, and, as a result, the pole tends to be located outside the observed patch longer, which then leads to higher polarization fractions. During a single rotation, in contrast, the pole comes closest to the origin at the rotational phase of $\psi = \pi$, and the polarization fraction becomes minimum at that point. Note that in the case of $\gamma = 0^{\circ}$, which is an example of case (8) given in Figure \ref{PF} (b), the polarization fraction is not changed by rotation, since the curve drawn by the pole is a circle with its center located at the origin. For $\eta = \gamma$, in contrast, the pole comes to the origin at $\psi = \pi$, and the polarization fraction vanishes completely by the cancellation.

Having understood the variety of the polarization angle and fraction as functions of the rotational phase without the mode conversion, we now look into how the mode conversion modifies them. In so doing, we also change the magnetic field strength. The results are exhibited for $B_p = 10^{13}, 5 \times 10^{13}, 10^{14}$, and $5 \times 10^{14} \mathrm{G}$ in Figures \ref{polarization_angle_1d13G}-\ref{polarization_fraction_5d14G}, which we consider in turn in the following.

The phase-resolved polarization angles and fractions shown for $B_p = 10^{13} \mathrm{G}$ in Figures \ref{polarization_angle_1d13G} and \ref{polarization_fraction_1d13G} are the mode-conversion counterparts of those given in Figures \ref{nmc_polarization_angle_1d13G} and \ref{nmc_polarization_fraction_1d13G} without the mode conversion (note that the color scales are different between Figures \ref{nmc_polarization_fraction_1d13G} and \ref{polarization_fraction_1d13G}). At $E \gtrsim E_{\mathrm{ad}} \simeq 1 \mathrm{keV}$, the mode conversion occurs adiabatically, and the $O$-mode photon becomes dominant. Then the polarization angle changes by $90^{\circ}$. Note again that the sharp boundary between blue and red is an artifact of the mod($\pi$) nature of the polarization angle, and there is nothing discontinuous there. In the case of $\eta = \gamma$, the $90^{\circ}$ change of the polarization angle occurs at much lower energies ($E \ll 1 \mathrm{keV}$) for $\psi \sim \pi$, which is particularly true of $\eta = \gamma = 5^{\circ}$. This is because the magnetic field is nearly aligned with the propagation direction of photons ($\theta_B \sim 0$), and $E_{\mathrm{ad}}$ becomes smaller (see Equation (\ref{adiabatic_energy})). Note also that the influences of the cyclotron energies of the protons are also apparent at $\lesssim 100 \mathrm{eV}$.

\begin{figure}
   \includegraphics[width=8.6cm]{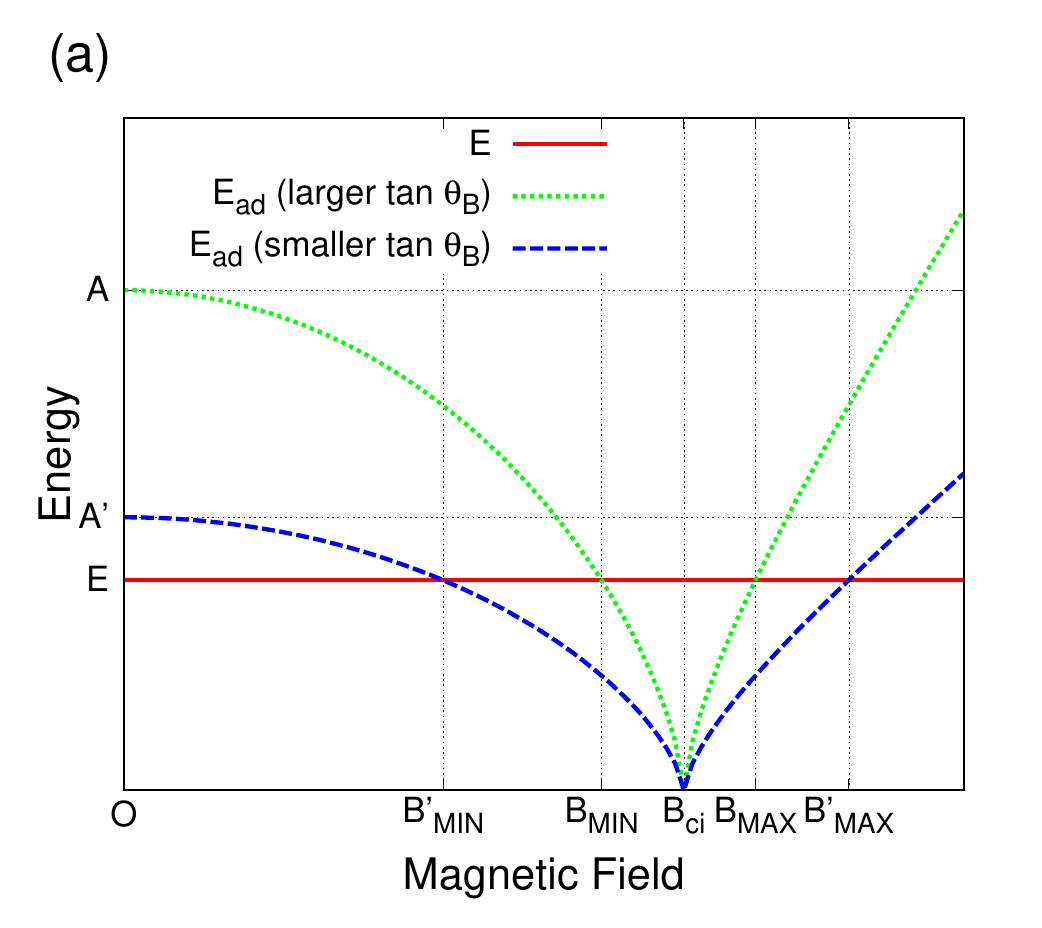}
  \begin{minipage}[t]{0.45\hsize}
   \includegraphics[width=4.3cm]{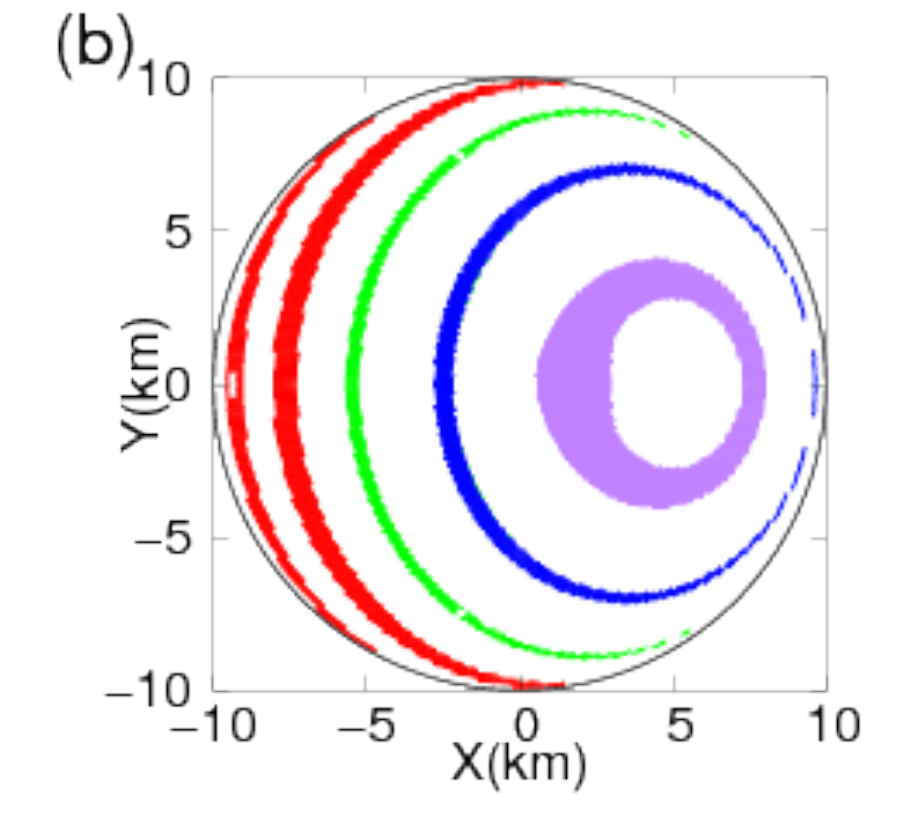}
  \end{minipage}
   \begin{minipage}[t]{0.45\hsize}
   \includegraphics[width=4.3cm]{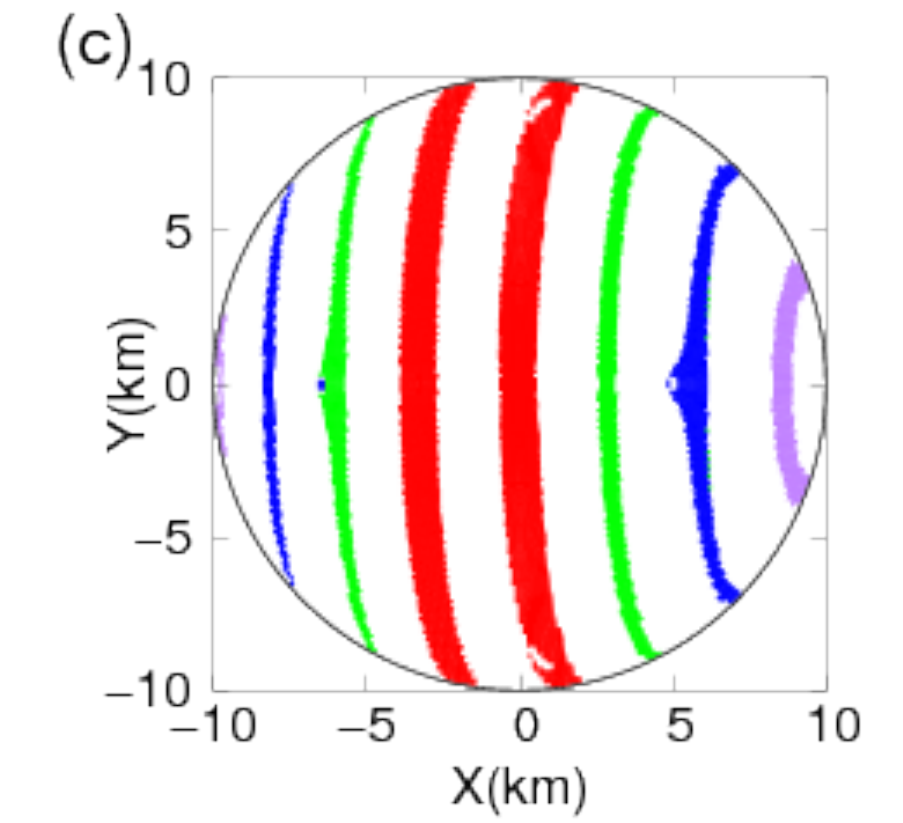}
  \end{minipage}
  \begin{center}
  \caption{(a) Schematic figure of the adiabatic energy $E_{\mathrm{ad}}$ as a function of the magnetic field strength $B$ for a given photon energy $E$. The green and blue dashed lines correspond to larger and smaller values of the angle between the photon momentum and the magnetic field, $\theta_B$. In the panel, $A$ and $A'$ are the corresponding $y$-intercepts. Note that $\theta _B$ is fixed in drawing these lines. The mode conversion occurs adiabatically when $B_{\mathrm{MIN}} < B < B_{\mathrm{MAX}}$ is satisfied, the condition corresponding to $E > E_\mathrm{ad}$. The adiabatic energy $E_\mathrm{ad}$ vanishes at $B_{ci}$. Panels (b) and (c) exhibit the regions where the mode conversion occurs for the photon energies of $E \simeq 33$eV (red), 40eV (green), 50eV (blue), and 60eV (purple), with different colors on the neutron star surface toward the observer, which is projected onto the $X$-$Y$ plane. In the two panels, the configurations of the neutron star are different: (b) $\eta = 0^{\circ}, \ \gamma = 30^{\circ}$ and (c) $\eta = 0^{\circ}, \ \gamma = 80^{\circ}$.}
   \label{cyclotron}
  \end{center}
\end{figure}

The polarization fraction is reduced by the mode conversion in general if it occurs at $E \sim E_{\mathrm{ad}} \sim 1 \mathrm{keV}$ and both the original $E$- and converted $O$-modes exist in some proportion, leading to partial cancellations. Note, however, that $E_{\mathrm{ad}}$ is in fact a function of the photon energy and is lowered remarkably at some energies. This is particularly the case for the cyclotron energies of the protons, as already mentioned earlier (see Figure \ref{cyclotron} (a)). At these energies, the photon is adiabatically converted from $E$-mode to $O$-mode completely. Since the cyclotron energy depends on the magnetic field strength, it is not constant on the neutron star surface. As a result, only those photons that have energies close to the local cyclotron energy and are propagating in certain directions are mode-converted and mixed with unconverted photons originating from different portions of the observed patch, which leads to the reduction of the polarization fraction as strips at $E \lesssim 100 \mathrm{eV}$. This issue will be considered more in detail in the following.

The mode conversion occurs adiabatically when $E \gtrsim E_{\mathrm{ad}} \propto ( \tan \theta _B |1 - u_i|)^{2/3}$, with $u_i = (E_{ci}/E)^2 \propto (B/E)^2$ (see Equation (\ref{adiabatic_energy})) and $\theta_B$ being the angle between the photon momentum and the magnetic field. Panel (a) schematically shows the dependence of $E_{\mathrm{ad}}$ on the magnetic field strength $B$ (green and blue dashed lines). It is seen that there is a region $B_{\mathrm{MIN}} < B < B_{\mathrm{MAX}}$, in which $E > E_\mathrm{ad}$ is satisfied and the mode conversion occurs for a given $\theta _B$. Note that $\theta_B$ is fixed to a certain nonzero value in drawing the dashed lines in the panel. Here $B_{\mathrm{ci}}$ is the magnetic field strength at which the cyclotron energy is equal to the photon energy $E$ and the adiabatic energy vanishes. This range is in fact dependent on $\theta_B$, the angle between the photon momentum and the magnetic field, through the adiabatic energy. It is found from the companion of the two dashed lines that the range gets wider as the $\theta _B$ becomes smaller. The adiabatic mode conversion occurs in wider ranges in $B$, as the magnetic field tends to be aligned with the $Z$-axis.

\begin{figure*}
\begin{minipage}[t]{0.5\hsize}
  \includegraphics[width=8cm]{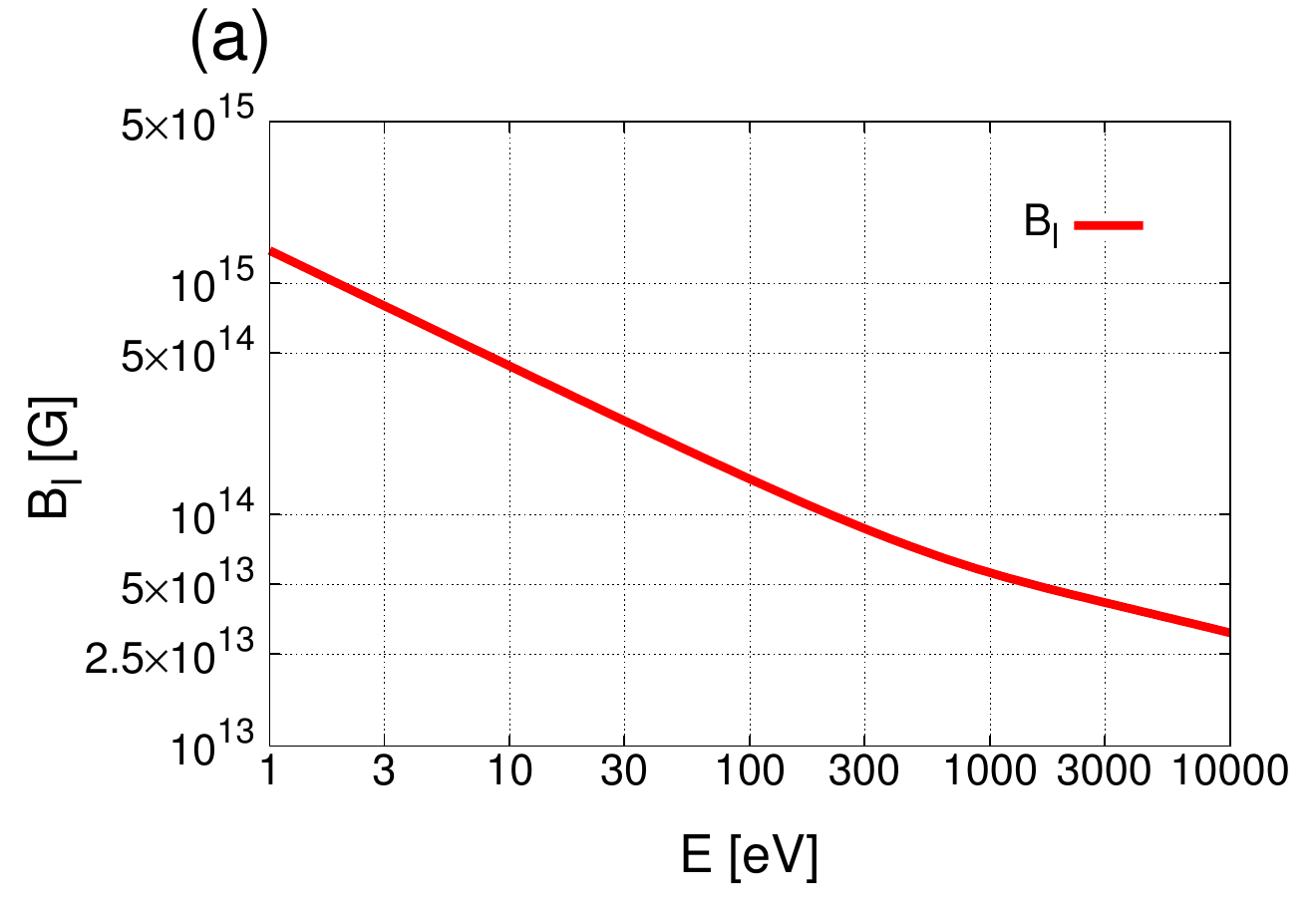} 
\end{minipage}
\begin{minipage}[t]{0.5\hsize}
   \includegraphics[width=8cm]{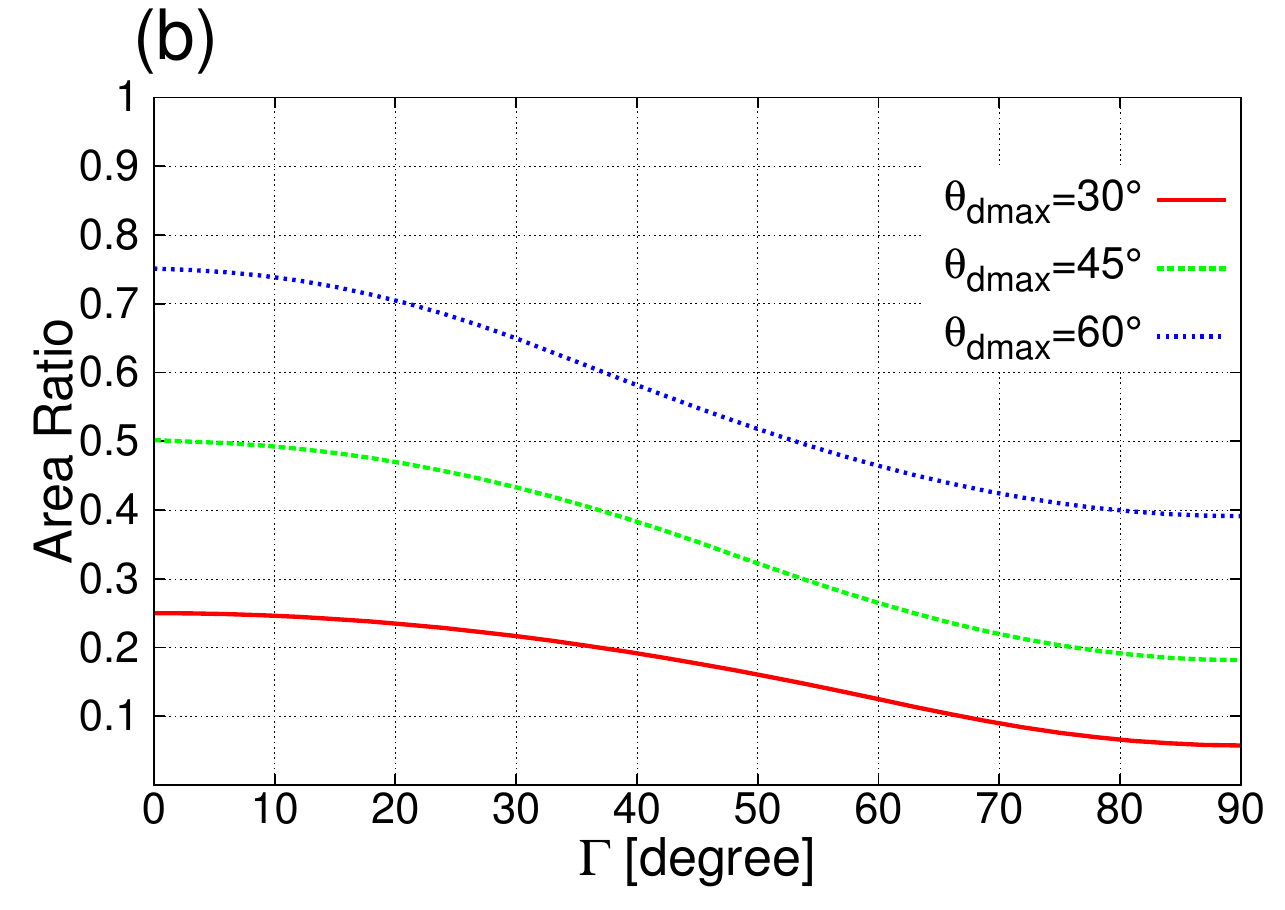}
\end{minipage}
   \begin{center}
   \includegraphics[width=17cm]{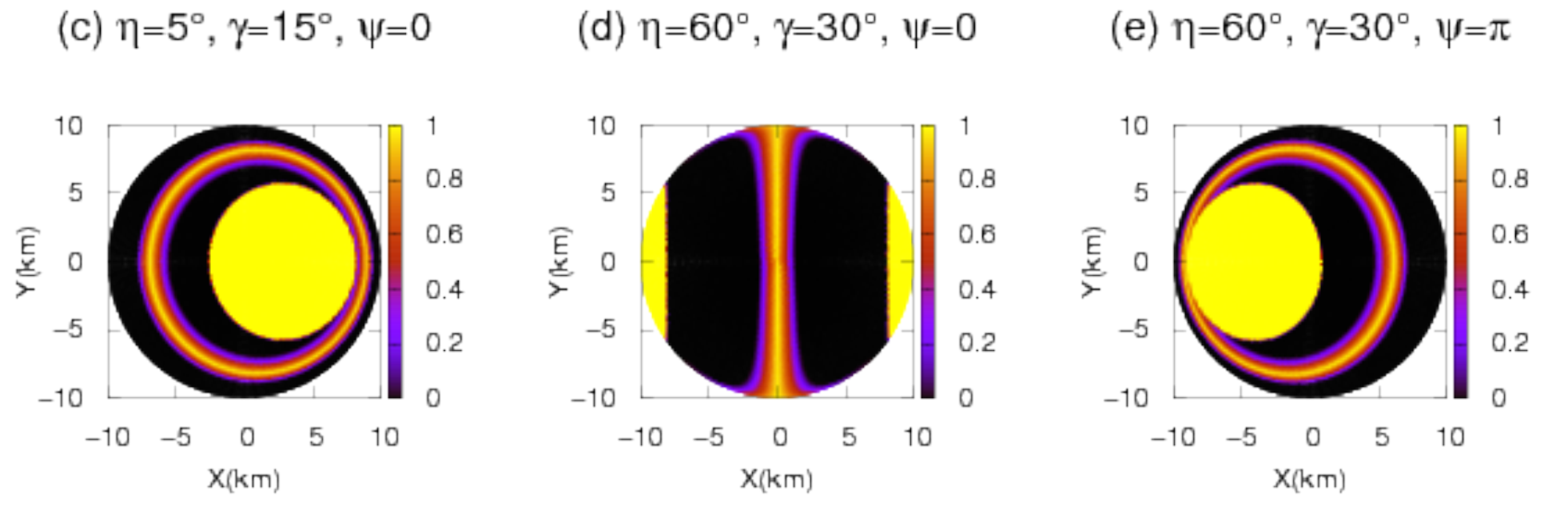}
  \caption{(a) Plot of $B_l$ as a function of the photon energy $E$. (b) Fraction of the projected area with $B > B_l$ as a function of $\Gamma$, the angle between the line of sight and the magnetic axis for three values of $\theta _{d \mathrm{max}}$: $\theta _{d \mathrm{max}} = 30^{\circ}$ (red solid line), $\theta _{d \mathrm{max}} = 45^{\circ}$ (green dashed line), and $\theta _{d \mathrm{max}} = 60^{\circ}$ (blue dotted line). (c)-(e) Snapshots of the fraction of the $E$-mode photon at $E = 2 \mathrm{keV}$ for three configurations: (c) $\eta = 5^{\circ}$, $\gamma = 15^{\circ}$, $\psi = 0$, (d) $\eta = 30^{\circ}$, $\gamma = 60^{\circ}$, $\psi = 0$; and (e) $\eta = 30^{\circ}$, $\gamma = 60^{\circ}$, $\psi = \pi$.}
   \label{B_l_graph}
   \end{center}
  \end{figure*}

The polarization fraction can be then understood from panels (b) and (c) in Figure \ref{cyclotron}, in which we show the areas where the mode-converted $O$-mode photons are emitted for different photon energies: $32 \mathrm{eV} \leq E \leq 34 \mathrm{eV}$ (red), $39 \mathrm{eV} \leq E \leq 41 \mathrm{eV}$ (green), $49 \mathrm{eV} \leq E \leq 51 \mathrm{eV}$ (blue), and $59 \mathrm{eV} \leq E \leq 61 \mathrm{eV}$ (purple). It should be mentioned here that there appear to be multiple strips with the same color in some cases; in fact, it may change with the rotational phase. In the two panels, we assume different configurations of the neutron star: (b) $\eta = 0^{\circ}, \ \gamma = 30^{\circ}$ and (c) $\eta = 0^{\circ}, \ \gamma = 80^{\circ}$. The dipole magnetic field strength is set to $B_p = 10^{13} \mathrm{G}$ for both cases.

The case in panel (b) is representative of the configurations in which the (projected) magnetic pole is near the origin of the $X$-$Y$ plane. It is found that the red and purple areas are larger on the projected surface than the green and blue ones. This leads to the fact that the polarization fraction is lower for $E \simeq 30$ and 60 eV than for other photon energies, and two distinct strips appear in the corresponding panel in Figure \ref{polarization_fraction_1d13G}. In contrast, panel (c) is a representative case, in which the magnetic pole is far from the origin and shows that each colored region has roughly the same area. As a result, the polarization fraction decreases almost uniformly for these photon energies, producing a single broad strip in the plot of the polarization fraction.

In Figures \ref{polarization_angle_5d13G} and~\ref{polarization_fraction_5d13G}, we present the results for one of the higher field strengths, $B_p = 5 \times 10^{13} \mathrm{G}$. It is apparent from Figure \ref{polarization_angle_5d13G} that the behaviors of the polarization angle are qualitatively different in some configurations from those for $B_p = 10^{13} \mathrm{G}$ given in Figure \ref{polarization_angle_1d13G}. In fact, in addition to the familiar result obtained for $\eta = 80^{\circ}$ and $\gamma = 5^{\circ}$, we find a case, e.g., with $\eta = 5^{\circ}$ and $\gamma = 15^{\circ}$, in which nothing occurs at all. For $\eta = 60^{\circ}$ and $\gamma = 30^{\circ}$ or $\eta = \gamma = 80^{\circ}$, the polarization angle changes only at some rotational phases; the result for $\eta = \gamma = 5^{\circ}$ has yet another pattern different from those in the above cases. The reason for all these phenomena is that the condition given in Equation (\ref{B_l}) is no longer satisfied at all rotational phases, and instead the condition in Equation (\ref{B_h}) holds at some phases. In the latter case, the mode conversion occurs inside the $O$-mode photosphere, and its effect is mainly to shift the photosphere of the $E$-mode photons outward.

The above explanations are substantiated in the following. We plot the values of $B_{l}$ as a function of the photon energy in panel (a) of Figure \ref{B_l_graph}, where the surface temperature is set to $kT = 0.4 \mathrm{keV}$ and $f = 1$ is assumed. Here $B_l$ decreases monotonically and the condition $B < B_l$ is satisfied everywhere on the neutron star surface at $E \lesssim 1 \mathrm{keV}$ for $B_p = 5 \times 10^{13} \mathrm{G}$. At higher photon energies, this is no longer the case. In fact, since $B_l < B_p$, the condition is violated near the magnetic pole, and the mode conversion occurs inside the $O$-mode photosphere. If the area with $B > B_l$ near the magnetic pole projected on the $X$-$Y$ plane is larger than the region with $B < B_l$, the $E$-mode is dominant and the polarization angle is unchanged from those of low-energy photons.

In the case of the dipole magnetic field, the condition of $B > B_l$ is equivalent to $\theta_d < \theta_{d \mathrm{max}}$ for the magnetic colatitude $\theta_d$. Then, the ratio between the (projected) area of the region satisfying $B > B_l$ and the (projected) star surface is a function of $\theta_{d \mathrm{max}}$ and $\Gamma$, the angle between the magnetic axis and the $Z$-axis. It is plotted as a function of the latter, with the former being fixed in panel (b) of Figure \ref{B_l_graph}. We choose three different values of $\theta_{d \mathrm{max}}$: $\theta_{d \mathrm{max}} = 30^{\circ}$ (red), $45^{\circ}$ (green), and $60^{\circ}$ (blue). Note that $\theta_{d \mathrm{max}}$ is a function of the photon energy and is larger for higher energies, as can be understood from panel (a) in Figure \ref{B_l_graph}.

We now revisit the results in Fig.~\ref{polarization_angle_5d13G}. In the case of $\eta = 5^{\circ}$ and $\gamma = 15^{\circ}$, the north pole, which has the strongest magnetic field and violates the condition given in Equation (\ref{B_l}) for $E \gtrsim 2 \mathrm{keV}$, stays close to the origin of the $X$-$Y$ plane: $\Gamma$ varies between $10^{\circ}$ and $20^{\circ}$. At $E = 2 \mathrm{keV}$, for example, the region, in which Equation (\ref{B_l}) is violated, corresponds to $\theta_{d \mathrm{max}} = 26^{\circ}$. This area alone is not sufficient to make the $E$-mode photons dominant, though. There is another region that predominantly emits the $E$-mode photons (the bright ring in panel (c) of Figure \ref{B_l_graph}). This happens not because of the violation of the condition in Equation (\ref{B_l}) but because of large values of $\tan \theta _B$, which narrows the region of $E > E_{\mathrm{ad}}$, where the mode conversion occurs adiabatically. The projected areas of both regions do not change much during the rotational period. As a result, the original $E$-mode is dominant at all rotational phases.

For $\eta = 60^{\circ}$ and $\gamma = 30^{\circ}$, in contrast, the rotational phase is important. The angle between the magnetic and $Z$-axis, $\Gamma$, ranges from $30^{\circ}$ to $90^{\circ}$, and the north pole comes close to the origin only at $\psi \sim \pi$ (see panels (d) and (e) in the same figure). Then, the $E$-mode photon is dominant at $30^{\circ} \leq \Gamma \lesssim 42^{\circ}$
or, equivalently, at $3 \pi /4 < \psi < 5 \pi /4$ (panel (e)).

In the case of $\eta = \gamma = 80^{\circ}$, the condition $B < B_l$ is not fulfilled at $\psi \sim 0$ and $2 \pi \ (\Gamma \sim 20^{\circ})$, where the south pole is located near the origin, as well as at $\psi \sim \pi \ (\Gamma \sim 0^{\circ})$, where the north pole faces the observer. At these phases, the mode conversion occurs inside the $O$-mode photosphere, and the polarization angle is unchanged.

In all of the above cases, Equation (\ref{B_l}) tends to be violated in wider regions on the neutron star surface for higher photon energies: $\theta_{d \mathrm{max}} = 39^{\circ}$ ($E = 3 \mathrm{keV}$), $\theta_{d \mathrm{max}} = 52^{\circ}$ ($E = 5 \mathrm{keV}$), and $\theta_{d \mathrm{max}} = 65^{\circ}$ ($E = 10 \mathrm{keV}$). Then, the values of $\tan \theta _B$ become smaller for the higher photon energies, narrowing the range of $E < E_{\mathrm{ad}}$ (see Figure \ref{cyclotron} (a)). This leads to narrower bright rings in panels (c)-(e) of Figure \ref{B_l_graph}.

Finally, we shift our attention to the case of $\eta = \gamma = 5^{\circ}$, which yields a distinct pattern in the polarization angle given in Figure \ref{polarization_angle_5d13G}. In fact, the jump of the polarization angle occurs at $\psi \sim \pi$ in the range of $100 \lesssim E \lesssim 1000 \mathrm{eV}$. This energy range corresponds to the vicinity of the cyclotron energy again. In contrast, the rotational phase $\psi \sim \pi$ is the phase at which $\tan \theta_B$ takes small values. At $E \gtrsim 2 \mathrm{keV}$, Equation (\ref{B_l}) is violated near the magnetic pole, which always faces the observer in this case, and the mode conversion occurs inside the $O$-mode photosphere.

The polarization fractions for $B_p = 5 \times 10^{13} \mathrm{G}$ are displayed in Figure \ref{polarization_fraction_5d13G}, with the mode conversion being taken into account. This should be compared with Figure \ref{polarization_fraction_1d13G}. Since the polarization-limiting radius is larger than that for $B_p = 10^{13} \mathrm{G}$ (see Equation (\ref{polarization_limiting_radius})), the magnetic north or south pole tends to be located outside the observed patch, and, as a result, the polarization fraction should be higher as long as the mode conversion is ignored. This is true at low energies, $E \lesssim 100 \mathrm{eV}$, where no conversion is expected from the beginning. The polarization fraction is lowered either when the partial conversion occurs nonadiabatically or when the observer sees not only the region in which the mode conversion occurs outside the photospheres of the two modes but also the region in which the mode conversion takes place between the two photospheres. The former occurs at $E \sim E_{\mathrm{ad}}$, while the latter is evident near the boundary between the $90^{\circ}$ change and the unchanging regimes of the polarization angle. The cyclotron energy of the proton in this case varies continuously from $\sim 300 \mathrm{eV}$ at the magnetic pole down to $\sim 150 \mathrm{eV}$ on the equator. In most cases, its effect is visible at $E \sim 300 \mathrm{eV}$. This is because for higher cyclotron energies, the adiabatic condition Equation (\ref{adiabatic_energy}) is satisfied for wider ranges of $\theta_B$, the angle between the magnetic field and photon momentum. This leads to the single vertical blue strip at $E \sim 300 \mathrm{eV}$ in Figure \ref{polarization_fraction_5d13G}.

The polarization fractions for $B_p = 10^{14} \mathrm{G}$ are presented in Figure \ref{polarization_fraction_1d14G}. Although not shown, the  polarization angles are essentially the same as those given in Figure \ref{nmc_polarization_angle_1d13G}, with the mode conversion being ignored entirely. This is because the condition given in Equation (\ref{B_l}) is not satisfied at $E \gtrsim E_{\mathrm{ad}} \sim 1 \mathrm{keV}$ in this case. This does not imply that the mode conversion occurs outside the $O$-mode photosphere in any region. In fact, the polarization fraction is reduced at the cyclotron energies of the proton, which range from $\sim 300 \mathrm{eV}$ on the equator to $600 \mathrm{eV}$ at the pole in the current case. It is added that the polarization fraction is increased as a whole owing to the larger polarization-limiting radius.

As shown in Figure \ref{polarization_fraction_5d14G}, if we raise the field strength further to $B_p = 5 \times 10^{14} \mathrm{G}$, then there remains no region that satisfies both $E \gtrsim E_{\mathrm{ad}}$ and Equation (\ref{B_l}) simultaneously, and the mode conversion occurs inside the $O$-mode photosphere even at the cyclotron energies of the proton. As a result, the polarization angle and fraction are identical to those in Figures \ref{nmc_polarization_angle_1d13G} and \ref{nmc_polarization_fraction_1d13G} except for an overall increase due to the larger size of the polarization-limiting surface. Note again that the $E$-mode photons emerging from the $E$-mode photosphere are affected by the mode conversion in the cases where $B > B_l$ is satisfied.

\begin{figure}
 \begin{tabular}{cc}
  \begin{minipage}[t]{0.45\hsize}
   \centering
   \includegraphics[width=4cm]{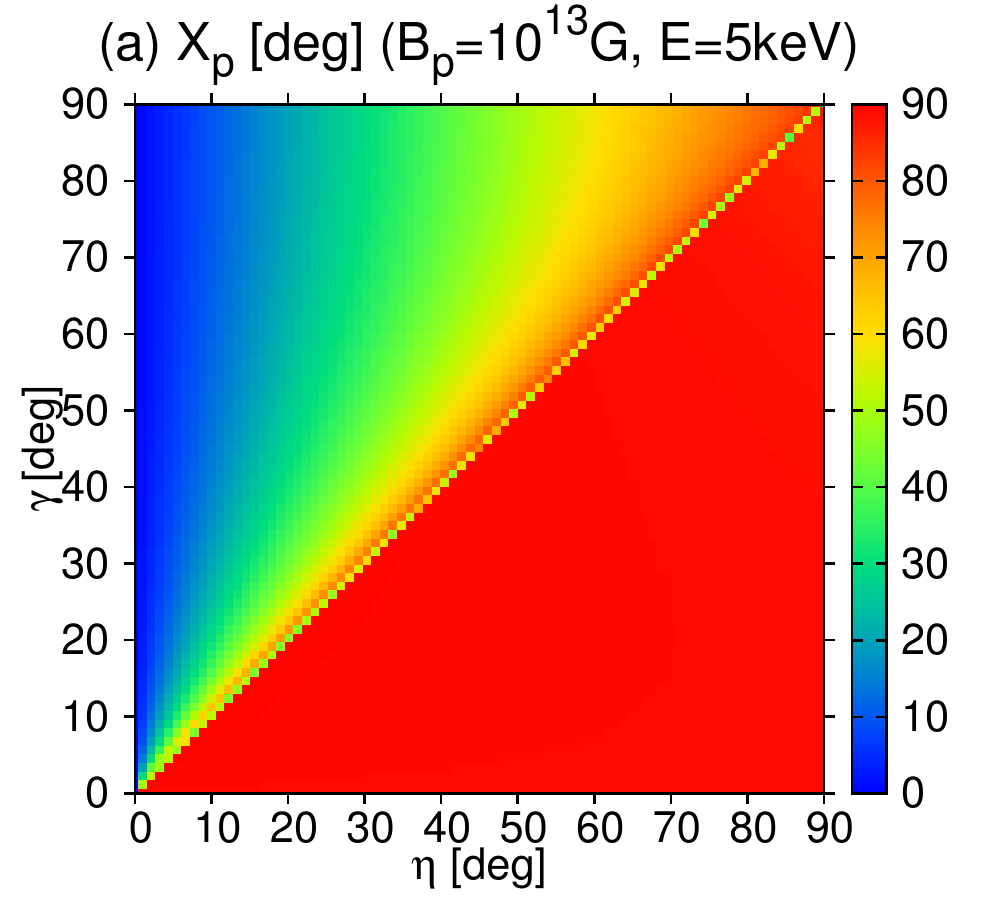}
  \end{minipage} &
  \begin{minipage}[t]{0.45\hsize}
   \centering
   \includegraphics[width=4cm]{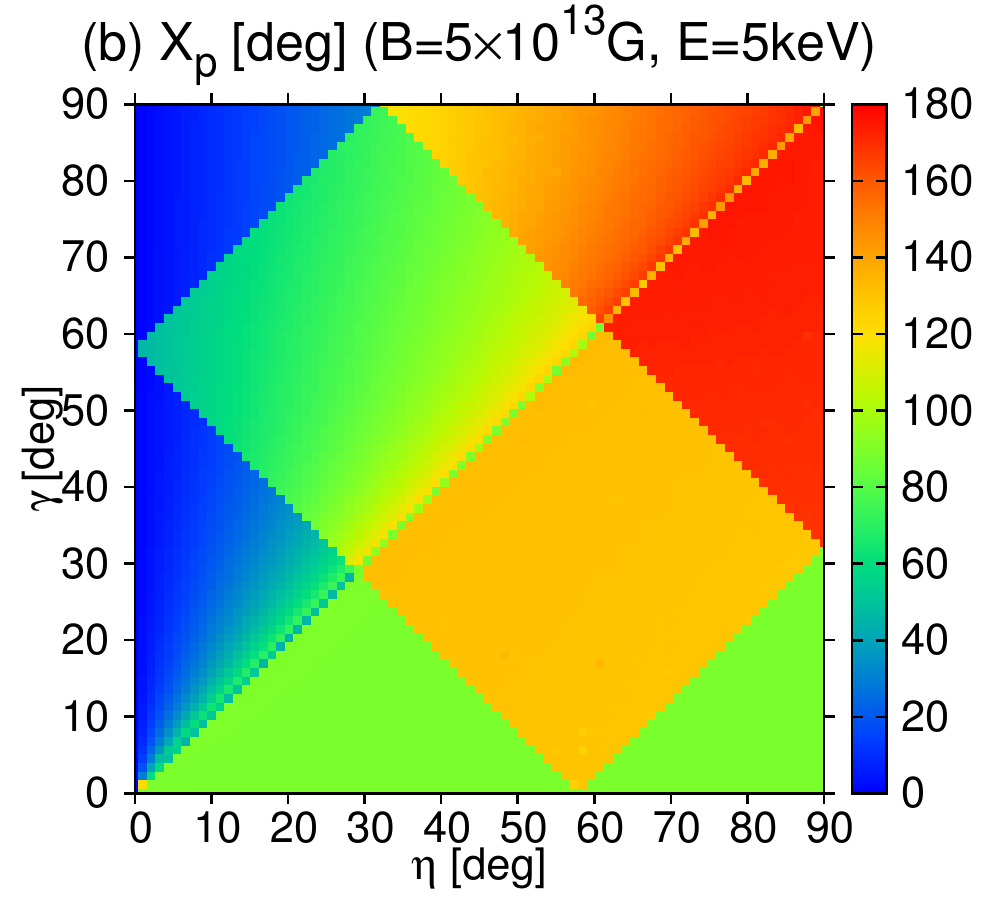}
  \end{minipage}
 \end{tabular}
 \caption{Semi-amplitudes of the polarization angle for two magnetic field strengths: (a) $B_p = 10^{13} \mathrm{G}$ and (b) $B_p = 5 \times 10^{13} \mathrm{G}$. The photon energy is set to $E = 5 \mathrm{keV}$.}
 \label{angle_variation}
\end{figure}

\begin{figure*}[htbp]
   \centering
   \includegraphics[width=17.2cm]{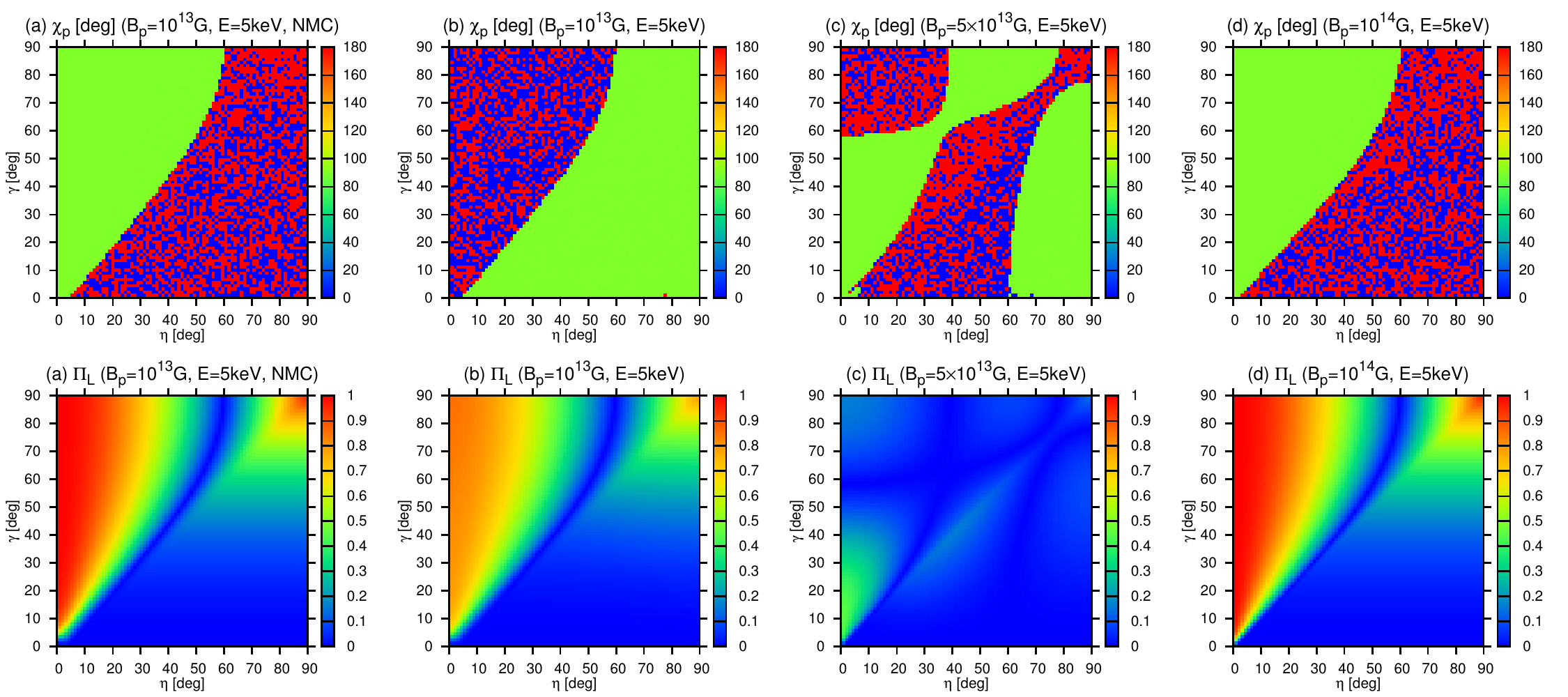}
\caption{Phase-averaged polarization angles $\chi_p$ (upper panels) and fractions $\Pi_L$ (lower panels) for four different cases. The photon energy is fixed to $E = 5 \mathrm{keV}$. The magnetic field strengths are assumed to be $B_p = 10^{13} \mathrm{G}$ in (a) and (b) and $B_p = 5 \times 10^{13}$ and $10^{14} \mathrm{G}$ for (c) and (d), respectively. The mode conversion is considered for all cases except case (a).} 
 \label{phase_averaged_field}
  \end{figure*}

We now consider the semi-amplitude defined in Equation (\ref{semi_amplitude}) for the three magnetic fields: $B_p = 10^{13}, \ 5 \times 10^{13}$ and $10^{14} \mathrm{G}$. The results are shown in Figure \ref{angle_variation} for the first two cases: (a) $B_p = 10^{13} \mathrm{G}$ and (b) $B_p = 5 \times 10^{13} \mathrm{G}$. The photon energy is set to $E = 5 \mathrm{keV}$ for both. In the first case, the features in the semi-amplitude are almost the same as those in Figure \ref{t+tvr_a} (a), in which the mode conversion is neglected. This is because the mode conversion occurs for all combinations of $\eta$ and $\gamma$ at this photon energy, and, as a result, the $O$-mode is always dominant. In the case of $B_p=10^{14} \mathrm{G}$, in contrast, the mode conversion occurs inside the $O$-mode photosphere and the $E$-mode always prevails, since Equation (\ref{B_l}) is not satisfied. Although the polarization angles are different by $90^{\circ}$, the semi-amplitude for $B_p = 10^{14} \mathrm{G}$ is almost identical to that for $B_p = 10^{13} \mathrm{G}$ and hence is not shown in the figure.

In contrast, the right panel for case (b) exhibits qualitatively different features with some discontinuous changes in the parameter space of $\eta$ and $\gamma$. The reason for these discontinuities is, of course, the mode conversion. In fact, the polarization angle changes by $90^{\circ}$ when the dominant mode is changed from $E$-mode to $O$-mode or vice versa. Such a change takes place twice or four times during a single rotation, as is understood from Figure \ref{polarization_angle_5d13G}. The semi-amplitude, which is the total variation of the polarization angle divided by four, may hence change by $45^{\circ}$ or $90^{\circ}$ at the discontinuities.

\subsection{Phase-averaged Quantities for Different configurations, Field Strengths and Photon Energies}

Turning to the phase-averaged quantities, we show in Figure \ref{phase_averaged_field} the four representative patterns of the polarization angle and fraction in the $\eta - \gamma$ plane: (a) $B_p = 10^{13} \mathrm{G}, \ E = 5 \mathrm{keV}$ with no mode conversion; (b) $B_p = 10^{13} \mathrm{G}, \ E = 5 \mathrm{keV}$; (c) $B_p = 5 \times 10^{13} \mathrm{G}, \ E = 5 \mathrm{keV}$; and (d) $B_p = 10^{14} \mathrm{G}, \ E = 5 \mathrm{keV}$. Note again that the phase-averaged quantities are calculated not as the averages of the corresponding phase-resolved quantities but from the integral of the Stokes parameters over the entire rotational phase.

It is seen from the upper panels that the phase-averaged polarization angle is either $\chi _p \simeq 0^{\circ}, \ 90^{\circ}$, or $180^{\circ}$, shown in blue, green, or red, respectively, and that the parameter space is divided into the regions either with $\chi_p \approx 90^{\circ}$ or with $\chi_p \approx 0, \ 180^{\circ}$. Note that the apparent discontinuity between $\chi _p \simeq 0^{\circ}$ and $180^{\circ}$ comes from the mod($\pi$) nature of the polarization angle and is spurious.

If the mode conversion is neglected (case (a)), the phase-averaged polarization angle can be understood from Figure \ref{PF} (b). In regions (1) and (3) in the figure, the polarization is roughly directed toward the $X$-axis at each rotational period, and $\chi _p \simeq 90^{\circ}$ is hence obtained. In regions (5)-(10), the average magnetic fields are oriented in the $Y$-axis more often than not, and, as a result, the polarization angle changes by $90^{\circ}$. In the boundary layer, i.e., regions (2) and (4), the polarization angle is still $\chi_p \approx 0^{\circ}$ or $180^{\circ}$,
but the polarization itself is suppressed. See the corresponding bottom panel.

It is apparent from Figure \ref{phase_averaged_field} that case (d), with the highest magnetic field strength, $B_p=10^{14} \mathrm{G}$, is quite similar to case (a). This is the case not only for the polarization angle but also for the polarization fraction and is simply because the mode conversion occurs in the $O$-mode photosphere in case (d), either, which is understood from Figures \ref{polarization_fraction_1d14G} and \ref{B_l_graph}. For the lower magnetic fields, $B_p = 10^{13}$ and $5 \times 10^{13} \mathrm{G}$, assumed in cases (b) and (c), the mode conversion is important. In case (b), photons are mostly in the $O$-mode at $E = 5 \mathrm{keV}$. as seen in Figures \ref{polarization_angle_1d13G} and \ref{polarization_fraction_1d13G}. As a consequence, the phase-averaged polarization angles are changed by $90^{\circ}$ from those of case (a). There are some $E$-mode photons emitted, though, from the region that satisfies $E < E_{\mathrm{ad}}$ because of large values of $\tan \theta _B$ in Equation (\ref{adiabatic_energy}). The polarization is partially canceled then, and the phase-averaged polarization fraction is lowered a bit in regions (1), (3), (5), (7), and (10).

In case (c) with $B_p = 5 \times 10^{13} \mathrm{G}$, in contrast, the observer will see not only the region in which the mode conversion occurs outside the photospheres of the two modes but also the region in which the mode conversion takes place between the two photospheres. The $E$-mode photons come from the latter region, at which Equation (\ref{B_l}) is not satisfied. It extends from the magnetic pole and covers approximately half the neutron star surface. The $O$-mode photons are originated at low magnetic latitudes, in contrast. As a result, the numbers of $E$-mode and $O$-mode photons are nearly equal in this case, and the phase-averaged polarization fractions are severely reduced. The phase-averaged polarization angles have different features in this case. In some parameter regions, the polarization angle is seen to change by $90^{\circ}$ because of the mode conversion.

\begin{figure}
 \centering
 \includegraphics[width=\columnwidth]{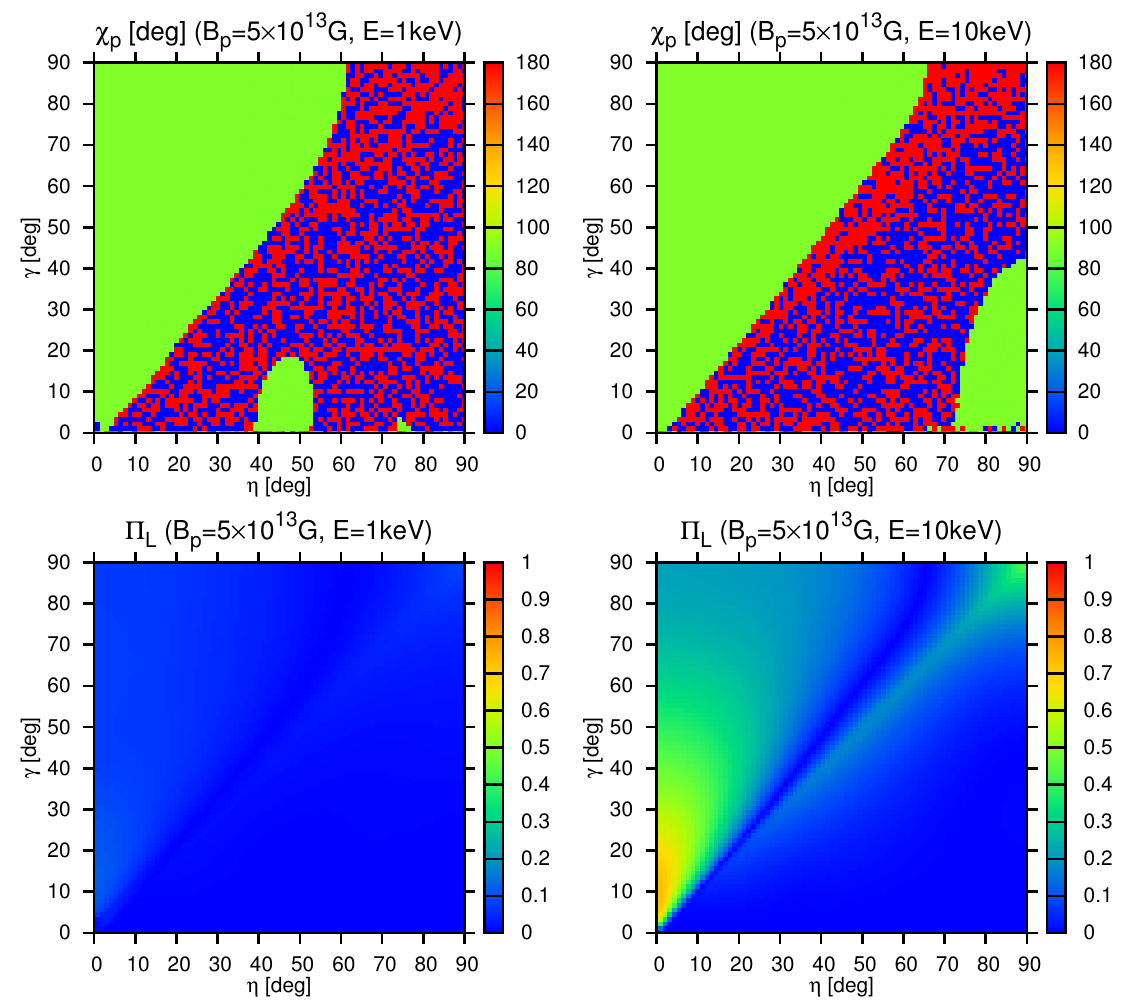}
 \caption{Phase-averaged polarization angles $\chi _p$ (upper panels) and fractions $\Pi _L$ (lower panels) for the photon energies of $E = 1 \mathrm{keV}$ (left) and $E = 10 \mathrm{keV}$ (right). The magnetic field strength is set to $B_p=5 \times 10^{13} \mathrm{G}$.}
 \label{field_averaged_energy}
\end{figure}

We have so far considered a single photon energy. We expect, however, that the results depend strongly on the photon energy. This is confirmed in Figure \ref{field_averaged_energy}, in which we show the phase-averaged polarization angle and fraction for $B_p = 5 \times 10^{13} \mathrm{G}$ but at $E = 1 \mathrm{keV}$ and $E = 10 \mathrm{keV}$ this time. It is indeed found that the phase-averaged polarization fraction is smaller at $E = 1 \mathrm{keV}$ than at $E = 5 \mathrm{keV}$. This is because $E = 1 \mathrm{keV}$ is much closer to the adiabatic energy $E_{\mathrm{ad}}$, nearly half the $E$-mode photons are converted to $O$-mode, and the polarization is almost canceled.

At $E = 10 \mathrm{keV}$, in contrast, the mode conversion occurs inside the $O$-mode photosphere in some regions because of the violation of Equation (\ref{B_l}) in this case (see Figures \ref{polarization_angle_5d13G} and~\ref{B_l_graph} (a)). The cancellation between the two modes is less severe than at $E = 5 \mathrm{keV}$, though. The phase-averaged polarization angles for different photon energies change by $90^{\circ}$ at different combinations of $\eta$ and $\gamma$. Such energy dependence of the mode conversion will be useful to distinguish the effects of the mode conversion from those of the configuration of the neutron star if they are observed at multiple energy bands in the future.

\begin{figure}
 \centering
 \includegraphics[width=\columnwidth]{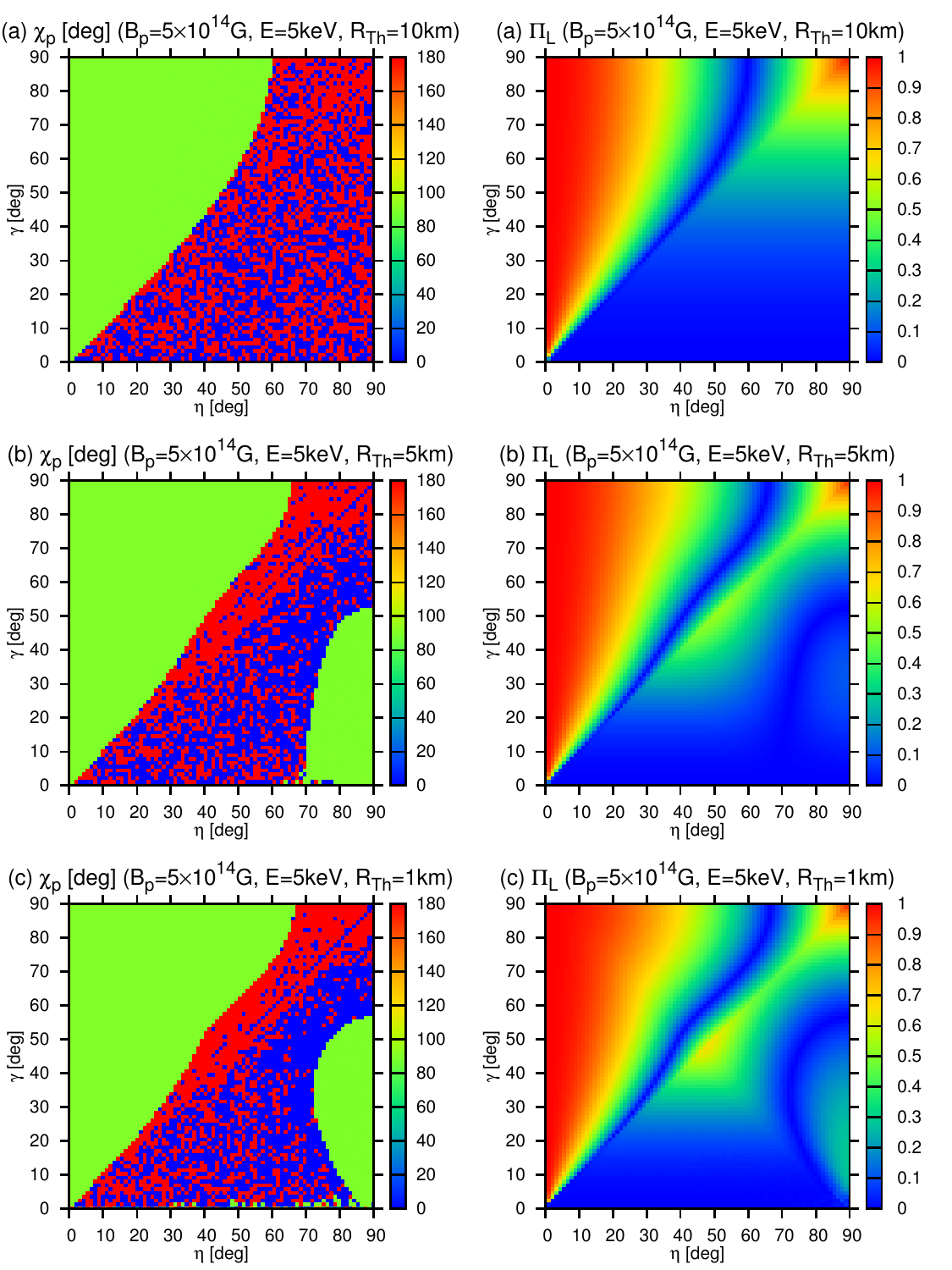}
 \caption{Phase-averaged polarization angles (left column) and fractions (right column) of a neutron star with hot spots. The dipole magnetic field strength is set to $B_p = 5 \times 10^{14} \mathrm{G}$. The spot size is assumed to be $R_{\mathrm{Th}} = 10, \ 5,$ and $1 \mathrm{km}$ from top to bottom.} 
 \label{pf_spot}
\end{figure}

\subsection{Hot Spot}

So far we have assumed that the temperature is uniform on the neutron star surface, but this may not be true. In fact, the observed energy spectra of the magnetar emissions are normally fitted with the composition of a blackbody radiation and a power-law emission and give us an estimate of the temperature and size of the region that produces the thermal emission as $T \sim 0.4 \mathrm{keV}$ and $R_{\mathrm{Th}} \sim 5 \mathrm{km}$ for anomalous X-ray pulsars (AXPs) and $T \sim 0.7 \mathrm{keV}$ and $R_{\mathrm{Th}} \sim 1 \mathrm{km}$ for soft gamma-ray repeaters (SGRs). These results suggest that the thermal-emission region does not cover the entire surface and may be associated with a hot spot.

We hence consider the possible effects of the existence of such a hot spot on the phase-averaged polarization angle and fraction. We actually assume that two hot spots of the same size cover both the magnetic polar regions. We set the magnetic field strength to $B_p = 5 \times 10^{14} \mathrm{G}$ so that the mode conversion should occur inside the $O$-mode photosphere. The results are shown in Figure \ref{pf_spot}. The photon energy is again fixed to $E = 5 \mathrm{keV}$. The phase-averaged polarization angles and fractions are presented in the left and right columns, respectively, for the hot spot radii of $R_{\mathrm{Th}} = 10, \ 5$, and $1 \mathrm{km}$. As for the polarization fraction, it is immediately apparent from the figure that the red region, where the polarization fraction is large, is not changed much by the variation in the spot size; it is the vicinity of $\eta = \gamma = 45^{\circ}$ that is most affected. The increase of the polarization fraction is also seen in the region near $\eta = 90^{\circ}, \ \gamma = 20^{\circ}$. The polarization angle also changes by $90^{\circ}$ in these parameter regions. As expected, the parameter regions with low polarization fractions tend to be affected \citep{2016MNRAS.459.3585G}.

\begin{figure}
 \centering
 \includegraphics[width=\columnwidth]{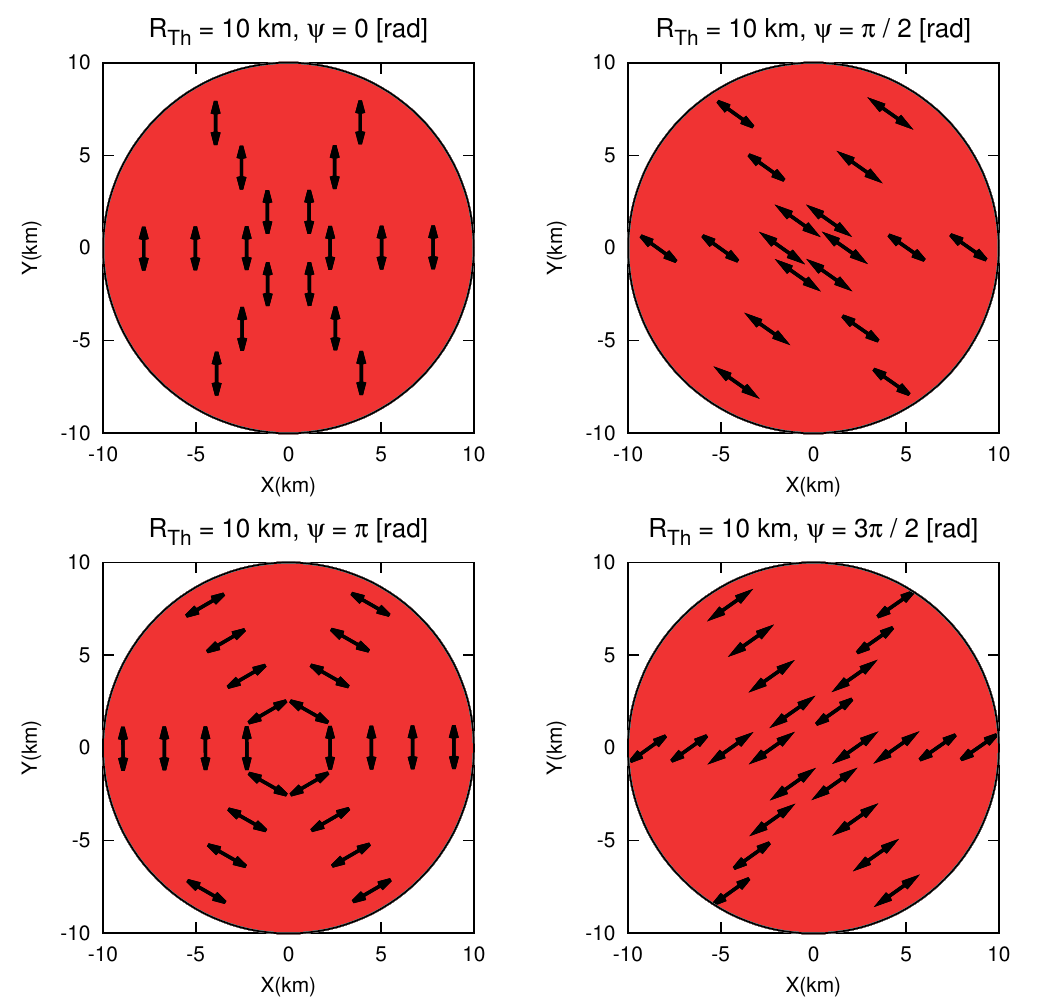}
\caption{Polarizations of photons at different rotational phases ($\psi = 0, \pi /2 , \pi , 3 \pi /2$) for the spot radius $R_{\mathrm{Th}}=10 \mathrm{km}$. The magnetic field strength is $B_p = 5 \times 10^{14} \mathrm{G}$ and $\eta = 45^{\circ}$,  $\gamma = 45 ^{\circ}$. The black circles indicate the observed patch on the polarization-limiting surface. The red areas indicate the hot spot, which covers the whole surface in this case.} 
 \label{ss10}
\end{figure}
\begin{figure}
 \centering
 \includegraphics[width=\columnwidth]{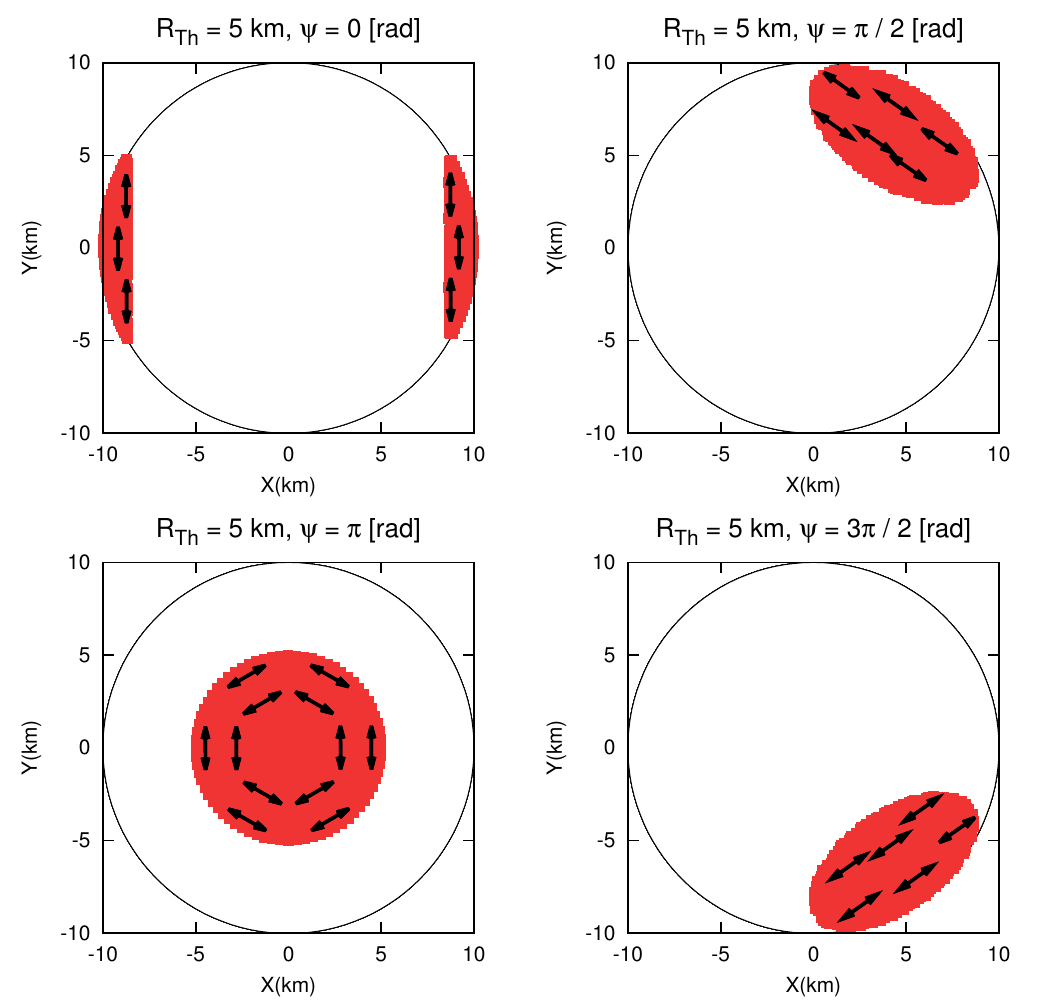}
 \caption{Same as Figure \ref{ss10} but for the spot radius of $R_{\mathrm{Th}} = 5 \mathrm{km}$.} 
 \label{ss5}
\end{figure}

\begin{figure}
 \centering
 \includegraphics[width=\columnwidth]{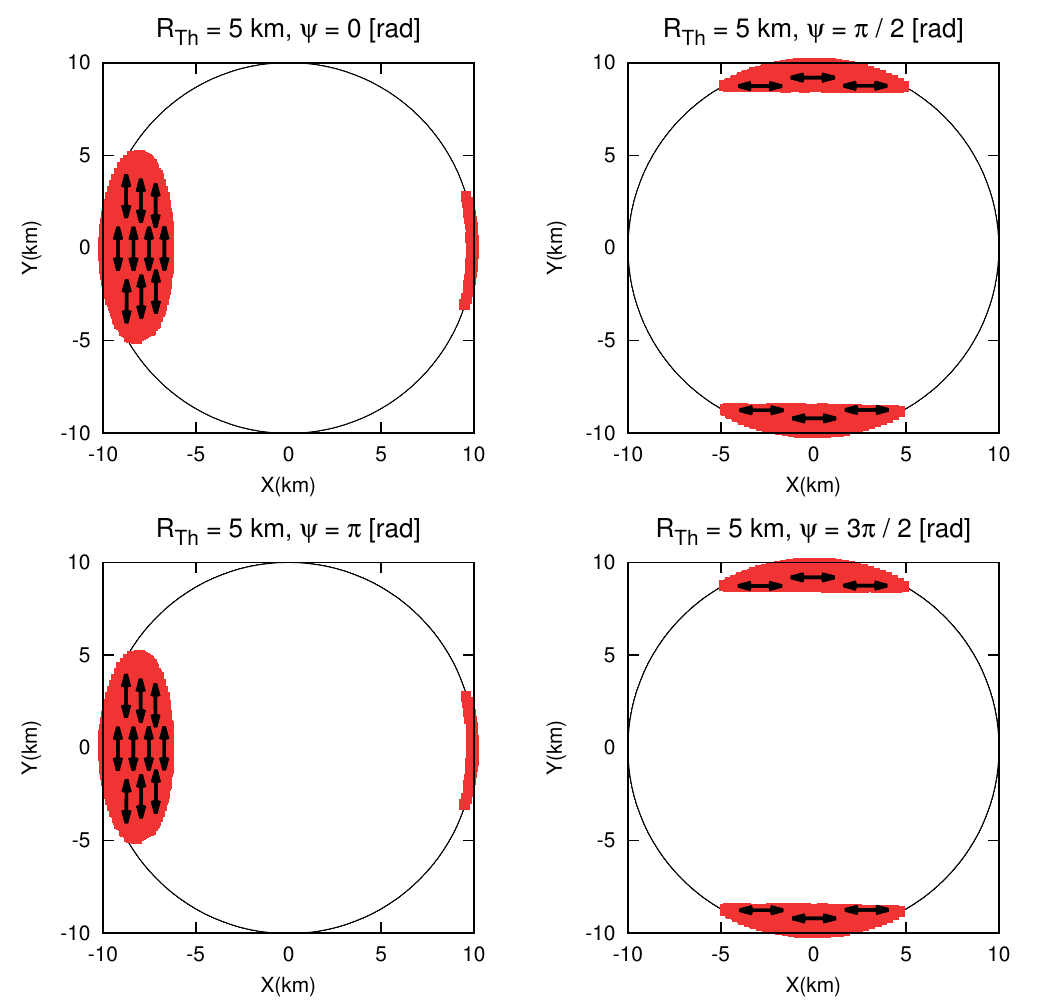}
 \caption{Same as Figure \ref{ss5} but for $\eta = 90^{\circ}, \ \gamma = 20^{\circ}$.} 
 \label{ss5_eta90_gamma20}
\end{figure}

\begin{figure*}
 \centering
 \includegraphics[width=17.2cm]{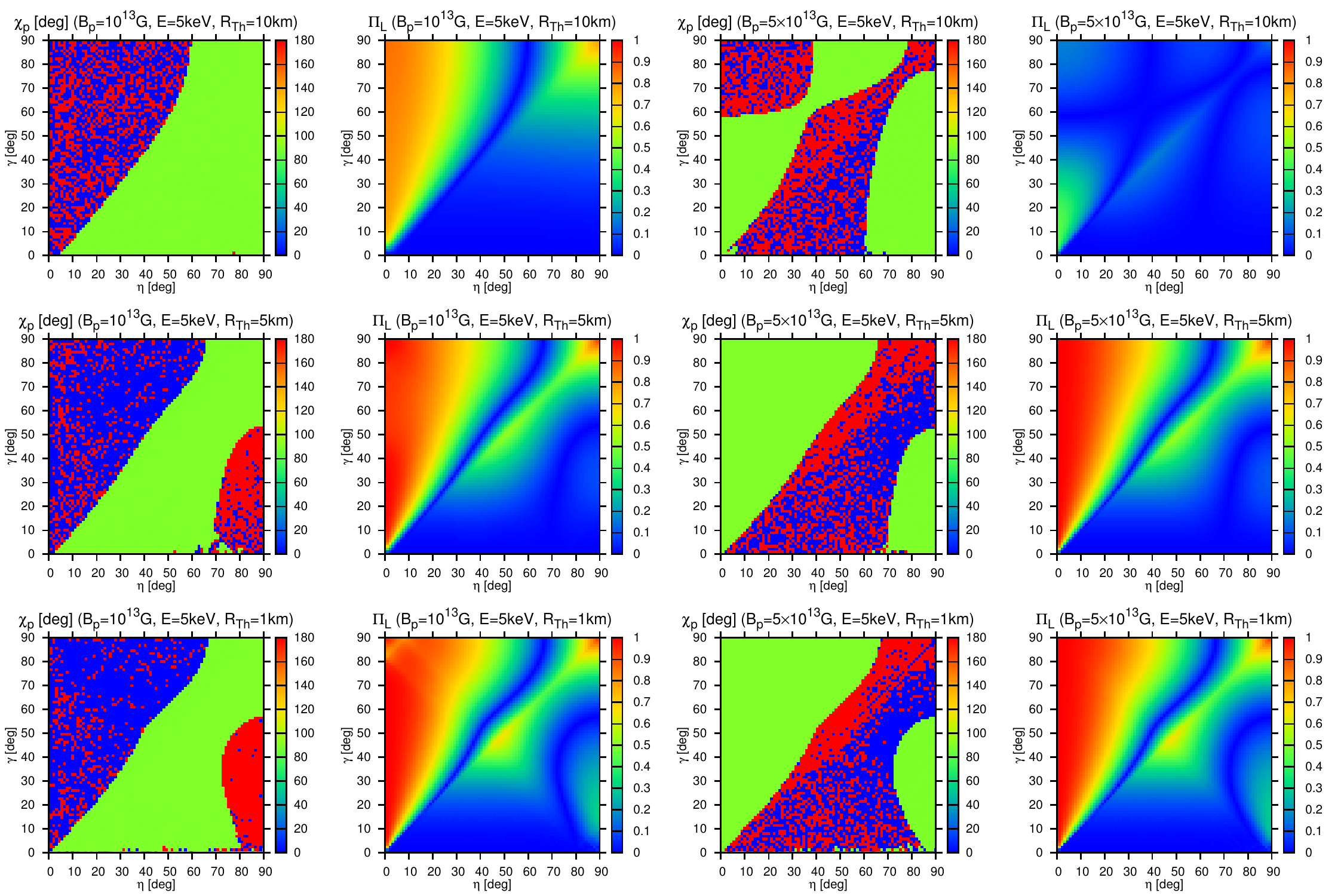}
 \caption{Phase-averaged polarization angles and fractions for different hot-spot sizes. The magnetic field strength is set to $B_p=10^{13} \mathrm{G}$ (left two columns) and $5 \times 10^{13} \mathrm{G}$ (right two columns). The top, middle, and bottom panels correspond to the spot sizes of $R_{\mathrm{Th}} = 10, \ 5$, and $1 \mathrm{km}$, respectively.}
 \label{field_spot}
\end{figure*}

In general, the polarization fraction tends to increase as the emission is limited to a smaller region, since the magnetic field becomes more uniform in this region. There is another reason, however, for the increases of the polarization fraction in the parameter regions mentioned above. This is understood from Figures \ref{ss10} and~\ref{ss5}, in which the snapshots of the polarization directions in the observed patch on the polarization-limiting surface are shown at different rotational phases for the spot radii of $R_{\mathrm{Th}} = 10 \mathrm{km}$ and $R_{\mathrm{Th}} = 5 \mathrm{km}$, respectively. The magnetic field strength is fixed to $B_p= 5 \times 10^{14} \mathrm{G}$, and $\eta = \gamma = 45^{\circ}$ is chosen. The localization of the emission region to the hot spot is evident in the latter case. At $\psi = 0$, hot spots at both the north and south poles are barely visible at the left and right edges of the observed patch, while at $\psi = \pi$, the hot spot at the north pole comes at the center. In the former case, the radiation is polarized in the $Y$-direction, whereas in the latter, the net polarization vanishes. At $\psi = \pi /2$ and $3 \pi /2$, in contrast, the polarization directions are tilted by about $45^{\circ}$ to the $Y$-axis. It is easily understood, then, that as the spot size gets smaller, the cancellation between the radiation at $\psi = 0$ and that at $\psi = \pi /2$ and $3 \pi /2$ becomes weaker, since the hot spots at $\psi = 0$ are less visible. This is the reason for the increase in the polarization fraction around $\eta = \gamma = 45^{\circ}$ with the decrease in the spot size exhibited in panels (a)-(c) of Figure \ref{pf_spot}. The change in the neighborhood of $\eta = 90^{\circ}, \ \gamma = 20^{\circ}$ is also understood in the same way (see Figure \ref{ss5_eta90_gamma20}).

The behaviors of the phase-averaged polarization angle should now be apparent. In the vicinity of $\eta = 45^{\circ}, \ \gamma = 45^{\circ}$, the contributions from the rotational phases around $\psi = 0$ are reduced as the spot radius gets smaller. Then, the phase-averaged polarization angle tends to be $\chi _p \simeq 0^{\circ}$, or $180^{\circ}$ for small spot sizes. In the case of $\eta \simeq 90^{\circ}, \gamma \simeq 20^{\circ}$, in contrast, it is evident from Figure \ref{ss5_eta90_gamma20} that the polarization angle tends to be $\chi _p \simeq 90^{\circ}$.

The mode conversion still occurs inside the $O$-mode photosphere on any part of the neutron star surface for $B_p=10^{14} \mathrm{G}$ at $E = 5 \mathrm{keV}$, since the condition $B > B_l$ is satisfied everywhere (see Figure \ref{B_l_graph} (a)). The phase-averaged polarization properties are hence essentially the same as those for $B_p=5 \times 10^{14} \mathrm{G}$, irrespective of the hot spot. As the magnetic field strength becomes even lower, the mode conversion starts to the place at low magnetic latitudes and lowers the polarization fractions in general if photons are emitted from the entire neutron star surface, as was demonstrated in the previous section. This is particularly the case for $B_p=5 \times 10^{13} \mathrm{G}$ (see the two top right panels of Figure \ref{field_spot}), since the surface is almost equally divided into the region where the mode conversion occurs inside the $O$-mode photosphere, near the pole, and the region where the mode conversion occurs outside the two photospheres, extended from the magnetic equator. In the case of $B_p=10^{13} \mathrm{G}$, the mode conversion always occurs, and photons are mostly in the $O$-modes, except in the region where $E < E_{\mathrm{ad}}$ is satisfied because of large values of $\tan \theta_B$. The latter effect is the reason why the phase-averaged polarization fractions are still somewhat reduced from those for the corresponding no mode conversion.

In Figure \ref{field_spot}, we show how the existence of hot spots modified the phase-averaged polarization angles and fraction for $B_p = 10^{13}$ and $5 \times 10^{13} \mathrm{G}$. The top, middle, and bottom panels correspond to the spot sizes of $R_{\mathrm{Th}} = 10, \ 5$ and $1 \mathrm{km}$, respectively. Note that the top four panels are essentially the same as those presented in Figure \ref{phase_averaged_field}. The two columns on the left show the results for $B_p=10^{13} \mathrm{G}$. In the case of $R_{\mathrm{Th}}= 10 \mathrm{km}$, as mentioned above, the polarization fractions decrease a little from those for no mode conversion, particularly in the region where it is high. The polarization angles are also changed by $90^{\circ}$.

As the size of the hot spot becomes smaller, the phase-averaged polarization fractions return to the higher values for no mode conversion. This is because the region with $E < E_{\mathrm{ad}}$ rarely enters the hot spot. The exceptional cases are limited to the configurations with $\gamma \sim \eta + 80^{\circ}$ for $R_{\mathrm{Th}} = 1 \mathrm{km}$. In these cases, $E < E_{\mathrm{ad}}$ is satisfied at some rotational phases, and the cancellation between the two modes lowers the phase-averaged polarization fractions slightly, as observed. In contrast, the $O$-mode is dominant for this magnetic field strength irrespective of the spot size, and the behavior of the polarization angles in the $\eta - \gamma$ plane is essentially the same as that for $B_p = 5 \times 10^{14} \mathrm{G}$, except for the overall difference by $90^{\circ}$ because of the mode conversion.

In the case of $B_p = 5 \times 10^{13} \mathrm{G}$, the effect of the hot spot is drastic, as can be immediately seen in the two right-hand columns in Figure \ref{field_spot}. In fact, the reduction of the phase-averaged polarization fraction by the mode conversion is nearly nullified when the spot size becomes as small as $5 \mathrm{km}$. This is easily understood as follows. Since the mode conversion occurs outside the two photospheres in the region near the equator, it is not included in the hot spot if its size is small. Then, the photons are mostly in the $E$-mode, just as in the case neglecting the mode conversion and the phase-averaged polarization fractions for $R_{\mathrm{Th}} = 1, \ 5 \mathrm{km}$ are almost the same as those for $B_p = 5 \times 10^{14} \mathrm{G}$. Since the dominant mode is the $E$-mode for all combinations of $\eta$ and $\gamma$ for these small hot-spot sizes, the polarization angles are identical to those for $B_p = 5 \times 10^{14} \mathrm{G}$ in Figure \ref{pf_spot}.

\begin{figure}
 \centering
 \includegraphics[width=9.0cm]{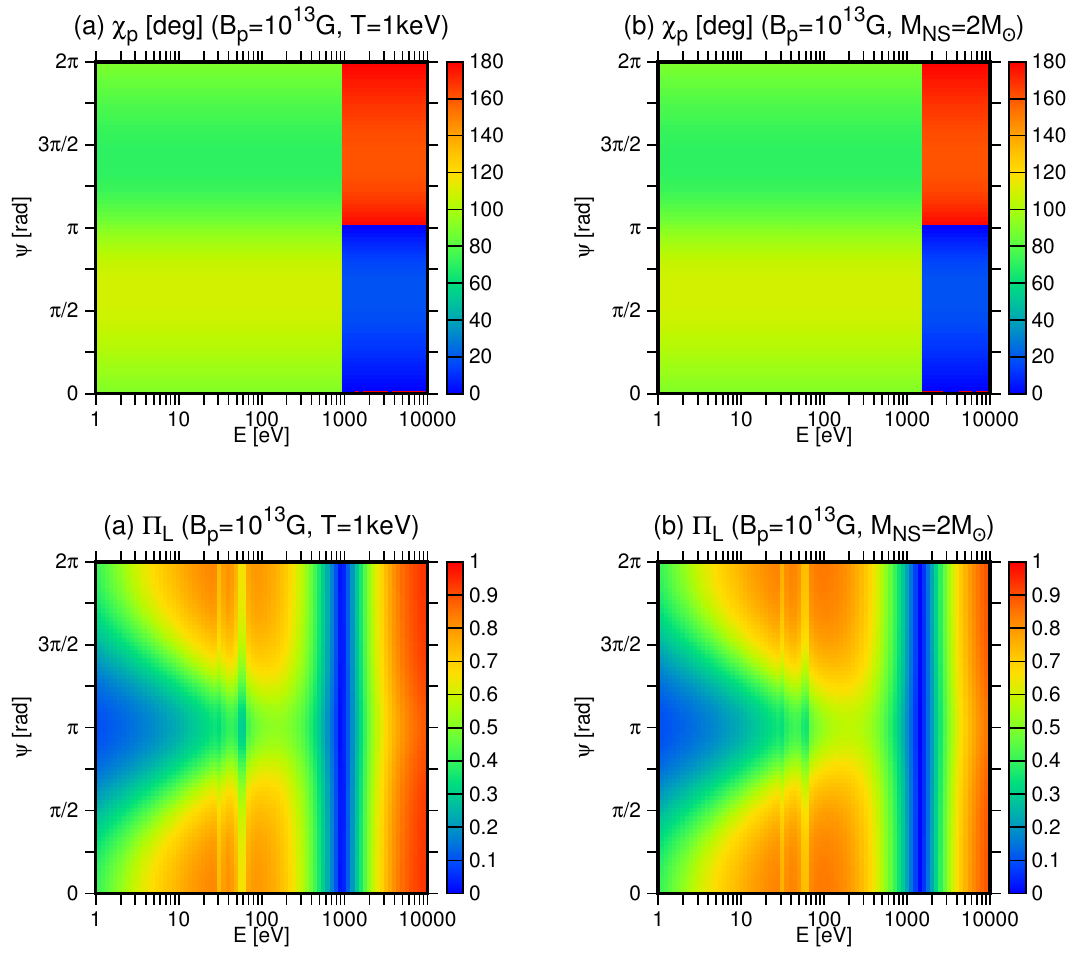}
 \caption{Same as for the case of $\eta = 5^{\circ}$ and $\gamma = 15^{\circ}$ in Figures \ref{polarization_angle_1d13G} and \ref{polarization_fraction_1d13G} but (a) for a different temperature $T=1 \mathrm{keV}$ or (b) for a different mass of neutron star $M_{\mathrm{NS}}=2 M_{\odot}$.}
\label{temp_mass} 
  \end{figure}

\subsection{Other Parameters}

We next discuss the dependence on the surface temperature $T$, neutron star mass $M_{\mathrm{NS}}$, and radius $R_{\mathrm{NS}}$. They affect the results mainly through the adiabatic energy for the vacuum resonance $E_{\mathrm{ad}}$, which depends on the scale height of the atmosphere $H_{\rho}$ in Equation (\ref{adiabatic_energy}). The latter is proportional to the temperature and the inverse of the surface gravity, $g = G M_{\mathrm{NS}} / R_{\mathrm{NS}}^2$. Recall that the adiabatic energy $E_{\mathrm{ad}}$ is the energy above which the mode conversion occurs adiabatically and the polarization angle changes by $90^{\circ}$, and near which the polarization fraction tends to be reduced.

The phase-resolved polarization angle $\chi _p$ and fraction $\Pi _L$ for $\eta = 5^{\circ}$ and $\gamma = 15 ^{\circ}$ are recalculated either with a higher temperature of $T=1 \mathrm{keV}$ or with a larger neutron star mass of $M_{\mathrm{NS}}=2 M_{\odot}$. They are $T=0.4 \mathrm{keV}$ and $M_{\mathrm{NS}}=1.4 M_{\odot}$, respectively, in the fiducial model. Note that it is the increase or decrease in the scale height that matters, and one can equally change the neutron star radius instead of the temperature or the neutron star mass, since the scale height is a function of the combination $T/(M_{\mathrm{NS}}/R_{\mathrm{NS}}^2)$. The magnetic field strength is set to $B_{p} = 10^{13} \mathrm{G}$. The results are shown in Figure \ref{temp_mass}. One can see that the difference between the models is almost indiscernible. This is just as expected, since the adiabatic energy depends on the scale height only weakly: $E_{\mathrm{ad}} \propto H_{\rho} ^{-1/3}$. We hence conclude that the results obtained so far are robust.

\subsection{Applications to Real Magnetars}

\begin{table}
 \caption{Observationally Inferred Magnetic Field Strengths, Surface Temperatures and the Radii of Hot Spots for Four of the Known Magnetars.}
 \label{magnetar_parameter}
 \begin{tabular}{cccc} \hline
  Magnetar \footnote{The obvious abbreviations are employed for 1E 2259+586, 4U 0142+61, SGR 0501+4516, and 1RXS J17089.0-400910. The values are taken from~\cite{2009PASJ...61..109N}.} & $B_p$ ($10^{14}$ G) & $T$ (keV) & $R_{\mathrm{Th}}$ (km) \\ \hline \hline
  2259+58 & 0.59 & 0.37 & 5.0 \\
  0142+61 & 1.3 & 0.36 & 9.4 \\
  0501+45 & 1.9 & 0.70 & 1.4 \\
  1708-40 & 4.7 & 0.48 & 4.5 \\ \hline
 \end{tabular}
\end{table}

\begin{figure}
 \centering
 \includegraphics[width=\columnwidth]{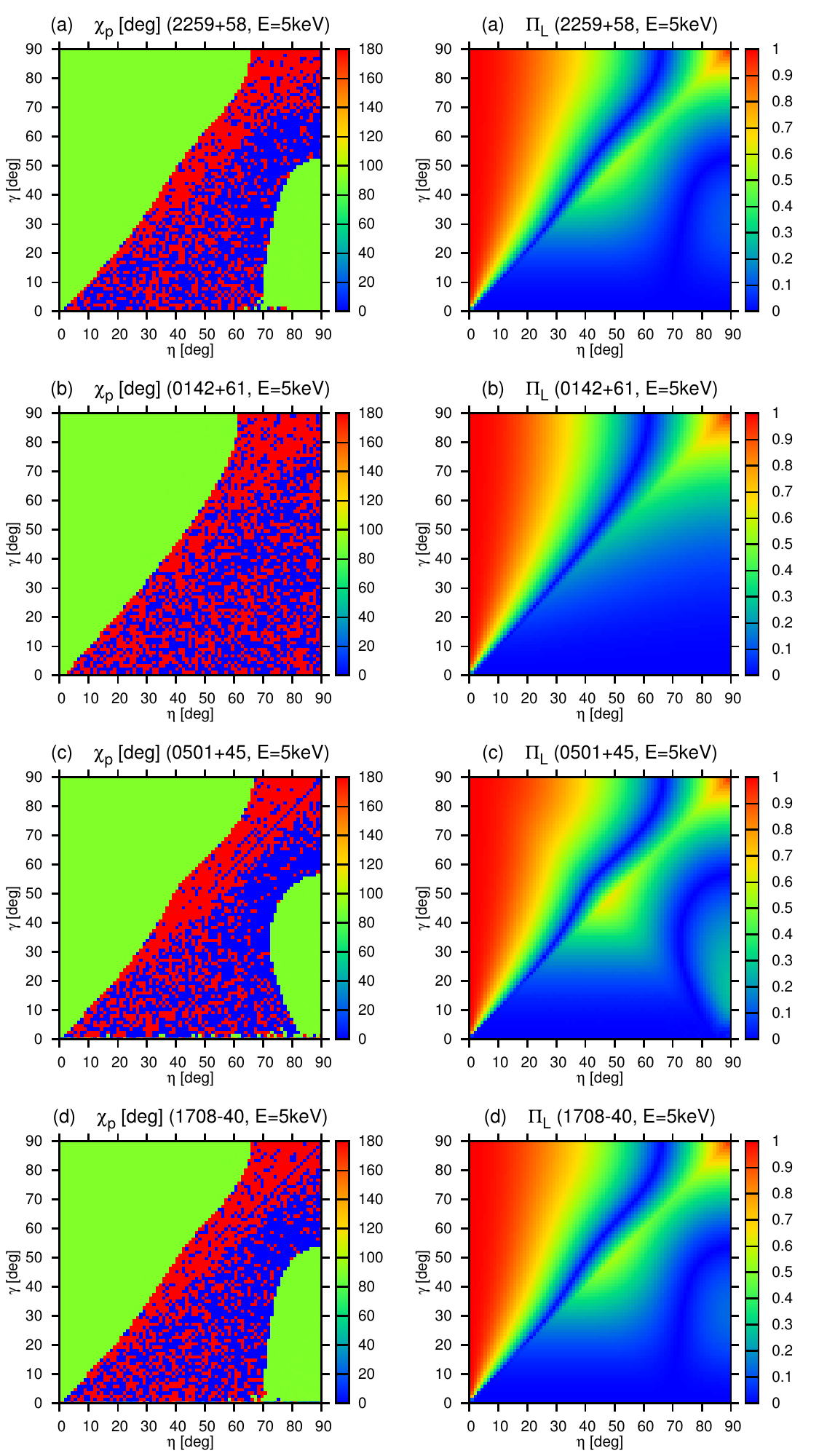}
 \caption{Phase-averaged polarization angles (left column) and fractions (right column) at $E = 5 \mathrm{keV}$ for the four magnetars: (a) 1E 2259+586, (b) 4U 0142+61, (c) SGR 0501+4516, and (d) 1RXS J17089.0-400910.}
 \label{magnetars_5keV}
\end{figure}

We finally apply the theory developed thus far to real magnetars. Our intention here is not to make a strong claim on the possibility to detect polarizations as envisaged in this paper from these magnetars, which would be impossible if one considers various uncertainties in theoretical interpretations and modelings of observations as explained below. Instead, we would like to get a rough idea of what the polarization angles and fractions would be like if our models were true. Here we deal with the four magnetars 1E 2259+586, 4U 0142+61, SGR 0501+4516, and 1RXS J17089.0-400910, since thermal radiation is identified observationally in the soft X-ray band \citep{2010ApJ...722L.162E}.

We employ the values of the dipole magnetic field strength $B_p$, the temperature $T$, and the radius of the emission region $R_{\mathrm{Th}}$ obtained from the spectral fittings by two blackbody components with different temperatures and radii by \cite{2009PASJ...61..109N}. They are summarized in Table \ref{magnetar_parameter}. Since the radius of the emission region for the high-temperature component is only about a tenth of that for the low-temperature component, and the former component gives a rather poor fit to the high-energy part of the spectrum, we assume in this paper that the low-temperature component is originated from the hot spot on the magnetar surface and do not consider the high-temperature component. In fact, the magnetars other than SGR 0501+4516 do not reproduce the apparent excesses at $> 7 \mathrm{keV}$ in their spectral fit \citep{2009PASJ...61..109N}. It should also be mentioned that the spectra of persistent emissions from these magnetars may be better fit by the superposition of a blackbody component plus a power-law tail \citep{2007MNRAS.381..293R,2007Ap&SS.308..505R,2009MNRAS.396.2419R,2014ApJ...789...75V}. The power-law tails become important already $\sim 3-4 \mathrm{keV}$ in some cases. It is important here, regardless of which model is better, that both of them indicate the existence of the thermal component and that the temperatures and radii of the emission regions inferred from the observed blackbody components are not much different between the two cases. Note, however, that Comptonization effects, which are supposed to be responsible for the formation of the high-energy tails in the spectra, are normally associated with flows of charged particles along magnetic field lines \citep{2002ApJ...574..332T}, which will hit the magnetar surfaces intensely \citep{2002ApJ...574..332T,2008MNRAS.386.1527N}. As a result, the atmospheric state may be different from what we have considered in this paper. As we know nothing of the mass and radius for these magnetars, we simply adopt the canonical values, $M = 1.4 M_{\odot}$ and $R_{\mathrm{NS}} = 10 \mathrm{km}$, for all of them.

\begin{figure}
 \centering
 \includegraphics[width=\columnwidth]{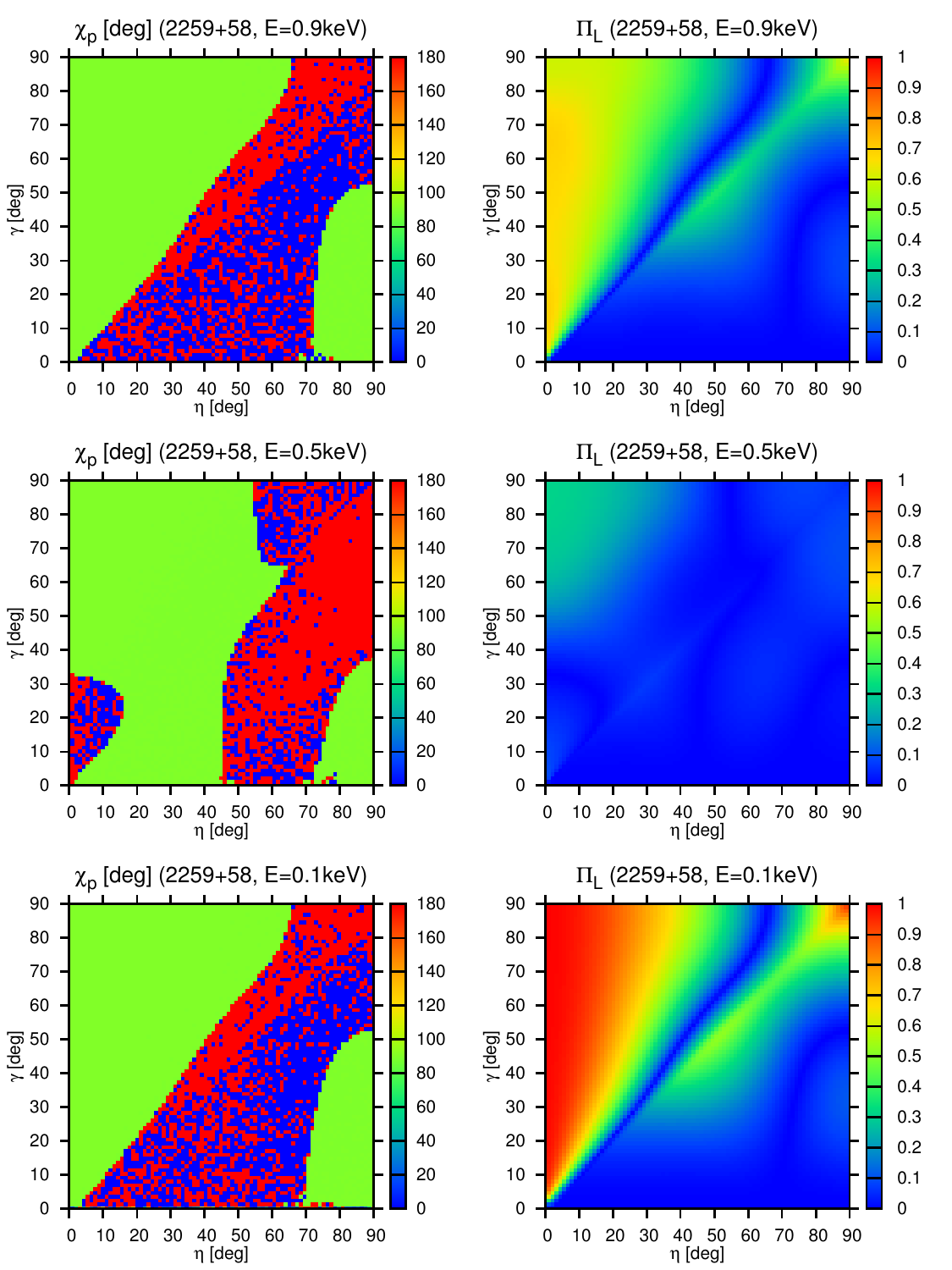}
 \caption{Phase-averaged polarization angles (left column) and fractions (right column) for magnetar 1E 2259+586 at $E = 0.9 \mathrm{keV}$ (top row), $E = 0.5 \mathrm{keV}$ (middle row), and $E = 0.1 \mathrm{keV}$ (bottom row).}
 \label{2259_low}
\end{figure}

With all of these caveats in mind, we present the phase-averaged polarization angles and fractions for $5 \mathrm{keV}$ photons in Figure \ref{magnetars_5keV}. As expected, the existence of the hot spot is recognized from the increase in the polarization fraction around $\eta = \gamma = 45^{\circ}$ for all of the cases except 4U 0142+61, in which the spot size is comparable to the neutron star radius. In fact, the smaller the spot is, the larger the enhancement becomes. These pictures are not changed qualitatively as long as the photon energy is higher than $\sim 1 \mathrm{keV}$. The effects of the small spot radii are also seen in the polarization angles in the parameter regions of $\eta \simeq 90^{\circ}, \ \gamma \simeq 20^{\circ}$, except for the case for 4U 0142+61.

At lower energies, $E \lesssim 1 \mathrm{keV}$, the phase-averaged polarization fraction may be reduced as a consequence of the partial mode conversion at $E \sim E_{\mathrm{ad}}$, and the polarization angle is also affected. This is demonstrated in Figure \ref{2259_low} for magnetar 1E 2259+586 at the photon energies of $E = 0.5$ and $0.9 \mathrm{keV}$. It is evident that at $E=0.9 \mathrm{keV}$ (top row) the reduction of the polarization fraction is already substantial, though the polarization angle is not so much affected. In contrast, at $E=0.5 \mathrm{keV}$ (middle row), the polarization angle is also modified in some region of $\eta$ and $\gamma$, and, as a matter of fact, the photons are essentially unpolarized for all configurations at this photon energy. At much smaller energies $E \sim 0.1 \mathrm{keV}$ (bottom row), however, the mode conversion is frozen, and the polarization angles and fractions return to those for nonconversion.

\section{Summary} \label{summary}

In this paper, we have systematically computed the phase-resolved polarization angles and fractions, which are one of the most important observables in future observations, for different photon energies and various configurations of the rotation axis and the dipole magnetic field to facilitate the interpretation of observational data. In so doing, we have accounted for the mode conversion, which was neglected in the previous study \citep{2015MNRAS.454.3254T}.

We have started with the reproduction of the previous results for $B_p=10^{13} \mathrm{G}$ \citep{2015MNRAS.454.3254T}. For that purpose, we have neglected the mode conversion intentionally. We have found a good agreement, although the bending of photon trajectories and modifications of the dipole magnetic field by general relativity are not considered in our calculations. This suggests that these effects are rather minor. We have then included the mode conversion and studied in detail how the results are modified. 

We have found that the adiabatic mode conversion occurs for high-energy photons with $E \gtrsim E_{\mathrm{ad}} \sim 2 \mathrm{keV}$ and the polarization angle changes by $90^{\circ}$. At $E \simeq E_{\mathrm{ad}}$, the mode conversion occurs nonadiabatically and the $E$- and $O$-modes are mixed, resulting in lower polarization fractions in general. At lower energies, the mode conversion is frozen, the photons are all in the original $E$-mode, and the polarization fraction returns to high values. The adiabatic energy $E_{\mathrm{ad}}$ is actually a function of photon energy, though, and vanishes at the cyclotron frequencies of the proton, $\sim 30-60 \mathrm{eV}$. The polarization fraction is somewhat reduced at these energies again, although the polarization angle is not affected.
At very low energies, the polarization fraction is lowered again, since the polarization-limiting surface gets much closer to the neutron star and the polarizations are largely canceled among photons coming from different parts of the neutron star surface.

We have also presented the semi-amplitude, i.e., the total variation of the polarization angle (divided by a factor of 4) and the phase-averaged polarization fraction following \cite{2015MNRAS.454.3254T}. We have divided the $\eta$-$\gamma$ plane into 10 regions and discussed the features in each region in detail. We have observed that high polarization fractions are obtained when $\eta \ll \gamma$. The semi-amplitude is small in that case. The mode conversion tends to reduce the phase-averaged polarization fractions.

We have then conducted more comprehensive investigations of both the phase-resolved and averaged quantities, varying not only the configuration of the rotation and magnetic axes but also the magnetic field strength and photon energy. We have also considered the effect of the possible existence of a hot spot on the neutron star surface. Although the dependence of the results on other parameters that specify the properties of the neutron star, i.e., the mass, radius, and surface temperature, has also been studied, we have found it minor, since they appear only in the adiabatic energy through the density scale height of the atmosphere of the neutron star.

We have shown that in the absence of the mode conversion, the behavior of the phase-resolved polarization angle in the $E$-$\psi$ (rotational phase) plane can be divided into three cases with $\eta < \gamma$, $\eta = \gamma$, and $\eta > \gamma$. In the first case, the polarization angle oscillates around $\chi_p=90^{\circ}$. In the second case, it changes by $180^{\circ}$, whereas in the third case, it changes more than $180^{\circ}$ during a single rotation of the neutron star. Without the mode conversion, the phase-resolved polarization fraction is large at high photon energies, as in the previous case. As the photon energy is lowered, the polarization-limiting surface comes closer to the neutron star, and the cancellation among photons originated from different parts of the neutron star surface tends to decrease the polarization fraction. This is particularly the case at the rotational phase of $\psi =\pi/2$.

Taking into account the mode conversion, we have demonstrated that the polarization angle is changed by $90^{\circ}$ at high photon energies $E \gtrsim E_{\mathrm{ad}}$. In the case of $\eta = \gamma$, $E_{\mathrm{ad}}$ becomes small at $\psi = \pi /2$, and the jump of the polarization angle occurs accordingly at much lower energies at this rotational phase. The phase-resolved polarization fraction is reduced by the mode conversion at $E \simeq E_{\mathrm{ad}}$, since it occurs nonadiabatically at these energies and the $E$- and $O$-modes are mixed in some proportions. At much lower energies, the mode conversion is frozen, and the results are essentially the same as those without the mode conversion except at the cyclotron energies of the proton $\sim 30-60 \mathrm{eV}$ for $B_p=10^{13} \mathrm{G}$, where $E_{\mathrm{ad}}$ vanishes and the resultant adiabatic mode conversion lowers the polarization fraction a bit.

For a bit stronger magnetic field, $B_p = 5 \times 10^{13} \mathrm{G}$, we have found that the $90^{\circ}$ change of the polarization angle can occur twice or four times at $E \gtrsim E_{\mathrm{ad}}$ during a single rotation of the neutron star. This happens because the neutron star surface is dominated at some rotational phases by the region that violates the condition given in Equation (\ref{B_l}), where the mode conversion occurs inside the $O$-mode photosphere, in addition to the region that has a large value of $\tan \theta_B$ and the effect of the mode conversion is suppressed. The phase-resolved polarization fraction is modified in two ways: since the polarization-limiting radius is larger, the polarization fraction tends to be higher as a whole; the cyclotron energy is raised to $\sim 150-300 \mathrm{eV}$, and the slight reductions of the polarization fraction have been observed at these photon energies. We have also seen some variations with the rotational phase at $E \gtrsim E _{\mathrm{ad}}$. We have found, in contrast, that the semi-amplitudes have an interesting pattern in the $\eta$-$\gamma$ plane according to the number of $90^{\circ}$ changes in the polarization angle during a single rotation of the neutron star.

For even stronger magnetic fields, the mode conversion tends to occur inside the $O$-mode photosphere. Although the reductions of the polarization fraction are still visible at the cyclotron energies of the proton for $B_p = 10^{14} \mathrm{G}$, even they are gone at $B_p=5 \times 10^{14} \mathrm{G}$, and the results are completely the same as those neglecting the mode conversion.

The phase-averaged polarization angles and fractions have been calculated for $B_p = 10^{13}, \ 5 \times 10^{13}$, and $10^{14} \mathrm{G}$. The mode conversion is important at $E = 5 \mathrm{keV}$ in the first two cases. For $B_p = 10^{13} \mathrm{G}$, the polarization fraction is reduced, and the polarization angle changes by $90^{\circ}$ by the mode conversion. For $B_p = 5 \times 10^{13} \mathrm{G}$, in contrast, the polarization fraction is lowered, because the observer sees not only the region in which the mode conversion takes place outside the photospheres of the two modes but also the region in which the mode conversion occurs between the two photospheres. A complicated pattern of the polarization angles in the $\eta$-$\gamma$ plane is also produced in this case. We have also demonstrated that the polarization angles and fractions depend strongly on the photon energy.

We have discussed the modifications that nonuniformities in the surface temperature may make in the polarization. In fact, we have considered the situation in which the thermal emissions are limited to the hot spots located at the magnetic poles. We have shown for $B_p= 5 \times 10^{14} \mathrm{G}$, at which the mode conversion occurs inside the $O$-mode photosphere, that the cancellation is somewhat relaxed, and the phase-averaged polarization fraction is increased in the vicinity of $\eta = \gamma = 45^{\circ}$ and $\eta = 90^{\circ}, \ \gamma = 20^{\circ}$. The smaller the spot is, the larger this effect becomes. The phase-averaged polarization angle also changes by $90^{\circ}$ in these parameter regions.

For $B_p = 10^{13} \mathrm{G}$ and $5 \times 10^{13} \mathrm{G}$, the mode conversion again becomes important. In fact, in the former case, the condition given in Equation (\ref{B_l}) is always satisfied at $E = 5 \mathrm{keV}$, and the mode conversion occurs except when $E < E_{\mathrm{ad}}$ holds because of large values of $\tan \theta_B$. We have found that as the hot-spot size becomes smaller, the latter condition is met only at some limited configurations, and the polarization fractions are raised in general. The phase-averaged polarization angle is further changed by the mode conversion in these cases.

At $B_p = 5 \times 10^{13} \mathrm{G}$, the neutron star surface is divided into two regions: the polar region, where the mode conversion occurs inside the $O$-mode photosphere and the $E$-mode photons are emitted, and the equatorial region, in which the mode conversion produces the $O$-mode photons. Since the two regions have nearly equal areas, the polarizations are canceled almost completely if the entire surface radiates these photons. In the presence of the hot spot, in contrast, we have demonstrated that the polarization fractions are increased, since the radiation is limited to the polar region and the cancellations tend to be suppressed. We have also shown that the polarization angles are little affected by the mode conversion.

We have finally considered four of the existing magnetars for which the magnetic field strength, surface temperature, and hot-spot size are estimated from observations. Realizing the possible caveats in our interpretation of the observations and modeling of the atmospheres of these magnetars, we have applied the theory to calculate the phase-averaged polarization angles and fractions for these objects. It is found that, under the assumption that our models are indeed applicable, the imprints of the mode conversion will manifest themselves only at low energies, $E \lesssim 1 \mathrm{keV}$, in magnetar 1E 2259+586, which has the lowest magnetic field strength, $B_p \sim 5 \times 10^{13} \mathrm{G}$, among the four, and that they will not be observed with the gas pixel detectors aboard {\it IXPE}~\citep{2013SPIE.8859E..08W}, {\it XIPE}~\citep{2016SPIE.9905E..15S} and {\it eXTP}~\citep{2016SPIE.9905E..1QZ}, which are all based on the photoelectric effect and of which operational energy is above 2keV. We will have to wait for polarimeters employing the Bragg reflections \citep{2013SPIE.8861E..1DM}.

In this paper, we ignore general relativistic effects such as ray bendings and possible modifications of dipole magnetic fields. Although they are likely to be minor, they have to be accounted for in the quantitative comparison with observations and the determination of the configuration of the neutron star thereby. We have also assumed for simplicity that the photons are all in the $E$-mode initially. In reality, there are some $O$-mode photons as well. In order to handle them properly, we need to solve the transport equations in the atmosphere of the neutron star. Circular polarizations that are entirely neglected in this paper are produced in principle \citep{2002PhRvD..66b3002H,2003PhRvL..91g1101L}, in addition to the linear polarizations considered in this paper. Hence, they have to be investigated quantitatively in the future. Although we have focused on the thermal emissions in this paper, nonthermal components are also known to exist at $E \lesssim 10 \mathrm{keV}$ in the magnetar radiation \citep{2007MNRAS.381..293R,2007Ap&SS.308..505R,2009MNRAS.396.2419R,2010ApJ...722L.162E,2014ApJ...789...75V}. Then, scatterings in the magnetosphere should be taken into account in considering these emissions \citep{2008MNRAS.386.1527N,2011ApJ...730..131F,2014MNRAS.438.1686T}. This is even more true at higher energies, $E \gtrsim 10 \mathrm{keV}$, where these nonthermal emissions are supposed to be dominant.

\vspace{10pt}

This work was supported by the Grants-in-Aid for the Scientific Research from the Ministry of Education, Culture, Sports, Science, and Technology (MEXT) of Japan (No. 16H03986) and MEXT Grant-in-Aid
for Scientific Research on Innovative Areas ``New Developments in Astrophysics Through Multi-Messenger Observations  of  Gravitational  Wave  Sources'' (Grant Number A05 24103006).

\appendix

\section{Derivation of Parameters in Equations (2) and (3)} \label{derivation_of_parameters}

We start with the Euler-Heisenberg Lagrangian
\begin{eqnarray}
 \mathcal{L} &=& - \frac{I}{4} + \frac{e^2}{8 \pi ^2} \int ^{\infty} _{0} \frac{d \eta}{\eta ^3} e^{- \eta} \left[ i \eta ^2 \frac{\sqrt{-K}}{4} \frac{ \cos (X_+) + \cos (X_-)}{\cos (X_+) - \cos (X_-)} + B_Q ^2 + \frac{\eta ^2}{6} I \right] , 
\end{eqnarray}
in which $X_{\pm}$ are given as 
\begin{eqnarray}
 X_+ &=& \frac{\eta}{B_Q} \sqrt{- \frac{I}{2} + i \frac{\sqrt{-K}}{2}}, \\
 X_- &=& \frac{\eta}{B_Q} \sqrt{- \frac{I}{2} - i \frac{\sqrt{-K}}{2}},
\end{eqnarray}
with two Lorentz invariants: $I = 2 ( |\vector{B}|^2 - |\vector{E}|^2)$, $K = - ( 4 \vector{E} \cdot \vector{B} )^2$.
It contains the one-loop level of the quantum correction to the classical Lagrangian of
electrodynamics. 

Normalizing these invariants as $I_{\mathrm{N}} = I / B_Q ^2$ and $K_{\mathrm{N}} = K / B_Q ^4$,
we can rewrite the Euler-Heisenberg Lagrangian as
 \begin{eqnarray}
  \mathcal{L} &=& - \frac{I_{\mathrm{N}} B_Q ^2}{4} + \frac{e^2}{8 \pi ^2} \int ^{\infty} _{0} \frac{d \eta}{\eta ^3} e^{- \eta} B_Q ^2  \left[ i \eta ^2 \frac{\sqrt{-K _{\mathrm{N}}}}{4} \frac{ \cos (X_{+\mathrm{N}}) + \cos (X_{-\mathrm{N}})}{\cos (X_{+\mathrm{N}}) - \cos (X_{-\mathrm{N}})} + 1 + \frac{\eta ^2}{6} I_{\mathrm{N}} \right] ,
 \end{eqnarray}
where $X_{+\mathrm{N}}$ and $X_{-\mathrm{N}}$ are $X_+$ and $X_-$ expressed in terms of
$I_\mathrm{N}$ and $K_\mathrm{N}$, respectively.
Then, the parameters $q$ and $m$ are derived from this form of the Lagrangian as
\begin{eqnarray}
q = - \frac{32}{B_Q ^4} \left. \frac{\partial \mathcal{L}}{\partial K_{\mathrm{N}}} \right|_{K_{\mathrm{N}}=0}, \ m = - \frac{16}{B_Q ^4} \left. \frac{\partial ^2 \mathcal{L}}{\partial I _{\mathrm{N}} ^2} \right|_{K_{\mathrm{N}}=0} .
\end{eqnarray}
Note that there is an alternative expression that appears different but is actually equivalent:
\begin{eqnarray}
 m = \int ^{\infty} _{0} d \eta \frac{e^2 e^{- \eta}}{8 b \pi ^2 \eta ^2} \left\{ 2 b^2 \eta^2 \frac{1}{\tanh ^3 ( b \eta)} - b \eta \frac{1}{\sinh ^2 ( b \eta)} - ( 1 + 2 b^2 \eta ^2 ) \frac{1}{\tanh (b \eta )} \right\} .
\end{eqnarray}
This can be obtained by using some formulae for the hyperbolic functions.

\providecommand{\noopsort}[1]{}\providecommand{\singleletter}[1]{#1}%

\end{document}